\documentclass[11pt]{article}
\pdfoutput=1
\usepackage[cp1250]{inputenc}
\usepackage{amsmath} 
\usepackage{amssymb} 
\usepackage{a4wide} 
\usepackage[left=3cm]{geometry} 
\usepackage{braket}
\usepackage{slashed}
\usepackage{bbm}
\usepackage{yfonts}
\usepackage{simplewick}
\usepackage{hyperref}
\usepackage[boxsize=0.7em,centertableaux]{ytableau}
\usepackage{empheq}
\usepackage{tikz}
\usetikzlibrary{matrix,arrows}
\usetikzlibrary{decorations.markings, decorations.pathreplacing}

\numberwithin{equation}{section}

\DeclareMathOperator{\Tr}{Tr}
\DeclareMathOperator{\ad}{ad}

\DeclareMathOperator{\res}{res}

\DeclareMathOperator{\rows}{rows}

\DeclareMathOperator{\hook}{hook}

\def\normord#1{\mathop{:}\nolimits\!#1\!\mathop{:}\nolimits}

\title{$\mathcal{W}$-symmetry, topological vertex and affine Yangian}
\author{Tomáš Procházka}

\begin{document}

\bibliographystyle{hieeetr}

\vskip 2.1cm

\centerline{\large \bf $\mathcal{W}$-symmetry, topological vertex and affine Yangian}
\vspace*{8.0ex}

\centerline{\large \rm Tom\'{a}\v{s} Proch\'{a}zka\footnote{Email: {\tt tomas.prochazka at lmu.de}}}

\vspace*{8.0ex}

\centerline{\large \it Arnold Sommerfeld Center for Theoretical Physics}
\centerline{\large \it Ludwig Maximilian University of Munich}
\centerline{\large \it Theresienstr. 37, D-80333 München, Germany}
\vspace*{2.0ex}
\centerline{\large \it Institute of Physics AS CR}
\centerline{\large \it Na Slovance 2, Prague 8, Czech Republic}
\vspace*{2.0ex}

\vspace*{6.0ex}

\centerline{\bf Abstract}
\bigskip

We discuss the representation theory of the non-linear chiral algebra $\mathcal{W}_{1+\infty}$ of Gaberdiel and Gopakumar and its connection to the Yangian of $\widehat{\mathfrak{u}(1)}$ whose presentation was given by Tsymbaliuk. The characters of completely degenerate representations of $\mathcal{W}_{1+\infty}$ are given by the topological vertex. The Yangian picture provides an infinite number of commuting charges which can be explicitly diagonalized in $\mathcal{W}_{1+\infty}$ highest weight representations. Many properties that are difficult to study in the $\mathcal{W}_{1+\infty}$ picture turn out to have a simple combinatorial interpretation, once translated to the Yangian picture.

 \vfill \eject

\tableofcontents

\setcounter{footnote}{0}

\newpage

\section{Introduction}
$\mathcal{W}$-algebras are extensions of the Virasoro algebra by currents of higher spin \cite{Zamolodchikov:1985wn, Bouwknegt:1992wg}. The most studied class of these is the family of algebras $\mathcal{W}_N$ obtained by extending the Virasoro algebra by currents of spin $3, 4, \ldots, N$ \cite{Fateev:1987zh, Lukyanov:1990tf}. They appear in various contexts in mathematical physics, for example in integrable hierarchies \cite{Gelfand:1995qu, Mathieu:1988pm}, matrix models \cite{Morozov:1994hh}, instanton counting and AGT correspondence \cite{Alday:2009aq, Mironov:2009by, Wyllard:2009hg} and recently the asymptotic symmetry algebras in $AdS_3/CFT_2$ holographic duality \cite{Campoleoni:2010zq, Henneaux:2010xg, Gaberdiel:2011wb, Campoleoni:2011hg}.

The original approach of Zamolodchikov \cite{Zamolodchikov:1985wn} for $\mathcal{W}_3$ which was then extended to $\mathcal{W}_4$ in \cite{Blumenhagen:1990jv,Kausch:1990bn} was to solve the OPE associativity conditions directly. Apart from that, most of the constructions of these algebras are indirect. There is a construction of $\mathcal{W}_N$ using the free field representation -- the Miura transform \cite{Fateev:1987zh, Lukyanov:1990tf} -- and constructions coming from the affine Lie algebra $\widehat{\mathfrak{su}(N)}$: the Casimir construction \cite{Bais:1987dc} and its generalization -- the coset construction \cite{Bais:1987zk} -- and the quantized Drinfeld-Sokolov reduction \cite{Bershadsky:1989mf, FigueroaO'Farrill:1990dz, Feigin:1990pn}.

Except for very special values of parameters, these algebras are non-linear, so they are not Lie algebras, except for the linear $\mathcal{W}_{\infty}$ and $\mathcal{W}_{1+\infty}$ algebras\footnote{In the older literature the linear $\mathcal{W}_{\infty}$ algebras were the only $\mathcal{W}_{\infty}$ algebras that were known. Here we use the adjective \emph{linear} when specializing to these.} constructed in \cite{Pope:1989sr,Pope:1990kc}. The linear $\mathcal{W}_{1+\infty}$ was studied further in \cite{Kac:1993zg,Kac:1995sk} generalizing the construction of the Virasoro algebra as a central extension of the algebra of vector fields on a circle to the algebra of all finite order differential operators on a circle.

Recently, there has been a renewed interest in $\mathcal{W}$-algebras which lead to a significant progress in their understanding. First, in the context of AGT correspondence \cite{Alday:2009aq}, Nekrasov instanton partition function \cite{Nekrasov:2002qd} in four-dimensional supersymmetric $SU(2)$ gauge theory was found to be related to conformal blocks in the two-dimensional Liouville theory. This was generalized to $SU(N)$ groups in \cite{Wyllard:2009hg}. From the mathematical point of view one is studying various algebraic structures acting on the equivariant cohomology of the instanton moduli spaces \cite{nakajima1994instantons}. In \cite{schifvas} a deformation of the linear $\mathcal{W}_{\infty}$ was obtained by deforming the $N\to\infty$ limit of the spherical degenerate double affine Hecke algebras (DDAHA) of $GL(N)$. The resulting family of algebras interpolates between algebras of $\mathcal{W}_N$ series. The equivariant cohomology of instanton moduli spaces was also studied in \cite{Maulik:2012wi} using the Yangian $\mathcal{Y}$ associated to Heisenberg algebra $\widehat{\mathfrak{gl}(1)}$. An explicit presentation of this algebra was later given in \cite{tsymbaliukrev}. This presentation is closely related to presentation of $q$-deformed version of $\mathcal{Y}$ whose variants go under names of Ding-Iohara algebra, elliptic Hall algebra, quantum continuous $\mathfrak{gl}(\infty)$ or the quantum toroidal algebra of type $\mathfrak{gl}(1)$ \cite{Ding:1996mq, Miki, feigin:2009, feigin:2011a, feigin:2011b, feigin:2011c, schifvas:2011, schifvas:2013, feigin2012quantum}. Many of the properties of $\mathcal{Y}$ were discovered in this $q$-deformed context first. One nice feature of these algebras is that they usually have a preferred infinite-dimensional abelian subalgebra, which acts diagonally in a large class of representations. It is not so difficult to find an infinite set of commuting charges also in the Virasoro algebra or in $\mathcal{W}_{3}$ \cite{Bazhanov:1994ft, Bazhanov:2001xm}, but it seems that for an explicit and simple diagonalization of these charges one needs to add a $\widehat{\mathfrak{u}(1)}$ current algebra \cite{Alba:2010qc} and thus consider $\mathcal{W}_{1+\infty}$.

In somewhat unrelated context of $AdS_3/CFT_2$ duality, Gaberdiel and Gopakumar \cite{Gaberdiel:2010pz,Gaberdiel:2011zw,Gaberdiel:2012ku} were also studying $\mathcal{W}_N$ series of algebras. Their motivation was to find a holographic dual description of three-dimensional higher spin theories. In three spacetime dimensions, the cosmological Einstein gravity can be reformulated as $SL(2)$ Chern-Simons theory \cite{Achucarro:1987vz, Witten:1988hc}. By extending the gauge group from $SL(2)$ to a larger gauge group one introduces interacting higher spin fields \cite{Blencowe:1988gj,Campoleoni:2010zq}. On geometries with a conformal boundary like for example the anti de Sitter spacetime there is a notion of asymptotic symmetry algebra introduced in \cite{Brown:1986nw} which in this case leads to the Virasoro algebra. Carrying out a similar analysis in the case of three dimensional cosmological higher spin gravity with gauge group $SL(N)$ one arrives at classical (non-linear) $\mathcal{W}_N$ algebra \cite{Campoleoni:2010zq,Henneaux:2010xg,Gaberdiel:2011wb,Campoleoni:2011hg}.

The Lie algebras $\mathfrak{sl}(N)$ defined for a positive integer $N$ have an associated interpolating family of algebras called $\mathfrak{hs}(\lambda)$ and defined for any $\lambda \in \mathbbm{C}$. One can think of them either as a one-parametric deformation of $N\to\infty$ limit of $\mathfrak{sl}(N)$ or as a deformation quantization of the algebra of the spherical harmonics on $S^2$. An important property of this family is the reduction to $\mathfrak{sl}(N)$: choosing $\lambda$ to be a positive integer $N$, the algebra $\mathfrak{hs}(N)$ develops an infinite-dimensional ideal and quotienting it out gives us $\mathfrak{sl}(N)$.

There is an analogous interpolating algebra for $\mathcal{W}_N$ series of algebras: the two-parametric non-linear $\mathcal{W}_{\infty}[\lambda,c]$ of \cite{Gaberdiel:2012ku}. It is a deformation of the linear $\mathcal{W}_{\infty}$ and it interpolates between all $\mathcal{W}_N$. It can be constructed by extending the Virasoro algebra by currents of spin $3,4,\ldots$. The most general solution of associativity conditions under assumptions of \cite{Gaberdiel:2012ku} is parametrized by two parameters: the rank-like parameter $\lambda$ and the central charge $c$. Considering the whole family at once instead of individual algebras $\mathcal{W}_N$ reveals an interesting discrete $\mathcal{S}_3$ symmetry called triality in \cite{Gaberdiel:2012ku}: for a fixed value of the central charge $c$, there are generically three different values of $\lambda_j \in \mathbbm{C}$ for which the algebras $\mathcal{W}_{\infty}[\lambda_j]$ are isomorphic. The three-fold symmetry shows up also in representation theory of $\mathcal{W}_{\infty}$. As one example, extending $\mathcal{W}_{\infty}$ to $\mathcal{W}_{1+\infty}$ by adding a $\widehat{\mathfrak{u}(1)}$ current, one finds that the vacuum character of $\mathcal{W}_{1+\infty}$ is equal to MacMahon function, which is the counting function of plane partitions (3 dimensional Young diagrams). The $\mathcal{S}_3$ acts in the space of plane partitions by permuting the coordinate axes.

We are now in a situation that we have two algebras, on the one hand the non-linear chiral algebra $\mathcal{W}_{1+\infty}$ of Gaberdiel and Gopakumar \cite{Gaberdiel:2012ku} and on the other hand the Yangian of $\widehat{\mathfrak{gl}(1)}$ of Maulik, Okounkov and Tsymbaliuk \cite{Maulik:2012wi,tsymbaliukrev} together with the Schiffmann-Vasserot algebra \cite{schifvas}. They share many properties: they are deformations of linear $\mathcal{W}_{1+\infty}$, they interpolate between $\mathcal{W}_N$ family of algebras and they have representations combinatorially related to plane partitions. It is natural to assume that they have not only related representation theory, but in fact that they represent the same structure. The goal of this work is to study this connection explicitly and use the properties that are easy to understand on the Yangian side to learn more about the representation theory of $\mathcal{W}_{1+\infty}$. The connection between Yangians and $\mathcal{W}$-algebras is something that is already known in the context of finite $\mathcal{W}$-algebras and classical Yangians, see \cite{Ragoucy:1998bv, Brundan:2004ca}.

\paragraph{Overview of this article}
In the section \ref{secyangian} we give presentation of the affine Yangian $\mathcal{Y}$ following \cite{tsymbaliukrev} and study its basic properties. The mode grading and spin filtration of $\mathcal{W}_{1+\infty}$ are found in $\mathcal{Y}$, followed by identification of the $\widehat{\mathfrak{u}(1)}$ and Virasoro subalgebras. The next section focuses on linear $\mathcal{W}_{1+\infty}$. Since this algebra is linear, many computations simplify and the mapping between linear $\mathcal{W}_{1+\infty}$ and $\mathcal{Y}$ can be described very explicitly. There is a simple representation of linear $\mathcal{W}_{1+\infty}$ on a pair of free fermions, which can be translated to a free fermionic representation of $\mathcal{Y}$. Bosonizing this representation we find a representation of $\mathcal{Y}$ on one free boson which is the basic building block of all other representations considered in this article.

The section \ref{secyangrep} studies representations of the Yangian. First, the translation of the usual highest weight representations of $\mathcal{W}_{1+\infty}$ is given, followed by a discussion of the quasi-finite representations. In the next part, the vacuum representation of $\mathcal{Y}$ is found. In section \ref{secplanepart}, the representations of $\mathcal{Y}$ on plane partitions and their generalizations are discussed and finally the free boson representation of $\mathcal{Y}$ for arbitrary values of parameters is given.

The last section focuses of $\mathcal{W}_{1+\infty}$ representations. First of all, by studying the characters of completely degenerate representations of $\mathcal{W}_{N}$ as $N \to \infty$, one finds that they correspond to a specialization of the topological vertex from the topological string theory to two non-trivial Young diagram asymptotics out of possible three. Together with the triality symmetry, one is lead to conjecture that there is an interesting class of representations of $\mathcal{W}_{1+\infty}$ parametrized by a triple of Young diagrams, whose character is given precisely by the topological vertex. Combinatorially, given an arbitrary triple of Young diagrams we can construct a minimal generalized plane partition with given asymptotics (this generalized plane partition has an infinite number of boxes unless the Young diagrams are trivial). This configuration is mapped to a primary state in $\mathcal{W}_{1+\infty}$ labelled by the same triple of Young diagrams. Descendants of this primary state combinatorially correspond to configurations obtained by adding a finite number of boxes to the minimal configuration, the number of boxes added being the level of a state. Next we shortly discuss the connection between $\mathcal{Y}$ and Schiffmann-Vasserot algebra. The following part discusses the explicit mapping between higher spin charges of the primary state in $U$-basis \cite{Prochazka:2014gqa} and Yangian $\psi_j$ charges of the same state. Next we discuss the free boson representations and constructs an explicit map between $\mathcal{Y}$ and $\mathcal{W}_{1+\infty}$. An explicit example of the representation $(\Box,\Box,\Box)$ is considered to illustrate how everything fits together. In the final part we discuss the specializations of the algebra, the null states in the vacuum representation and the irreducible characters of the Lee-Yang minimal model.

\section{Affine Yangian of $\mathfrak{gl}(1)$}
\label{secyangian}

In this section we introduce the algebra $\mathcal{Y}$, the Yangian of $\widehat{\mathfrak{gl}(1)}$ following \cite{tsymbaliukrev} and we study its basic properties.

In order to find a map between $\mathcal{W}_{1+\infty}$ and $\mathcal{Y}$, the zeroth step is to understand an analogue of grading of $\mathcal{W}_{1+\infty}$ by mode and filtration by spin. $\mathcal{Y}$ has also a rescaling invariance that is absent in $\mathcal{W}_{1+\infty}$ so it must be fixed before comparing $\mathcal{W}_{1+\infty}$ and $\mathcal{Y}$. As usual in Yangians in general, there is a certain automorphism of the algebra which is related to the shift of the spectral parameter. The corresponding symmetry of $\mathcal{W}_{1+\infty}$ is the trivial shift of $\widehat{\mathfrak{u}(1)}$ zero mode.

In both $\mathcal{Y}$ and $\mathcal{W}_{1+\infty}$, the coproduct lets us construct representations of algebras on $N$ free bosons starting from representations on one free boson. Because of this property, it plays a crucial role in understanding the map between $\mathcal{Y}$ and $\mathcal{W}_{1+\infty}$.

The last part of this section identifies the $\widehat{\mathfrak{u}(1)}$ and the Virasoro subalgebra of $\mathcal{Y}$. In $\mathcal{W}_{1+\infty}$ these subalgebras are the starting point of the construction of the algebra, while $\mathcal{Y}$ in Tsymbaliuk presentation contains them in more complicated way. Although we don't show that all $\widehat{\mathfrak{u}(1)}$ or Virasoro commutation relations are satisfied, we verify many of these and find the mapping between parameters $\lambda_j$ of $\mathcal{W}_{1+\infty}$ and $h_j$ parameters of $\mathcal{Y}$ and in particular find the expression for the central charge $c$ in terms of $h_j$ parameters. The computations of this section illustrate the important role of the Serre relations (Y6) and (Y7).

\subsection{Presentation}
The affine Yangian of $\mathfrak{gl}(1)$ is an associative algebra with generators $e_j, f_j$ and $\psi_j$, $j = 0, 1, \ldots$ and relations \cite{tsymbaliukrev, tsymbaliuk2014affine}
\begin{align}
0 & = \left[ \psi_j, \psi_k \right] \tag{Y0} \\
\nonumber
0 & = \left[ e_{j+3}, e_k \right] - 3 \left[ e_{j+2}, e_{k+1} \right] + 3\left[ e_{j+1}, e_{k+2} \right] - \left[ e_j, e_{k+3} \right] \\
& \quad + \sigma_2 \left[ e_{j+1}, e_k \right] - \sigma_2 \left[ e_j, e_{k+1} \right] - \sigma_3 \left\{ e_j, e_k \right\} \tag{Y1} \\
\nonumber
0 & = \left[ f_{j+3}, f_k \right] - 3 \left[ f_{j+2}, f_{k+1} \right] + 3\left[ f_{j+1}, f_{k+2} \right] - \left[ f_j, f_{k+3} \right] \\
& \quad + \sigma_2 \left[ f_{j+1}, f_k \right] - \sigma_2 \left[ f_j, f_{k+1} \right] + \sigma_3 \left\{ f_j, f_k \right\} \tag{Y2} \\
0 & = \left[ e_j, f_k \right] - \psi_{j+k} \tag{Y3} \\
\nonumber
0 & = \left[ \psi_{j+3}, e_k \right] - 3 \left[ \psi_{j+2}, e_{k+1} \right] + 3\left[ \psi_{j+1}, e_{k+2} \right] - \left[ \psi_j, e_{k+3} \right] \\
& \quad + \sigma_2 \left[ \psi_{j+1}, e_k \right] - \sigma_2 \left[ \psi_j, e_{k+1} \right] - \sigma_3 \left\{ \psi_j, e_k \right\} \tag{Y4} \\
\nonumber
0 & = \left[ \psi_{j+3}, f_k \right] - 3 \left[ \psi_{j+2}, f_{k+1} \right] + 3\left[ \psi_{j+1}, f_{k+2} \right] - \left[ \psi_j, f_{k+3} \right] \\
& \quad + \sigma_2 \left[ \psi_{j+1}, f_k \right] - \sigma_2 \left[ \psi_j, f_{k+1} \right] + \sigma_3 \left\{ \psi_j, f_k \right\} \tag{Y5}
\end{align}
together with ``boundary'' conditions
\begin{align}
\left[ \psi_0, e_j \right] & = 0, & \left[ \psi_1, e_j \right] & = 0, & \left[ \psi_2, e_j \right] & = 2 e_j \tag{Y4'} \\
\left[ \psi_0, f_j \right] & = 0, & \left[ \psi_1, f_j \right] & = 0, & \left[ \psi_2, f_j \right] & = -2f_j \tag{Y5'}
\end{align}
and a generalization of Serre relations
\begin{align}
\mathrm{Sym}_{(j_1,j_2,j_3)} \left[ e_{j_1}, \left[ e_{j_2}, e_{j_3+1} \right] \right] & = 0 \tag{Y6} \\
\mathrm{Sym}_{(j_1,j_2,j_3)} \left[ f_{j_1}, \left[ f_{j_2}, f_{j_3+1} \right] \right] & = 0 \tag{Y7}.
\end{align}
Here $\mathrm{Sym}$ is the complete symmetrization over all indicated indices ($6$ terms).

The algebra is parametrized by three complex numbers $h_1, h_2$ and $h_3$ constrained by
\begin{equation}
\label{sigma1zero}
\sigma_1 \equiv h_1+h_2+h_3=0
\end{equation}
and these parameters appear symmetrically in the definition of the algebra through elementary symmetric polynomials
\begin{equation}
\sigma_2 = h_1 h_2 + h_1 h_3 + h_2 h_3 \quad \quad \quad \mathrm{and} \quad \quad \quad \sigma_3 = h_1 h_2 h_3.
\end{equation}

Let us emphasize that we define $\mathcal{Y}$ as an associative algebra. The defining commutation relations involve both commutators and anticommutators and cubic Serre relations. These anticommutators are not associated to any $\mathbbm{Z}_2$ grading. In certain situations (see example the linear case discussed in the section \ref{seclinear}) we will consider a Lie algebra whose universal enveloping algebra will be a certain specialization of $\mathcal{Y}$. In this sense, we can think of $\mathcal{Y}$ as being a deformation of the universal enveloping algebra of linear $\mathcal{W}_{1+\infty}$.

\paragraph{Generating functions}
It is useful to combine the generators into the generating fields which will not only simplify the defining relations of $\mathcal{Y}$ but also let us understand some properties of $\mathcal{Y}$ more easily. Following Tsymbaliuk \cite{tsymbaliukrev, tsymbaliuk2014affine}, (see also \cite{Bourgine:2015szm}), we introduce the following generating functions:
\begin{eqnarray}
\nonumber
e(u) & = & \sum_{j=0}^{\infty} \frac{e_j}{u^{j+1}} \\
\label{genfunc}
f(u) & = & \sum_{j=0}^{\infty} \frac{f_j}{u^{j+1}} \\
\nonumber
\psi(u) & = & 1 + \sigma_3 \sum_{j=0}^{\infty} \frac{\psi_j}{u^{j+1}}.
\end{eqnarray}
Unlike in the usual situation in the conformal field theory, where the parameter of generating function has the interpretation of a space coordinate and the sum is over ``modes'' of the field, here the summation is over what would be the ``spins'' of the fields and the parameter $u$ has no simple space-time interpretation. In what follows we will call $u$ the spectral parameter.

Multiplying (Y1) and (Y2) by $u^{-j-1} v^{-k-1}$ and summing over $j$ and $k$, we find that the relations are equivalent to
\begin{align}
\nonumber
(u-v-h_1)(u-v-h_2)(u-v-h_3) e(u) e(v) & \sim (u-v+h_1)(u-v+h_2)(u-v+h_3) e(v) e(u) \\
\nonumber
(u-v+h_1)(u-v+h_2)(u-v+h_3) f(u) f(v) & \sim (u-v-h_1)(u-v-h_2)(u-v-h_3) f(v) f(u) \\
\nonumber
(u-v-h_1)(u-v-h_2)(u-v-h_3) \psi(u) e(v) & \sim (u-v+h_1)(u-v+h_2)(u-v+h_3) e(v) \psi(u) \\
\label{genfunsimrel}
(u-v+h_1)(u-v+h_2)(u-v+h_3) \psi(u) f(v) & \sim (u-v-h_1)(u-v-h_2)(u-v-h_3) f(v) \psi(u)
\end{align}
where we omit the lowest terms (involving $e_j, f_j$ or $\psi_j$ with $j \leq 2$) that are regular at $u=0$ or regular $v=0$. Because of these terms, we should be careful when deriving things directly from the commutation relations written in terms of generating functions (\ref{genfunsimrel}). See \cite{tsymbaliukrev} for an alternative way of writing the commutation relations in terms of the generating functions. Introducing a rational function parametrizing the algebra
\begin{empheq}[box=\fbox]{align}
\label{magicfunction}
\varphi(u) \equiv \frac{(u+h_1)(u+h_2)(u+h_3)}{(u-h_1)(u-h_2)(u-h_3)}
\end{empheq}
the previous relations can be written as
\begin{align}
\nonumber
e(u) e(v) & \sim \varphi(u-v) e(v) e(u) \\
\nonumber
f(u) f(v) & \sim \varphi(v-u) f(v) f(u) \\
\nonumber
\psi(u) e(v) & \sim \varphi(u-v) e(v) \psi(u) \\
\psi(u) f(v) & \sim \varphi(v-u) f(v) \psi(u)
\end{align}
Here the $\sim$ sign represents the equality up to terms omitted in (\ref{genfunsimrel}). If we interpreted $\sim$ sign as an equality, we would obtain stronger relations which do not hold in $\mathcal{Y}$.

The function (\ref{magicfunction}) seems to be the most non-trivial ingredient in $\mathcal{Y}$ and we will see its important role many times in what follows. The relation (Y3) can be written using the generating functions as \footnote{See also equation (2.6) in \cite{Bourgine:2015szm} for the analogous formula in the context of $SH^c$ algebra.}
\begin{equation}
e(u) f(v) - f(v) e(u) = -\frac{1}{\sigma_3} \frac{\psi(u)-\psi(v)}{u-v}.
\end{equation}
Finally, the relation (Y6) can be written in terms of generating fields as
\begin{equation}
\sum_{\pi \in S_3} (u_{\pi(1)} - 2u_{\pi(2)} + u_{\pi(3)}) e(u_{\pi(1)}) e(u_{\pi(2)}) e(u_{\pi(3)}) = 0
\end{equation}
and similarly for (Y7).

\subsection{Basic properties}
Let us now discuss the simplest properties of $\mathcal{Y}$ that we can see immediately from the defining relations.

\paragraph{Scaling symmetries}

There are two scaling symmetries of $\mathcal{Y}$, with charges given as follows:
\begin{center}
\begin{tabular}{|c|c|c|c|c|c|c|c|c|}
\hline
scaling & $\psi_j$ & $e_j$ & $f_j$ & $h_j$ & $u$ & $\psi(u)$ & $e(u)$ & $f(u)$ \\
\hline
$\alpha$ & $j-2$ & $j-1$ & $j-1$ & $1$ & $1$ & $0$ & $-2$ & $-2$ \\
\hline
$\beta$ & $0$ & $1$ & $-1$ & $0$ & $0$ & $0$ & $1$ & $-1$ \\
\hline
\end{tabular}
\end{center}
Having chosen the values of parameters $h_j$, we see that changing the value of the central element $\psi_0$ is equivalent to rescaling parameters $h_j$. We could have fixed this symmetry by choosing a suitable value of $\psi_0$, such as $\psi_0 = 1$, but in the following it will be convenient to consider different ways of fixing it. Fixing this scaling symmetry is necessary when comparing to $\mathcal{W}_{1+\infty}$ which does not have such a scaling symmetry.

The second scaling symmetry is less interesting. It is similar to the usual possibility of rescaling harmonic oscillator creation and annihilation operators and can be fixed if we introduce a bilinear form with respect to which these operators are conjugate.

We also have a discrete exchange symmetry
\begin{align}
\nonumber
e_j & \leftrightarrow f_j \\
\psi_j & \leftrightarrow -\psi_j \\
\nonumber
h_j & \leftrightarrow -h_j.
\end{align}

\paragraph{Central elements}
It is easy to see that the elements $\psi_0$ and $\psi_1$ are central. Because of the scaling symmetry $\alpha$, when comparing quantities with $\mathcal{W}_{1+\infty}$ algebra, what appears in expressions are the scaling-invariant combinations $\psi_0 \sigma_2$ and $\psi_0^3 \sigma_3^2$. This means that there is no invariant meaning of $\psi_0$ alone as it can be rescaled away. The $\psi_1$ will be later identified as the charge associated to $\widehat{\mathfrak{u}(1)}$ subalgebra.

\paragraph{Mode grading}
The algebra $\mathcal{Y}$ is naturally graded, with generators $e_j$ having degree $+1$, generators $\psi_j$ being of degree $0$ and $f_j$ having degree $-1$. The corresponding grading of $\mathcal{W}_{1+\infty}$ is the grading by (minus) the mode number.

\paragraph{Spin filtration}
Apart from the mode grading, there is also a filtration which assigns generators $\psi_j$, $e_j$ and $f_j$ the filtration degree $j$. This filtration corresponds roughly to filtration of $\mathcal{W}_{1+\infty}$ by spin. At the special point $h_j = 0$ the terms of different spin degree vanish and spin filtration becomes grading. This can be useful for example for the state counting.

\paragraph{Triangular decomposition}
There are several interesting subalgebras of $\mathcal{Y}$. The first, which we will denote by $\mathcal{Y}^+$ , is the so called \emph{shuffle algebra} studied by Negut \cite{Negut01012014} and is the subalgebra generated by generators $e_j$ and relations (Y1) and (Y6). Similarly, we have the conjugate shuffle subalgebra $\mathcal{Y}^-$ generated by $f_j$ and relations (Y2) and (Y7). The commutative subalgebra with generators $\psi_j$ is often called the Bethe or Baxter subalgebra $\mathcal{B}$ \cite{Maulik:2012wi}. When studying the representations of $\mathcal{Y}$, it will be the subalgebra that will act diagonally on the basis of the representation space. The whole algebra $\mathcal{Y}$ as a vector space admits a triangular decomposition
\begin{equation}
\label{triang}
\mathcal{Y} = \mathcal{Y}^+ \oplus \mathcal{B} \oplus \mathcal{Y}^-
\end{equation}
which will later be important when studying highest weight representations of $\mathcal{W}_{1+\infty}$. We also have two intermediate subalgebras, the subalgebra $\mathcal{Y}^+_0$ generated by $\mathcal{Y}^+$ and $\mathcal{B}$ and analogous subalgebra $\mathcal{Y}^-_0$ generated by $\mathcal{Y}^-$ and $\mathcal{B}$.

\subsection{Spectral shift}
\label{spectralshift}
$\mathcal{Y}$ admits the following automorphism of the algebra parametrized by $q \in \mathbbm{C}$:
\begin{align}
\nonumber
e_j & \to \sum_{k=0}^j {j \choose k} q^{j-k} e_k \\
\label{spectralshifteq}
f_j & \to \sum_{k=0}^j {j \choose k} q^{j-k} f_k \\
\nonumber
\psi_j & \to \sum_{k=0}^j {j \choose k} q^{j-k} \psi_k
\end{align}
In terms of the generating functions, this amounts to a shift in the spectral parameter,
\begin{align}
\nonumber
e(u) & \to e(u-q) \\
f(u) & \to f(u-q) \\
\nonumber
\psi(u) & \to \psi(u-q).
\end{align}
It is obvious that (Y0), (Y4') and (Y5') are invariant under this transformation. (Y3) is easily verified using the binomial identity
\begin{equation}
\sum_{j=0}^t {r \choose j} {s \choose t-j} = {r + s \choose t}
\end{equation}
Relations (Y1), (Y2), (Y4) and (Y5) can be also verified, in fact if we denote one of these relations $I(j,k)$, we find that under this automorphism
\begin{equation}
I(j,k) \to \sum_{l=0}^{\infty} q^l \sum_{m=0}^l {j \choose m} {k \choose l-m} I(j-m,k-l+m).
\end{equation}
We will see later that this automorphism in the language of $\mathcal{W}_{1+\infty}$ amounts to shift of the charge operator $J_0$ by a constant which is indeed an automorphism of $\widehat{\mathfrak{u}(1)}$ subalgebra, since $J_0$ is a central element that does not appear on the right hand side of any commutation relation. $\mathcal{W}_{1+\infty}$ admits splitting as $\widehat{\mathfrak{u}(1)} \times \mathcal{W}_{\infty}$ and the shift in $J_0$ leaves $\mathcal{W}_{\infty}$ invariant.

\subsection{Affine $\mathfrak{u}(1)$ and Virasoro subalgebras}
\label{secvirasoro}
Our main goal is to identify $\mathcal{Y}$ with the chiral algebra $\mathcal{W}_{1+\infty}$. By definition, $\mathcal{W}_{1+\infty}$ is obtained from the Virasoro algebra by adding primary generating fields of spin $3, 4, \ldots$ and spin $1$ field \cite{Gaberdiel:2012ku}. In order to identify the higher spin fields, one first needs to find and fix a Virasoro algebra. From the commutation relations (Y0)-(Y7) it is not at all obvious that $\mathcal{Y}$ contains $\widehat{\mathfrak{u}(1)}$ or Virasoro subalgebra. The purpose of this section is to map a set of elements generating $\widehat{\mathfrak{u}(1)} \oplus \mathfrak{Vir}$ to $\mathcal{Y}$ and verify that they satisfy the correct commutation relations (at least for lower mode numbers).

Let us start with the following ansatz for embedding of $\widehat{\mathfrak{u}(1)}$ and Virasoro subalgebra in $\mathcal{Y}$ that is compatible with the mode grading and spin filtration described above:
\begin{align}
\nonumber
J_0 & \leftrightarrow \psi_1 \\
\nonumber
J_{-1} & \leftrightarrow e_0 \\
\nonumber
J_{+1} & \leftrightarrow -f_0 \\
L_{-1} & \leftrightarrow e_1 + \alpha e_0 \\
\nonumber
L_{+1} & \leftrightarrow -f_1 - \alpha f_0 \\
\nonumber
L_{-2} & \leftrightarrow \frac{1}{2} \left[ e_2, e_0 \right] - \beta \frac{\sigma_3 \psi_0}{2} \left[ e_1, e_0 \right] \\
\nonumber
L_{+2} & \leftrightarrow -\frac{1}{2} \left[ f_2, f_0 \right] + \beta \frac{\sigma_3 \psi_0}{2} \left[ f_1, f_0 \right]
\end{align}
Here $\alpha$ and $\beta$ are so far free parameters that we want to fix. We will require the following commutation relations to be satisfied (in particular we require $J(z)$ to be a primary field)
\begin{align}
\nonumber
\left[ J_m, J_n \right] & = \kappa \, m \, \delta_{m+n,0} \\
\left[ L_m, L_n \right] & = (m-n) L_{m+n} + \frac{(m-1)m(m+1)}{12} c\, \delta_{m_n,0} \\
\nonumber
\left[ L_m, J_n \right] & = -n J_{m+n}
\end{align}
The simplest commutator that we can take is
\begin{equation}
\psi_0 = \left[e_0, f_0 \right] = -\left[ J_{-1}, J_{+1} \right] = \kappa
\end{equation}
that lets us identify the normalization constant of $J_n$ generators $\kappa$ with the central element $\psi_0$. Next we can determine $L_0$,
\begin{equation}
2L_0 = \left[ L_{+1}, L_{-1} \right] = \psi_2 + 2\alpha \psi_1 + \alpha^2 \psi_0.
\end{equation}
Now we can determine $J_{\pm 2}$
\begin{align}
\nonumber
J_{-2} & = \left[ L_{-1}, J_{-1} \right] = \left[ e_1, e_0 \right] \\
J_{+2} & = \left[ J_{+1}, L_{+1} \right] = \left[ f_0, f_1 \right]
\end{align}
and $J_{\pm 3}$,
\begin{align}
\nonumber
J_{-3} & = \left[ L_{-2}, J_{-1} \right] = \frac{1}{2} \left[ e_0, \left[ e_0, e_2 \right] \right] \\
J_{+3} & = \left[ J_{+1}, L_{+2} \right] = -\frac{1}{2} \left[ f_0, \left[ f_0, f_2 \right] \right].
\end{align}
The term proportional to $\beta$ vanishes due to Serre relations (Y6) and (Y7) with $j_1 = j_2 = j_3 = 0$,
\begin{equation}
\left[ J_{-1}, J_{-2} \right] = \left[ e_0, \left[ e_1, e_0 \right] \right] = 0.
\end{equation}
We could have also determined $J_{\pm 3}$ using
\begin{equation}
J_{-3} = \frac{1}{2} \left[ L_{-1}, J_{-2} \right] = \frac{1}{2} \left[ e_1, \left[ e_1, e_0 \right] \right]
\end{equation}
which is consistent with the previous result thanks to (Y6) with $j_1 = 1, j_2 = j_3 = 0$:
\begin{equation}
0 = \left[ e_0, \left[ e_0, e_2 \right] \right] + \left[ e_1, \left[ e_0, e_1 \right] \right]
\end{equation}

\paragraph{Determination of parameters $\alpha, \beta$}
To find the value of $\beta$, we compute
\begin{equation}
-2J_0 = \left[ L_{-2}, J_{+2} \right] = -2\psi_1 - \sigma_3 \psi_0^2 + \beta \sigma_3 \psi_0^2
\end{equation}
where we also used
\begin{align}
\label{psi3ej}
\nonumber
\left[ \psi_3, e_j \right] & = 6e_{j+1} + 2\sigma_3 \psi_0 e_j \\
\left[ \psi_3, f_j \right] & = -6f_{j+1} - 2\sigma_3 \psi_0 f_j
\end{align}
We see that the compatibility with the previous identifications requires taking $\beta = 1$. Considering now the commutator
\begin{equation}
3e_1 + 3\alpha e_0 = 3 L_{-1} = \left[ L_1, L_{-2} \right] = 3 e_1 + \alpha e_0 + (1-\beta) \sigma_3 \psi_0 e_0
\end{equation}
we see that if $\beta = 1$, we must choose $\alpha = 0$ in order to satisfy the commutation relations of the Virasoro algebra.

\paragraph{Central charge}
Now we can determine the central charge of the Virasoro algebra \footnote{See also a derivation of the central charge in the context of $SH^c$ in appendix C of \cite{Kanno:2013aha}.} from
\begin{equation}
4 L_0 + \frac{c}{2} = \left[ L_2, L_{-2} \right] = 2\psi_2 - \frac{1}{2} \sigma_2 \psi_0 - \frac{1}{2} \sigma_3^2 \psi_0^3
\end{equation}
using
\begin{align}
\nonumber
\left[ \psi_4, e_j \right] & = 12 e_{j+2} + 6 \sigma_3 \psi_0 e_{j+1} - 2 \sigma_2 e_j + 2\sigma_3 \psi_1 e_j \\
\left[ \psi_4, f_j \right] & = -12 f_{j+2} - 6 \sigma_3 \psi_0 f_{j+1} + 2 \sigma_2 f_j - 2\sigma_3 \psi_1 f_j.
\end{align}
We can thus read off the central charge,
\begin{equation}
c = -\sigma_2 \psi_0 -\sigma_3^2 \psi_0^3
\end{equation}
This can be identified with $c_{1+\infty}$ of $\mathcal{W}_{1+\infty}$ \cite{Prochazka:2014gqa}
\begin{equation}
c_{1+\infty} = 1 + (\lambda_1-1)(\lambda_2-1)(\lambda_3-1)
\end{equation}
if we identify the parameters via
\begin{empheq}[box=\fbox]{align}
\label{lambdatoh}
\lambda_1 = -\psi_0 h_2 h_3, \quad \quad \lambda_2 = -\psi_0 h_1 h_3, \quad \quad \lambda_3 = -\psi_0 h_1 h_2.
\end{empheq}
Note that this is invariant under the scaling symmetry $\alpha$ of $\mathcal{Y}$, which is consistent with the fact that $\mathcal{W}_{1+\infty}$ has no such symmetry so $\lambda_j$ should be scaling invariant. We are not claiming that the map between parameters (\ref{lambdatoh}) follows uniquely from the identification of the central charge (since we actually need to match $2$ parameters, $c$ and $N$, in order to have such a map), but the identification (\ref{lambdatoh}) is the one that is also consistent with the structure of null states in Verma modules as we will see later in section \ref{specializations}.

\paragraph{Identification of $L_{-3}$}
The generator $L_{-3}$ can be obtained from
\begin{equation}
L_{-3} = \left[ L_{-1}, L_{-2} \right] = \frac{1}{2} \left[ e_1, \left[ e_2, e_0 \right] \right] - \frac{\sigma_3 \psi_0}{2} \left[ e_1, \left[ e_1, e_0 \right] \right]
\end{equation}
The second term is proportional to $J_{-3}$ while the first term can be rewritten in other ways. From (Y6) with $(j_1,j_2,j_3)=(0,1,1)$ we find
\begin{equation}
\left[ e_0, \left[ e_1, e_2 \right] \right] = \left[ e_1, \left[ e_2, e_0 \right] \right].
\end{equation}
This combined with the Jacobi identity for $(j_1,j_2,j_3)=(0,1,2)$ implies
\begin{equation}
\left[ e_2, \left[ e_0, e_1 \right] \right] = -2 \left[ e_0, \left[ e_1, e_2 \right] \right] = -2 \left[ e_1, \left[ e_2, e_0 \right] \right].
\end{equation}
Using (Y6) for $(j_1,j_2,j_3)=(0,0,2)$ we may verify that this nested commutator is in fact proportional to $[e_0,[e_0,e_3]]$.

\paragraph{Commutator $\left[J_{-1},J_{-3}\right]$}
Using the Jacobi identity we see that
\begin{equation}
\left[ J_{-1}, J_{-3} \right] = \frac{1}{2} \left[ e_0, \left[ e_1, \left[ e_1, e_0 \right] \right] \right] = - \frac{1}{2} \left[ e_1, \left[ \left[ e_1, e_0 \right], e_0 \right] \right] - \frac{1}{2} \left[ \left[ e_1, e_0 \right], \left[ e_0, e_1 \right] \right] = 0.
\end{equation}
The first term on the right-hand side vanishes as follows from (Y6), while the second term vanishes because of the antisymmetry of the commutator.

\paragraph{Evaluation of $J_{-4}$}
As a last example of the computations of this section, we show that $J_{-4}$ obtained in various ways gives the same result after using the relations (Y6) and Jacobi identities. There are now three different ways to get $J_{-4}$,
\begin{align}
\nonumber
J_{-4} & = \frac{1}{3} \left[ L_{-1}, J_{-3} \right] = \frac{1}{6} \left[ e_1, \left[ e_1, \left[ e_1, e_0 \right] \right] \right] \\
& = \frac{1}{2} \left[ L_{-2}, J_{-2} \right] = \frac{1}{4} \left[ \left[ e_2, e_0 \right], \left[ e_1, e_0 \right] \right] \\
\nonumber
& = \left[ L_{-3}, J_{-1} \right] = \frac{1}{2} \left[ \left[ e_1, \left[ e_2, e_0 \right] \right], e_0 \right].
\end{align}
Using the Jacobi identity once in the second line
\begin{equation}
\left[\left[ e_2, e_0 \right], \left[ e_1, e_0 \right] \right] = - \left[ \left[ e_0, \left[ e_1, e_0 \right] \right], e_2 \right] - \left[ \left[ \left[ e_1, e_0 \right], e_2 \right], e_0 \right] = 2 \left[ \left[ \left[ e_0, e_2 \right], e_1 \right], e_0 \right]
\end{equation}
we arrive at the third line. Using the Jacobi identity in the other way, we find
\begin{equation}
\left[ \left[ e_2, e_0 \right], \left[ e_1, e_0 \right] \right] = \left[ e_1, \left[ e_1, \left[ e_1, e_0 \right] \right] \right] + \left[ \left[ \left[ e_2, e_0 \right], e_1 \right], e_0 \right]
\end{equation}
and
\begin{equation}
\left[ \left[ e_2, e_0 \right], \left[ e_1, e_0 \right] \right] = \frac{2}{3} \left[ e_1, \left[ e_1, \left[ e_1, e_0 \right] \right] \right]
\end{equation}
from which the equality of first two expressions for $J_{-4}$ follows.

\paragraph{Higher mode numbers} At this point is should be clear that using the defining relations of $\mathcal{Y}$ (especially the Serre relations (Y6) and (Y7)) becomes quite impractical as the mode number increases. We can write down a closed-form expression for Heisenberg modes,
\begin{align}
\label{jmad}
\nonumber
J_{-m} & = \frac{1}{(m-1)!} \ad_{e_1}^{m-1} e_0 \quad \quad m \geq 1 \\
J_{+m} & = -\frac{1}{(m-1)!} \ad_{f_1}^{m-1} f_0 \quad \quad m \geq 1,
\end{align}
but to prove that $\left[ J_m, J_n \right] = 0$ for $m,n > 0$ seems to be difficult and one has to deal with complicated nested commutators. This is one of disadvantages of the Yangian presentation of the algebra. The purpose of this section was to illustrate the non-trivial interplay between various defining relations of $\mathcal{Y}$ and especially the role of the cubic Serre relations (Y6) and (Y7). The main difficulty lies in the subalgebras $\mathcal{Y}^+$ and $\mathcal{Y}^-$. The commutation relations (Y0), (Y3), (Y4), (Y4'), (Y5) and (Y5') are strong enough to reduce any commutator in $\mathcal{Y}$ to a computation in shuffle subalgebras $\mathcal{Y}^+$ or $\mathcal{Y}^-$. It is the relations (Y1) and (Y6) in $\mathcal{Y}^+$ and their analogue in $\mathcal{Y}^-$ that cause most of the difficulties.

\subsection{Coproduct}
In the following the existence of coproduct in $\mathcal{Y}$ will play an important role. Naively, we could consider the analogy of the quantum group coproduct which is obtained by twisting the usual cocommutative coproduct by $\psi(u)$:
\begin{align}
\label{delta0}
\nonumber
\Delta_0(\psi(u)) & = \psi(u) \otimes \psi(u) \\
\Delta_0(e(u)) & = e(u) \otimes \mathbbm{1} + \psi(u) \otimes e(u) \\
\nonumber
\Delta_0(f(u)) & = \mathbbm{1} \otimes f(u) + f(u) \otimes \psi(u)
\end{align}
If we used (wrongly) the relations (\ref{genfunsimrel}) with equalities, it would seem that this comultiplication is compatible with the multiplication in $\mathcal{Y}$. But in fact one can check that $\Delta_0$ is not compatible with the defining relations of the Yangian (Y0)-(Y7) if $\sigma_3 \neq 0$. Nevertheless, there is a modification of the naive coproduct $\Delta_0$ which is compatible with the defining relations and is equivalent to the coproduct in $\mathcal{W}_{1+\infty}$. The closed form formula for this coproduct is not known, but we can define it on the generators of $\mathcal{Y}$
\begin{eqnarray}
\nonumber
\Delta(e_0) & = & \Delta_0(e_0) = e_0 \otimes \mathbbm{1} + \mathbbm{1} \otimes e_0 \\
\nonumber
\Delta(f_0) & = & \Delta_0(f_0) = f_0 \otimes \mathbbm{1} + \mathbbm{1} \otimes f_0 \\
\label{newcoproduct}
\Delta(\psi_3) & = & \Delta_0(\psi_3) + 6\sigma_3 \sum_{m>0} m J_m \otimes J_{-m} \\
\nonumber
& = & \psi_3 \otimes \mathbbm{1} + \mathbbm{1} \otimes \psi_3 + \sigma_3 \psi_2 \otimes \psi_0 + \sigma_3 \psi_1 \otimes \psi_1 + \sigma_3 \psi_0 \otimes \psi_2 \\
\nonumber
& & + 6 \sigma_3 \sum_{m>0} m J_m \otimes J_{-m}
\end{eqnarray}
The additional mixing term in $\psi_3$ involving the $\hat{\mathfrak{u}}(1)$ currents (\ref{jmad}) will play an important role in the following. Because of the infinite sum, one should consider certain completion of $\mathcal{Y}$ to properly define this coproduct. But since we are mainly interested in the highest weight representations of $\mathcal{Y}$ in which $J_m$ acts locally nilpotently, considering the formal infinite sums won't cause any problems in the following.

Later we will in addition use the coproducts of the following generators:
\begin{eqnarray}
\nonumber
\Delta(e_1) & = & e_1 \otimes \mathbbm{1} + \mathbbm{1} \otimes e_1 + \sigma_3 \psi_0 \otimes e_0 \\
\nonumber
\Delta(f_1) & = & f_1 \otimes \mathbbm{1} + \mathbbm{1} \otimes f_1 + \sigma_3 f_0 \otimes \psi_0 \\
\label{newcoproducts}
\Delta(\psi_0) & = & \psi_0 \otimes \mathbbm{1} + \mathbbm{1} \otimes \psi_0 \\
\nonumber
\Delta(\psi_1) & = & \psi_1 \otimes \mathbbm{1} + \mathbbm{1} \otimes \psi_1 + \sigma_3 \psi_0 \otimes \psi_0 \\
\nonumber
\Delta(\psi_2) & = & \psi_2 \otimes \mathbbm{1} + \mathbbm{1} \otimes \psi_2 + \sigma_3 \psi_1 \otimes \psi_0 + \sigma_3 \psi_0 \otimes \psi_1
\end{eqnarray}
These can be derived from (\ref{newcoproduct}) and from the fact that the coproduct is a homomorphism of algebras. The asymmetry in $\Delta(e_1)$ and $\Delta(f_1)$ is due to the mixing term.

\section{Linear $\mathcal{W}_{1+\infty}$ and $\mathcal{Y}$}
\label{seclinear}

In the special case $\sigma_3 = 0$ (which happens if one of $h_j$ is zero) the anticommutators in (Y0)-(Y7) vanish and the algebra simplifies - it becomes the universal enveloping algebra of certain Lie algebra. On the $\mathcal{W}_{1+\infty}$ side, at $\lambda = 0$ we have a linear basis of generating fields and the resulting Lie algebra was the first case of $\mathcal{W}_{1+\infty}$ that was studied in \cite{Pope:1989sr, Pope:1990kc, Kac:1993zg} long time before its non-linear deformations were found. Because of these many simplifications, it is easier to study the map between $\mathcal{Y}$ and $\mathcal{W}_{1+\infty}$ first in this simplified setting and this is what we shall do in this section.

The linear $\mathcal{W}_{1+\infty}$ has simple free fermionic representation which can be bosonized. In this way we obtain a representation of $\mathcal{W}_{1+\infty}$ on one free boson. Deformation of this bosonic representation to representation of $\mathcal{Y}$ at $\sigma_3 \neq 0$ will be important step in finding the map between $\mathcal{W}_{1+\infty}$ and $\mathcal{Y}$ in the following sections.

Let us emphasize that only in this section we specialize parameters $h_j$ of $\mathcal{Y}$ to get the linear version of $\mathcal{W}_{1+\infty}$. Everywhere else the full non-linear two-parametric family of algebras $\mathcal{W}_{1+\infty}[\lambda]$ of \cite{Gaberdiel:2012ku} is considered.

\subsection{Differential operators on circle}
In this section we review the construction of linear $\mathcal{W}_{1+\infty}$ by Kac and Radul \cite{Kac:1993zg, Frenkel:1994em, Kac:1995sk} and which is also reviewed in \cite{Awata:1994tf}.

Instead of considering the algebra of the first order linear differential operators on a circle (vector fields) as one usually does when introducing the Virasoro algebra, we can study a larger algebra of all differential operators of finite order. One possible basis of this algebra is given by the operators of the form
\begin{equation}
x^j \partial_x^k
\end{equation}
where $j \in \mathbbm{Z}$ and $k \geq 0$. It is however more convenient to introduce the Euler operator
\begin{equation}
E \equiv x \partial_x
\end{equation}
and study the operators of the form
\begin{equation}
x^m q(E)
\end{equation}
where $q(E)$ is a polynomial in $E$. With this choice, the natural mode grading in this algebra by eigenvalues of the Euler operator is manifest. The commutation relations between two operators of this form are
\begin{equation}
\left[ x^m f(E), x^n g(E) \right] = x^{m+n} \left( f(E+n)g(E) - f(E)g(E+m) \right).
\end{equation}
We obtain the linear $\mathcal{W}_{1+\infty}$ of Kac and Radul by centrally extending this algebra of differential operators. Let us define a two-cocycle
\begin{equation}
\Psi\left(x^m f(E), x^n g(E)\right) \equiv \delta_{m+n,0} \left[ \theta(m > 0) \sum_{j=1}^m f(-j) g(m-j) - \theta(n > 0) \sum_{j=1}^n f(n-j) g(-j) \right]
\end{equation}
(which can be shown to be cohomologically non-trivial) and let us label the generators of the centrally extended algebra by $W(x^m f(E))$ to emphasize that the commutation relations are modified. The Kac-Radul $\mathcal{W}_{1+\infty}$ algebra commutation relations are thus
\begin{equation}
\left[ W(x^m f(E)), W(x^n g(E)) \right] = W\left( \left[ x^m f(E), x^n g(E) \right] \right) + c_{1+\infty} \Psi(x^m f(E), x^n g(E)).
\end{equation}
Here $c_{1+\infty}$ is the central element of the algebra.

\subsection{Affine $\mathfrak{u}(1)$ and Virasoro subalgebras}
Kac-Radul $\mathcal{W}_{1+\infty}$ contains both affine $\mathfrak{u}(1)$ and Virasoro subalgebras. The current algebra generators are just those corresponding to the multiplication operators
\begin{equation}
V_{1,m} \equiv W(x^m).
\end{equation}
They satisfy commutation relations
\begin{equation}
\left[ V_{1,m}, V_{1,n} \right] = c_{1+\infty} m \delta_{m+n,0}
\end{equation}
which are those of affine $\mathfrak{u}(1)$. Similarly, the first order differential operators
\begin{equation}
V_{2,m} = -\frac{1}{2} W\left(x^{m+1} \partial_x + \partial_x x^{m+1}\right) = -\frac{1}{2} W\left( 2 x^m E + (m+1) x^m \right)
\end{equation}
satisfy the Virasoro commutation relations
\begin{equation}
\left[ V_{2,m}, V_{2,n} \right] = (m-n) V_{2,m+n} + c_{1+\infty} \frac{(m-1)m(m+1)}{12} \delta_{m+n,0}
\end{equation}
with central charge $c_{1+\infty}$. Finally, we have
\begin{equation}
\left[ V_{2,m}, V_{1,n} \right] = -n V_{1,m+n}
\end{equation}
which tells us that $V_{1,n}$ are the mode operators of primary spin $1$ field with respect to Virasoro subalgebra.

\subsection{Pope-Shen-Romans basis}
The $\mathcal{W}_{1+\infty}$ studied by Kac \& Radul is in fact the same algebra as $\mathcal{W}_{1+\infty}$ studied earlier by Pope, Shen and Romans \cite{Pope:1989sr, Pope:1990kc}. Following \cite{Awata:1994tf}, we can introduce another basis of the algebra given by
\begin{equation}
V_{s,m} = \frac{1}{{2s-2 \choose s-1}} \sum_{j=0}^{s-1} (-1)^j {s-1 \choose j}^2 W\left(x^m \left[-E-m-1\right]_{s-j-1} \left[E\right]_j \right)
\end{equation}
These operators correspond to the mode operators of the quasi-primary fields of spin $s$,
\begin{equation}
\left[ V_{2,m}, V_{s,n} \right] = ((s-1)m-n) V_{s,m+n} \\
\end{equation}
for $m = -1,0,1$. Once we have a basis of the quasi-primary fields, the mode index dependence of commutators is completely fixed (see for instance \cite{Bowcock:1990ku}), so the commutation relations of $\mathcal{W}_{1+\infty}$ can be written in the form
\begin{equation}
\left[ V_{j,m}, V_{k,n} \right] = \sum_{\substack{0 \leq l \leq j+k-2 \\j+k-l \; \text{even}}} C_{jk}^l N_{jk}^l(m,n) V_{l,m+n}
\end{equation}
where the coefficients $N_{jk}^l$ are universal (determined only by the global conformal invariance and spins)
\begin{align}
\nonumber
N_{jk}^0(m,n) & = {m+j-1 \choose j+k-1} \delta_{m+n,0} \\
N_{jk}^l(m,n) & = \sum_{s=0}^{j+k-l-1} \frac{(-1)^s}{(j+k-l-1)! (2l)_{j+k-l-1}} {j+k-l-1 \choose s} \times \\
\nonumber
& \times \left[ j+m-1 \right]_{j+k-l-s-1} \left[ j-m-1 \right]_s \left[ k+n-1 \right]_s \left[k-n-1 \right]_{j+k-l-s-1}
\end{align}
while the structure constants $C_{jk}^l$ characterize the algebra itself
\begin{align}
\nonumber
C_{jk}^0 & = \frac{(j-1)!^2 (2j-1)!}{4^{j-1} (2j-1)!! (2j-3)!!} \delta_{jk} c_{1+\infty} \\
C_{jk}^l & = \frac{1}{2 \times 4^{j+k-l-2}} (2l)_{j+k-l-1} {}_4 F_3 \; \left( \begin{array}{cc} \frac{1}{2}, \frac{1}{2}, -\frac{1}{2} (j+k-l-2), -\frac{1}{2} (j+k-l-1) \\ \frac{3}{2}-j, \frac{3}{2}-k, \frac{1}{2}+l \end{array} ; 1 \right)
\end{align}
and were determined in \cite{Pope:1989sr, Pope:1990kc}\footnote{We use a slightly different normalization following \cite{Awata:1994tf}. The relation is $(V_j)_{\text{PRS}} = 4^{j-2} (V_j)_{\text{here}} = 4^{j-2} (W_j)_{\text{AFMO}}$ and correspondingly for $C_{jk}^l$.}.

\subsection{Linear $\mathcal{W}_{1+\infty}$ and $\mathcal{Y}$}
Let us now define a map from $\mathcal{Y}$ with one of $h_j$-parameters equal to zero to linear $\mathcal{W}_{1+\infty}$ discussed in the previous section. Since we have identified the $\widehat{\mathfrak{u}(1)}$ subalgebra of both algebras, we can define
\begin{align}
\nonumber
e_0 & \leftrightarrow V_{1,-1} \\
f_0 & \leftrightarrow -V_{1,1}
\end{align}
Furthermore, we identify $\psi_3$ which we can use to generate all higher $e_j$ and $f_j$,
\begin{equation}
\psi_3 \leftrightarrow 3 V_{3,0}.
\end{equation}
One could consider more general map with $V_{3,0}$ mapping to a combination of $\psi_3$ with lower $\psi_j$. But $\psi_2$ can be eliminated by spectral shift and $\psi_0$ and $\psi_1$ are central so they wouldn't modify the $e_j$ and $f_j$ computed using $\psi_3$.

These three elements already generate full $\mathcal{Y}$ so one must only verify that they satisfy the correct defining relations. First of all, we have
\begin{equation}
\psi_0 = \left[ e_0, f_0 \right] = c_{1+\infty}
\end{equation}
which relates the value of central charge to $\psi_0$. To determine parameters $\sigma_j$, we first we determine $e_j$ up to $j=3$ using (\ref{psi3ej}) and then check consistency with (Y1). We find
\begin{equation}
(\sigma_2+1) V_{1,-2} = \left[ e_3, e_0 \right] - 3\left[ e_2, e_1 \right] + \sigma_2 \left[ e_1, e_0 \right] = \sigma_3 e_0^2 = \sigma_3 V_{1,-1}^2.
\end{equation}
For this equation to be consistent, we need
\begin{equation}
\sigma_3 = 0, \quad \quad \quad \sigma_2 = -1
\end{equation}
which corresponds to $h_j$-parameters equal to $(0,1,-1)$. This is as expected, since for $\sigma_3 = 0$ the defining relations of $\mathcal{Y}$ linearize. Assuming these values of $\sigma_j$, we can iteratively find the expressions for first few generators
\begin{align}
\nonumber
e_1 & = V_{2,-1} \\
\nonumber
f_1 & = -V_{2,1} \\
\nonumber
e_2 & = V_{3,-1} \\
f_2 & = -V_{3,1} \\
\nonumber
e_3 & = V_{4,-1} - \frac{1}{5} V_{2,-1} \\
\nonumber
f_3 & = -V_{4,1} + \frac{1}{5} V_{2,1}
\end{align}
and
\begin{align}
\nonumber
\psi_1 & = V_{1,0} \\
\nonumber
\psi_2 & = 2V_{2,0} \\
\psi_3 & = 3 V_{3,0} \\
\nonumber
\psi_4 & = 4 V_{4,0} - \frac{2}{5} V_{2,0} \\
\nonumber
\psi_5 & = 5 V_{5,0} - \frac{15}{7} V_{3,0}.
\end{align}
Note that the generators of the Baxter subalgebra $\mathcal{B}$ of $\mathcal{Y}$ are mapped to zero modes of the linear basis $V_{s,0}$ in linear $\mathcal{W}_{1+\infty}$. This is one of nice properties of this linear basis which will have no analogue when $\sigma_3 \neq 0$.

In general, a closed-formula for the transformation between $\psi_j$ and $V_{s,j}$ can be derived using the representation theory. Since we will not need this in what follows, let us state just the result: the expression for $V_{s,0}$ is obtained by expanding
\begin{equation}
\frac{1}{s{2s-2 \choose s-1}} \sum_{j=0}^{s-1} {s-1 \choose j}^2 \left[ x+j \right]_s = \frac{1}{s{2s-2 \choose s-1}} \sum_{l=0}^s {s \choose l} \frac{(s+l-2)!}{(l-1)!^2} \left[ x \right]_l
\end{equation}
as a polynomial in $x$ and replacing
\begin{equation}
x^k \to \psi_k.
\end{equation}
The mapping is much simpler in terms of symbols of differential operators used by Kac and Radul. We find that
\begin{align}
\nonumber
\psi_k & \leftrightarrow (-1)^{k+1} W \Big( (E+1)^k - E^k \Big) = (-1)^{k+1} W\Big( (\partial x)^k - (x \partial)^k \Big) \\
e_k & \leftrightarrow (-1)^k W\Big( x^{-1} E^k \Big) \\
\nonumber
f_k & \leftrightarrow (-1)^{k+1} W\Big( E^k x \Big)
\end{align}
(see also \cite{tsymbaliuk2014affine}). The inverse mapping is given by mapping $W(E^k)$ to
\begin{equation}
\frac{B_{k+1}(0)-B_{k+1}(-x)}{k+1}
\end{equation}
with $x^j$ again replaced by $\psi_j$. Here $B_k(x)$ are the Bernoulli polynomials. To characterize the highest weight representations, Kac and Radul introduce a generating function of charges of the highest weight state $\ket{\Lambda}$. The highest weight state satisfies
\begin{align}
\nonumber
W(E^k) \ket{\Lambda} & = \Delta_k \ket{\Lambda} \\
W(x^m E^k) \ket{\Lambda} & = 0, \quad \quad \quad m > 0
\end{align}
which defines the charges $\Delta_k$. These charges can be encoded in the exponential generating function
\begin{equation}
\phi_{\mathrm{KR}}(t) = -\sum_{k=0}^{\infty} \frac{t^k}{k!} \Delta_k.
\end{equation}
Now we can easily express this generating function in terms of $\psi_j$ charges of the highest weight state,
\begin{equation}
\phi_{\mathrm{KR}}(t) = \sum_{k=0}^{\infty} \frac{t^k B_{k+1}(-x)}{(k+1)!} + const. = \frac{1}{e^t-1} \sum_{j=1}^{\infty} \frac{(-1)^j t^j \psi_j}{j!}.
\end{equation}
This results lets us translate between $\psi_j$ charges of $\mathcal{Y}$ and Kac-Radul generating function $\phi_{\mathrm{KR}}(t)$.

\subsection{Free fermion representation}
We can find a simple representation of linear $\mathcal{W}_{1+\infty}$ with $c=1$ in terms of a complex fermion \cite{Bergshoeff:1990yd,Sezgin:1990ee}
\begin{equation}
\bar{\Psi}(z) \Psi(w) \sim \frac{1}{z-w}.
\end{equation}
This OPE is equivalent to anticommutation relations between modes
\begin{equation}
\left\{ \bar{\Psi}_m, \Psi_n \right\} = \delta_{m+n,0}.
\end{equation}
where
\begin{equation}
\Psi(z) = \sum_{m \in \mathbbm{Z}+\frac{1}{2}} z^{-m-\frac{1}{2}} \Psi_m.
\end{equation}
The mode index runs over half-integers, so $\Psi(z)$ is periodic in the complex plane and antiperiodic when mapped to the cylinder. The quasiprimary generators of $\mathcal{W}_{1+\infty}$ are bilinears in fermion fields
\begin{align}
\nonumber
V_1(z) & = (\bar{\Psi}\Psi)(z) \\
V_2(z) & = \frac{1}{2} \Big( \bar{\Psi}^\prime \Psi \Big)(z) - \frac{1}{2} \Big( \bar{\Psi} \Psi^\prime \Big)(z) \\
\nonumber
V_3(z) & = \frac{1}{6} \Big( \bar{\Psi}^{(2)} \Psi\Big)(z) - \frac{2}{3} \Big( \bar{\Psi}^\prime \Psi^\prime \Big)(z) + \frac{1}{6} \Big( \bar{\Psi} \Psi^{(2)} \Big)(z)
\end{align}
and in general
\begin{equation}
V_s(z) = \frac{(s-1)!}{4^{s-1} \left( \frac{1}{2} \right)_{s-1}} \sum_{k=0}^{s-1} (-1)^k {s-1 \choose k}^2 \Big( \bar{\Psi}^{(s-1-k)} \Psi^{(k)} \Big)(z).
\end{equation}
Using the map between $\mathcal{Y}$ and linear $\mathcal{W}_{1+\infty}$, we can identify the generating elements
\begin{align}
\nonumber
e_0 & = V_{1,-1} \leftrightarrow \sum_{m \in \mathbbm{Z}+\frac{1}{2}} :\bar{\Psi}_m \Psi_{-m-1}: \\
f_0 & = -V_{1,1} \leftrightarrow -\sum_{m \in \mathbbm{Z}+\frac{1}{2}} :\bar{\Psi}_m \Psi_{-m+1}: \\
\nonumber
\psi_3 & = 3V_{3,0} \leftrightarrow \sum_{m \in \mathbbm{Z}+\frac{1}{2}} \left( 3m^2 + \frac{1}{4} \right) :\bar{\Psi}_m \Psi_{-m}:
\end{align}
and in general
\begin{align}
\nonumber
\psi_k & \leftrightarrow \frac{(-1)^{k+1}}{2^k} \sum_{m \in \mathbbm{Z}+\frac{1}{2}} \left[ (2m+1)^k - (2m-1)^k \right] :\bar{\Psi}_m \Psi_{-m}: \\
e_k & \leftrightarrow \sum_{m \in \mathbbm{Z}+\frac{1}{2}} \left(m-\frac{1}{2}\right)^k :\bar{\Psi}_{-m} \Psi_{m-1}: \\
\nonumber
f_k & \leftrightarrow -\sum_{m \in \mathbbm{Z}+\frac{1}{2}} \left(m+\frac{1}{2}\right)^k :\bar{\Psi}_{-m} \Psi_{m+1}:
\end{align}
These maps can be compactly written in terms of generating functions
\begin{align}
\nonumber
e(u) & \leftrightarrow \sum_{m\in\mathbbm{Z}} \frac{\bar{\Psi}_{m-1/2} \Psi_{-m-1/2}}{u+m} \\
f(u) & \leftrightarrow - \sum_{m\in\mathbbm{Z}} \frac{\bar{\Psi}_{m+1/2} \Psi_{-m+1/2}}{u+m}
\end{align}
We cannot write an analogous expression for $\psi(z)$ because the conventional definition of this generating function has $\sigma_3$ multiplying $\psi_k$ which in the case of linear $\mathcal{W}_{1+\infty}$ is equal to zero. Verification that Yangian relations (Y1)-(Y7) are satisfied is simple.

For completeness, we can write down the fermion representation of $\mathcal{W}_{1+\infty}$ in Kac-Radul basis. We have
\begin{align}
W\left[ x^m E^k \right] & \leftrightarrow \sum_{n\in\mathbbm{Z}+\frac{1}{2}} \left( n - \frac{1}{2} \right)^k :\bar{\psi}_{m+n} \psi_{-n}:
\end{align}

\subsection{Free boson representation}
We can bosonize the fermionic representation found in the previous section to a representation of linear $\mathcal{W}_{1+\infty}$ in terms of one free boson. The basic formula is the bosonization map
\begin{align}
\nonumber
\bar{\Psi}(z) & \leftrightarrow : e^{i\phi(z)} : \\
\Psi(z) & \leftrightarrow : e^{-i\phi(z)} :
\end{align}
The vertex operators $e^{i\alpha\phi(z)}$ have the operator product
\begin{equation}
: e^{i\alpha\phi(z)}: \, : e^{i\beta\phi(w)} : = (z-w)^{\alpha\beta} : e^{i\alpha\phi(z)} e^{i\beta\phi(w)} :.
\end{equation}
When translating between fermions and bosons, the main formula for fermionic bilinears is
\begin{equation}
\frac{1}{z-w} + : \bar{\Psi}(z) \Psi(w) : = \bar{\Psi}(z) \Psi(w) \leftrightarrow : e^{i\phi(z)}: \, : e^{-i\phi(w)} : = \frac{1}{z-w} : e^{i\phi(z)} e^{-i\phi(w)} :
\end{equation}
For example,
\begin{equation}
V_1(w) \leftrightarrow \lim_{z \to w} \left[ \bar{\Psi}(z) \Psi(w) - \frac{1}{z-w} \right] = \lim_{z \to w} \frac{ : e^{i\phi(z)} e^{-i\phi(w)}: -1}{z-w} = i \partial \phi(w) \equiv J(w)
\end{equation}
and analogously
\begin{align}
\nonumber
V_2(w) & \leftrightarrow \frac{1}{2} :J(w)^2: \\
V_3(w) & \leftrightarrow \frac{1}{3} :J(w)^3: \\
\nonumber
V_4(w) & \leftrightarrow \frac{1}{4} :J(w)^4: - \frac{3}{20} :J^\prime(w)^2: + \frac{1}{10} :J^{\prime\prime}(w)J(w):
\end{align}
In terms of the bosonic mode oscillators
\begin{equation}
J(z) = i\partial\phi(z) = \sum_{m \in \mathbbm{Z}} \frac{a_m}{z^{m+1}}
\end{equation}
where the oscillators $a_m$ satisfy the commutation relations
\begin{equation}
\left[ a_m, a_n \right] = m \, \delta_{m+n,0}
\end{equation}
we have
\begin{align}
\nonumber
e_0 & \leftrightarrow a_{-1} \\
f_0 & \leftrightarrow -a_{1} \\
\nonumber
\psi_3 & \leftrightarrow \sum_{j+k+l=0} : a_j a_k a_l : = a_0^3 + 6a_0 \sum_{j>0} a_{-j} a_j + 3 \sum_{j,k>0} \left(a_{-j} a_{-k} a_{j+k} + a_{-j-k} a_j a_k \right)
\end{align}
The operator $\psi_3$ realized in terms of the free boson is often called the cut-and-join operator \cite{Mironov:2009cj}. In principle it is possible to write a closed-form formula for the generators \cite{Sezgin:1990ee,Fukuma:1990yk}, but these expressions are far from being as simple as in the fermionic case. In particular, the non-linearity grows with increasing spin.

\section{Representations of $\mathcal{Y}$}
\label{secyangrep}

In this section we will discuss the representations of $\mathcal{Y}$. First of all, we use the triangular decomposition of the algebra to define the analogue of the usual highest weight representations of $\mathcal{W}_{1+\infty}$. To study the irreducible representations, it is useful to have the Shapovalov form and this is what we do next. Using these two ingredients, we can generalize the notion of the quasi-finite representations of \cite{Kac:1993zg,Kac:1995sk} to $\mathcal{Y}$. It turns out that the characterization of the quasi-finite representations in $\mathcal{Y}$ is very simple in terms of the action of the generating field $\psi(u)$ on the highest weight state: the irreducible representation has only a finite number of states at each level if $\psi(u)$ eigenvalue of the highest weight state is a ratio of two polynomials in $u$ of the same degree (and with the same leading coefficient). In the next section, we find the charges $\psi(u)$ of the vacuum representation, which for generic values of the parameters $h_j$ has the character given by the MacMahon function. We discuss the structure of the vacuum representation from the point of view of $\mathcal{Y}$ up to level $2$.

The following section discusses more general actions of $\mathcal{Y}$ on generalized plane partitions with certain combinatorial restrictions. The representations discussed in \cite{tsymbaliukrev} as well as representations of the $q$-deformed version of $\mathcal{Y}$ -- the toroidal algebra of \cite{feigin2012quantum} are all of this form. The vector space basis of the representation space is given by the generalized plane partitions, possibly allowing for a non-trivial asymptotics along the coordinate axes and for certain geometric restrictions. The highest weight state corresponds to plane partition with the minimum number of boxes consistent with the given asymptotics. Generators $\psi_j$ of the Baxter subalgebra $\mathcal{B}$ act diagonally in the basis of the generalized plane partitions. Operators $e_j$ on the other hand act as box addition operators and operators $f_j$ act as box removal operators. In certain sense, these representations are three-dimensional generalization of the harmonic oscillator algebra, if one considers the states $\ket{n}$ in the harmonic oscillator Hilbert space to be $1$-dimensional partitions. Next it is shown how useful is the complementary picture of the generalized plane partitions, which is that of the rhombic plane tilings.

In the last section we discuss the $\sigma_3 \neq 0$ deformation of the free boson representation found in the linear $\mathcal{W}_{1+\infty}$. The generating elements $e_0$ and $f_0$ are undeformed, the non-trivial task is to determine the deformed cut-and-join generator $\psi_3$. The resulting expression does not seem to be a zero mode of any local operator constructed from $\widehat{\mathfrak{u}(1)}$ fields, but it appeared in the studies of the Benjamin-Ono equation.

\subsection{Highest weight representations}
The triangular decomposition of $\mathcal{Y}$ (\ref{triang}) lets us study the highest weight representations similarly to the highest weight representations of $\mathcal{W}_{1+\infty}$. We can define the Verma module by choosing the highest weight state that is an eigenstate of $\psi_j$ generators with eigenvalues $\psi_{\Lambda,j}$ and annihilated by raising operators $f_j$,
\begin{align}
\nonumber
\psi_j \ket{\Lambda} & = \psi_{\Lambda,j} \ket{\Lambda} \\
f_j \ket{\Lambda} & = 0.
\end{align}
The subalgebra $\mathcal{Y}^+$ acts on this vector with the only relations being those of the $\mathcal{Y}^+$ itself (these relations are generated by (Y1) and (Y6)). Generically, the Verma module constructed in this way is irreducible, but for special choices of the highest weight charges $\psi_{\Lambda,j}$ or parameters $h_j$ it becomes reducible.

We may similarly introduce the dual Verma module with the highest weight state $\bra{\Lambda}$ such that
\begin{align}
\nonumber
\bra{\Lambda} \psi_j & = \psi_{\Lambda,j} \bra{\Lambda} \\
\bra{\Lambda} e_j & = 0
\end{align}
and right action of $\mathcal{Y}^-$.

\paragraph{Bilinear Shapovalov form}
The usual way of studying when does the Verma module become reducible proceeds by defining the Shapovalov bilinear form. Denoting the anti-automorphim exchanging $e_j \leftrightarrow f_j$ by $\mathrm{a}$,
\begin{align}
\nonumber
\mathrm{a}(e_j) & = f_j \\
\mathrm{a}(f_j) & = e_j \\
\nonumber
\mathrm{a}(\psi_j) & = \psi_j
\end{align}
the Shapovalov form is the quadratic form on $\mathcal{Y}^+$ defined by
\begin{equation}
B(x,y) \braket{\Lambda|\Lambda} = \bra{\Lambda} \mathrm{a}(x) y \ket{\Lambda}.
\end{equation}
It is symmetric and becomees degenerate when the Verma module becomes reducible. Furthermore, similarly to the case of the Virasoro algebra, due to the highest weight vector $\ket{\Lambda}$ being $\psi_2$ eigenvector and relations (Y4') and (Y5'), the Shapovalov form is block diagonal with blocks corresponding to vectors of definite $\psi_2$ eigenvalue. In fact, since we have an infinite set of commuting charges $\psi_j$, we can also diagonalize these and we end up with one-dimensional eigenspaces (at least at lower levels).

\subsection{Quasi-finite representations}
A general problem in algebras like $\mathcal{Y}$ or $\mathcal{W}_{1+\infty}$ is that the generic Verma module has an infinite number of states at level $1$,
\begin{equation}
e_j \ket{\Lambda}, \quad j \geq 0.
\end{equation}
This leads to a question under which conditions there is only a finite number of states at level $1$. In the linear $\mathcal{W}_{1+\infty}$ the answer was given in \cite{Kac:1993zg, Frenkel:1994em, Kac:1995sk} where the authors introduced the so-called quasi-finite representations.

In the case of the highest weight representations of $\mathcal{Y}$ it is quite easy to find the right characterization of these special representations. The commutation relation (Y3) immediately tells us that the Shapovalov form at level $1$ has matrix elements
\begin{equation}
\bra{\Lambda} f_j e_k \ket{\Lambda} = - \bra{\Lambda}\psi_{j+k}\ket{\Lambda} = -\psi_{\Lambda,j+k} \braket{\Lambda|\Lambda}.
\end{equation}
We will find a finite number of states at level $1$ in the irreducible representation if the rank of this infinite matrix is finite. If the rank is less than or equal to $r$, the first $r+1$ rows will satisfy a relation
\begin{equation}
\sum_{j=0}^{r} \alpha_j \psi_{j+k} = 0, \quad \quad k \geq 0.
\end{equation}
with not all $\alpha_j$ vanishing. Multiplying this equation by $\sigma_3 u^{-k-1}$ and summing over $k$, we find
\begin{align}
\nonumber
0 & = \alpha_0 \big( \psi_{\Lambda}(u) - 1 \big) + \alpha_1 \big( u\psi_{\Lambda}(u) - u - \sigma_3 \psi_{\Lambda,0} \big) + \alpha_2 \big( u^2\psi_{\Lambda}(u) - u^2 - u \sigma_3 \psi_{\Lambda,0} - \sigma_3 \psi_{\Lambda,1} \big) + \cdots \\
& = \psi_{\Lambda}(u) \sum_{j=0}^{r} \alpha_j u^j - \sum_{j=0}^{r} \alpha_j u^j - \sigma_3 \sum_{j=1}^{r} \alpha_j \sum_{k=0}^{j-1} u^{j-1-k} \psi_{\Lambda,k}
\end{align}
so that $\psi_{\Lambda}(u)$ is a ratio of two polynomials,
\begin{equation}
\psi_{\Lambda}(u) = \frac{\sum_{j=0}^{r} \alpha_j u^j + \sigma_3 \sum_{j=1}^{r} \alpha_j \sum_{k=0}^{j-1} u^{j-1-k} \psi_{\Lambda,k}}{\sum_{j=0}^{r} \alpha_j u^j}
\end{equation}
which is analogous to the Drinfeld polynomial studied in the theory of the quantum groups. In order to have $\psi_{\Lambda}(u) \to 1$ as $u \to \infty$, we need to have the same degree of the polynomials in both numerator and denominator and the same leading coefficient. The rank of the level $1$ Shapovalov form gives us an upper bound on this degree. Once we have a finite number of states at level $1$, we can check by induction using the commutation relation (Y1) that there is also a finite number of states at higher levels.

\subsection{Vacuum MacMahon representation}
\label{vacrep}

In this section we want to study the (charged) MacMahon vacuum representation of $\mathcal{Y}$, i.e. the highest weight representation of $\mathcal{Y}$ where the states are in one-to-one correspondence with the plane partitions, directly using the defining relations of $\mathcal{Y}$.

\paragraph{Level 1}
The vacuum representation of $\mathcal{Y}$ should have the character given by the MacMahon function. In particular, there should be only one state at level $1$. This information is enough to determine the values of $\psi_j$ charges of the highest weight state $\ket{0}$. The level $1$ Shapovalov form
\begin{equation}
\bra{0} f_k e_j \ket{0} = - \bra{0} \psi_{j+k} \ket{0}
\end{equation}
must have rank $1$. This is true if and only if
\begin{equation}
\psi_j \ket{0} = q^j \psi_0 \ket{0}, \quad \quad \quad \psi_0 \neq 0.
\end{equation}
This means that the generating function $\psi(z)$ acts on the highest weight state as
\begin{equation}
\label{vacuumcharges}
\psi(u) \ket{0} = \frac{u-q+\psi_0 \sigma_3}{u-q} \ket{0}.
\end{equation}
As required, all the level $1$ states are proportional,
\begin{equation}
e_j \ket{0} = q^j e_0 \ket{0}.
\end{equation}
In the following, we will specialize to the case $q=0$ to simplify the computations. This is no loss of generality, since we can always use the spectral shift transformation of section \ref{spectralshift} to reintroduce the parameter $q$ (for $\psi_0 \neq 0$ which is satisfied in this case). For $q=0$ the highest weight state $\ket{0}$ is annihilated by $e_j, j \geq 1$,
\begin{align}
\nonumber
e_0 \ket{0} & = J_{-1} \ket{0} \neq 0 \\
e_j \ket{0} & = 0, \quad \quad j \geq 0.
\end{align}
and the only nonzero $\psi_j$ charge of the vacuum state is $\psi_0$.

We now compute the $\psi_j$ charges of the unique level $1$ state
\begin{equation}
\ket{\Box} \sim e_0 \ket{0}.
\end{equation}
Using (Y4) we find
\begin{align}
\nonumber
\psi_1 \ket{\Box} & = 0 \\
\psi_2 \ket{\Box} & = 2 \ket{\Box} \\
\nonumber
\psi_3 \ket{\Box} & = 2\sigma_3\psi_0 \ket{\Box}
\end{align}
and the recurrence relation
\begin{equation}
\psi_{j+3} \ket{\Box} + \sigma_2 \psi_{j+1} \ket{\Box} - \sigma_3 \psi_j \ket{\Box} = 0, \quad \quad j \geq 4.
\end{equation}
This is a third order homogeneous linear difference equation with characteristic equation
\begin{equation}
0 = \alpha^3 + \sigma_2 \alpha - \sigma_3 = (\alpha-h_1)(\alpha-h_2)(\alpha-h_3).
\end{equation}
The unique solution satisfying our initial conditions is
\begin{equation}
\psi_j \ket{\Box} = \left(\frac{2(h_2 h_3 \psi_0 + 1) h_1^j}{(h_1-h_2)(h_1-h_3)} + \frac{2(h_1 h_3 \psi_0 + 1) h_2^j}{(h_2-h_1)(h_2-h_3)} + \frac{2(h_1 h_2 \psi_0 + 1) h_3^j}{(h_3-h_1)(h_3-h_2)}\right) \ket{\Box}.
\end{equation}
The generating function of these charges is
\begin{equation}
\psi(u) \ket{\Box} = \frac{u+\psi_0 \sigma_3}{u} \frac{(u+h_1)(u+h_2)(u+h_3)}{(u-h_1)(u-h_2)(u-h_3)} \ket{\Box} = \frac{u+\psi_0\sigma_3}{u} \varphi(u) \ket{\Box}.
\end{equation}
Notice the appearance of the function $\varphi(u)$.

\paragraph{Level 2}
At level $2$ we have three independent states which we can choose to be the states
\begin{equation}
e_0 e_0 \ket{0}, \quad\quad e_1 e_0 \ket{0} \quad \quad \text{and} \quad e_2 e_0 \ket{0}.
\end{equation}
The relation (Y1) lets us compute all states $e_j e_k \ket{0}$ in terms of this basis - it reduces to a simple recurrence relation
\begin{equation}
e_{j+3} e_0 \ket{0} + \sigma_2 e_{j+1} e_0 \ket{0} - \sigma_3 e_j e_0 \ket{0} = 0
\end{equation}
which is the same recurrence relation as the one for $\psi_j$ charges at level 1. This means that the general solution is a linear combination
\begin{equation}
\ytableausetup{mathmode, boxsize=0.7em}
e_j \ket{\Box} \sim e_j e_0 \ket{0} = h_1^j \ket{\ydiagram{2}} + h_2^j \ket{\ydiagram{1,1}} + h_3^j \ket{\begin{ytableau} \scriptstyle 2 \end{ytableau}}
\end{equation}
where the three independent states $\ket{\ydiagram{2}}, \ket{\ydiagram{1,1}}$ and $\ket{\begin{ytableau} \scriptstyle 2 \end{ytableau}}$ (the reason for this notation will be clear soon) can be determined from the initial conditions
\begin{align}
\nonumber
e_0 e_0 \ket{0} & = \ket{\ydiagram{2}} + \ket{\ydiagram{1,1}} + \ket{\begin{ytableau} \scriptstyle 2 \end{ytableau}} \\
e_1 e_0 \ket{0} & = h_1 \ket{\ydiagram{2}} + h_2 \ket{\ydiagram{1,1}} + h_3 \ket{\begin{ytableau} \scriptstyle 2 \end{ytableau}} \\
\nonumber
e_2 e_0 \ket{0} & = h_1^2 \ket{\ydiagram{2}} + h_2^2 \ket{\ydiagram{1,1}} + h_3^2 \ket{\begin{ytableau} \scriptstyle 2 \end{ytableau}}
\end{align}
to be
\begin{align}
\nonumber
\ket{\ydiagram{2}} & = \frac{1}{(h_1-h_2)(h_1-h_3)} \left[ e_2 + h_1 e_1 + h_2 h_3 e_0 \right] e_0 \ket{0} \\
\ket{\ydiagram{1,1}} & = \frac{1}{(h_2-h_1)(h_2-h_3)} \left[ e_2 + h_2 e_1 + h_1 h_3 e_0 \right] e_0 \ket{0} \\
\nonumber
\ket{\begin{ytableau} \scriptstyle 2 \end{ytableau}} & = \frac{1}{(h_3-h_1)(h_3-h_2)} \left[ e_2 + h_3 e_1 + h_1 h_2 e_0 \right] e_0 \ket{0}.
\end{align}
Expressed in terms of the generating functions, the result is
\begin{equation}
e(u) \ket{\Box} = \frac{\ket{\ydiagram{2}}}{u-h_1} + \frac{\ket{\ydiagram{1,1}}}{u-h_2} + \frac{\ket{\begin{ytableau} \scriptstyle 2 \end{ytableau}}}{u-h_3}.
\end{equation}
Next we compute the $\psi(u)$ eigenstates at level $2$. We first diagonalize $\psi_3$ and find that states $\ket{\ydiagram{2}},\ket{\ydiagram{1,1}}$ and $\ket{\begin{ytableau} \scriptstyle 2 \end{ytableau}}$ are precisely the eigenstates of $\psi_3$ with eigenvalues
\begin{align}
\nonumber
\psi_3 \ket{\ydiagram{2}} & = 2h_1(2h_2 h_3 \psi_0 + 3) \ket{\ydiagram{2}} \\
\psi_3 \ket{\ydiagram{1,1}} & = 2h_2(2h_1 h_3 \psi_0 + 3) \ket{\ydiagram{1,1}} \\
\nonumber
\psi_3 \ket{\begin{ytableau} \scriptstyle 2 \end{ytableau}} & = 2h_3(2h_1 h_2 \psi_0 + 3) \ket{\begin{ytableau} \scriptstyle 2 \end{ytableau}}.
\end{align}
Since all other $\psi_j$ commute with $\psi_3$ and these eigenvalues are generically distinct, these vectors are the simultaneous eigenvectors of all $\psi_j$ generators. The values of higher $\psi_j$ charges on these states can again most conveniently be expressed in terms of the generating function,
\begin{align}
\nonumber
\psi(u) \ket{\ydiagram{2}} & = \frac{u+\sigma_3 \psi_0}{u} \varphi(u) \varphi(u-h_1) \ket{\ydiagram{2}} \\
\psi(u) \ket{\ydiagram{1,1}} & = \frac{u+\sigma_3 \psi_0}{u} \varphi(u) \varphi(u-h_2) \ket{\ydiagram{1,1}} \\
\nonumber
\psi(u) \ket{\begin{ytableau} \scriptstyle 2 \end{ytableau}} & = \frac{u+\sigma_3 \psi_0}{u} \varphi(u) \varphi(u-h_3) \ket{\begin{ytableau} \scriptstyle 2 \end{ytableau}}.
\end{align}
Another useful property of the basis of level $2$ states it their orthogonality with respect to the Shapovalov form introduced above. We find that
\begin{align}
\label{vaclevel2inner}
\nonumber
\braket{\ydiagram{2} | \ydiagram{2}} & = \frac{2 \psi_0 (1 + h_2 h_3 \psi_0)}{(h_1-h_2)(h_1-h_3)} \braket{0|0} \\
\braket{\ydiagram{1,1} | \ydiagram{1,1}} & = \frac{2 \psi_0 (1 + h_1 h_3 \psi_0)}{(h_2-h_1)(h_2-h_3)} \braket{0|0} \\
\nonumber
\braket{\begin{ytableau} \scriptstyle 2 \end{ytableau} | \begin{ytableau} \scriptstyle 2 \end{ytableau}} & = \frac{2 \psi_0 (1 + h_1 h_2 \psi_0)}{(h_1-h_3)(h_2-h_3)} \braket{0|0}
\end{align}
with all other products vanishing. This will be useful later when discussing the null states and minimal models.

\paragraph{Level 3}
At level $3$, we should have $6$ linearly independent states. One possible basis of level $3$ states is $e_j e_0 e_0 \ket{0}, j=0,\ldots,5$. Similarly to the situation at level $2$, states of the form
\begin{equation}
v_j \equiv e_j e_0^2 \ket{0}
\end{equation}
satisfy linear recurrence relation with constant coefficients
\begin{equation}
8\sigma_3^2 v_j -12\sigma_2\sigma_3 v_{j+1} + 4\sigma_2^2 v_{j+2} - 9\sigma_3 v_{j+3} + 5\sigma_2 v_{j+4} + v_{j+6} = 0
\end{equation}
Characteristic equation of this difference equation has roots
\begin{equation}
h_1, h_2, h_3, 2h_1, 2h_2, 2h_3.
\end{equation}
This means that there are $6$ states such that
\begin{align}
\nonumber
e_j e_0^2 \ket{0} & = h_1^j \left( \ket{\ydiagram{2,1}} + \ket{\begin{ytableau} \scriptstyle 2 & \end{ytableau}} \right) + h_2^j \left( \ket{\ydiagram{2,1}} + \ket{\begin{ytableau} \scriptstyle 2 \\ \phantom{1} \end{ytableau}} \right) + h_3^j \left( \ket{\begin{ytableau} \scriptstyle 2 & \end{ytableau}} + \ket{\begin{ytableau} \scriptstyle 2 \\ \phantom{1} \end{ytableau}} \right) \\
& + (2h_1)^j \ket{\ydiagram{3}} + (2h_2)^j \ket{\ydiagram{1,1,1}} + (3h_3)^j \ket{\begin{ytableau} \scriptstyle 3 \end{ytableau}}.
\end{align}

Already at this point there is a structure that is clearly starting to emerge, but proceeding to higher levels in this way becomes difficult. One of the problems is the fact that the relations (Y1) and (Y6) should be used to express the vectors of the form $e_j e_k e_l \ket{0}$ in terms of a chosen basis, but the form of these relations does not let us do this in a simple and canonical way like for instance in the case of a Lie algebra - at the current stage we don't have an analogue of Poincar\'{e}-Birkhoff-Witt theorem for $\mathcal{Y}^+$.

But we will see that what we found here generalizes to the full vacuum representation: there is a basis of the representation space given by the plane partitions. The generating function $\psi(u)$ is diagonal and explicitly diagonalizable in this basis. The generators $e_j$ act by adding a single box while $f_j$ act by remove a box. The dependence on the spin index $j$ will involve combinatorial information obtained from the given plane partition. Finally, the Shapovalov form will be diagonal in this basis.

\paragraph{Notation - plane partitions}

\begin{figure}
\centering
\begin{minipage}{5cm}
\includegraphics[scale=0.4]{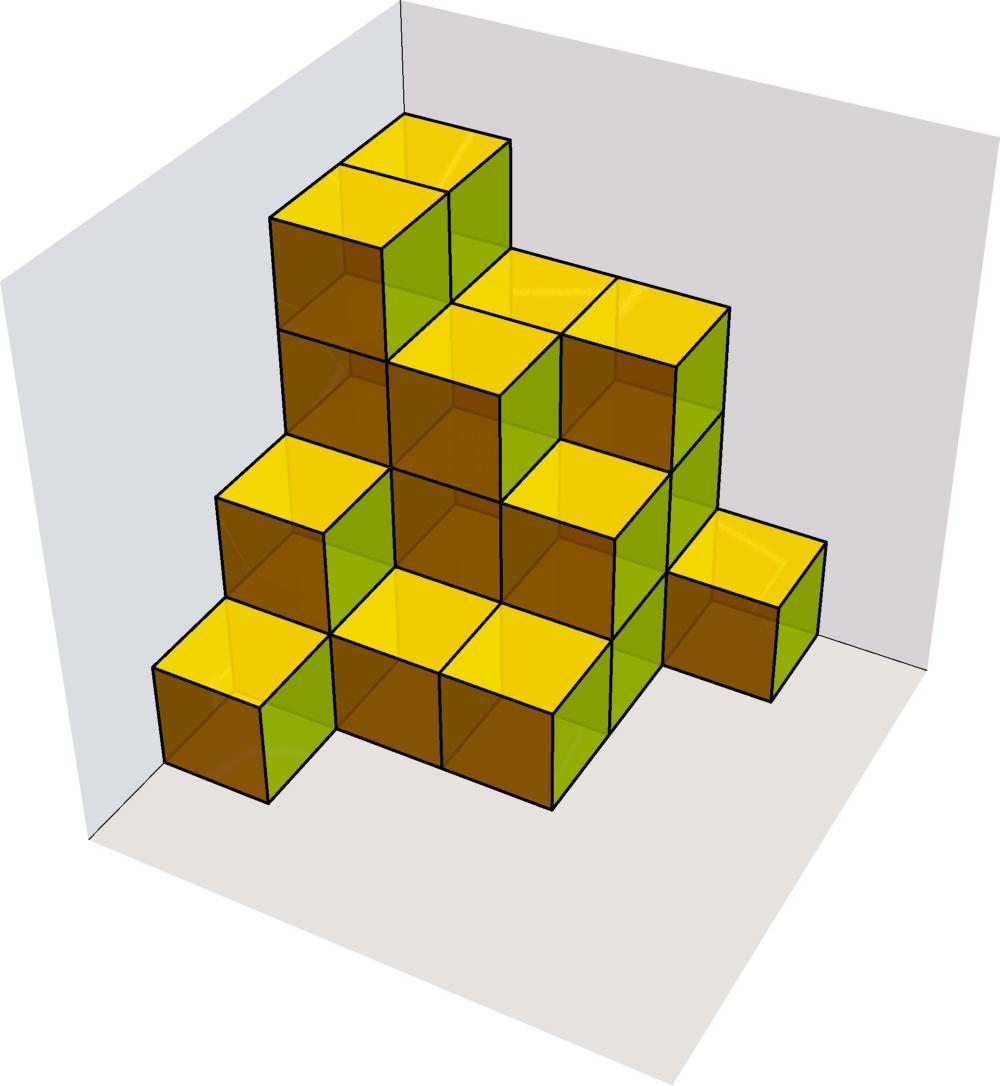}
\end{minipage}
\qquad \qquad
\begin{minipage}{5cm}
\begin{tikzpicture}[scale=0.50]
\filldraw[fill=white,thick] (0,0) rectangle (1,1);
\filldraw[fill=white,thick] (0,1) rectangle (1,2);
\filldraw[fill=white,thick] (1,1) rectangle (2,2);
\filldraw[fill=white,thick] (2,1) rectangle (3,2);
\filldraw[fill=white,thick] (0,2) rectangle (1,3);
\filldraw[fill=white,thick] (1,2) rectangle (2,3);
\filldraw[fill=white,thick] (2,2) rectangle (3,3);
\filldraw[fill=white,thick] (0,3) rectangle (1,4);
\filldraw[fill=white,thick] (1,3) rectangle (2,4);
\filldraw[fill=white,thick] (2,3) rectangle (3,4);
\filldraw[fill=white,thick] (3,3) rectangle (4,4);
\node at (0.5,1.5) {2};
\node at (0.5,2.5) {4};
\node at (1.5,2.5) {3};
\node at (2.5,2.5) {2};
\node at (0.5,3.5) {4};
\node at (1.5,3.5) {3};
\node at (2.5,3.5) {3};
\end{tikzpicture}
\end{minipage}
\caption{An example of a plane partition and its projection from the direction of the third axis. It is customary not to fill in $1$	for the columns of height one.}
\label{figplanepart}
\end{figure}

Let us explain now the notation that we started using already in this section. The plane partitions are a generalization of the usual (two-dimensional) partitions of integers. Just as the partitions are in one-to-one correspondence with the Young diagrams, the plane partitions correspond to the configurations of boxes in the corner of a room - for one such configuration see the figure \ref{figplanepart}. By choosing a direction, say $x_3$ and projecting along it, we may associate to each plane partition a Young diagram filled with integers which denote the heights of the projected columns. See figure \ref{figplanepart} for an example. In this way, the plane partitions correspond to the Young diagrams filled in with integers that are non-increasing along the rows and along the columns. In the following, we will use both notations interchangeably. The direction $x_3$ will be the one that we will project along and in the 3 dimensional diagrams it will go upwards. The $x_1$ coordinate will increase as we go to the right, so the rows of the Young diagram have fixed $x_2$ coordinate and increasing $x_1$ coordinate.

Once we move on to more complicated representations, we will consider a generalization of the plane partitions where an infinite number of boxes is allowed along the coordinate axes. We will require the plane partition to asymptote a Young diagram along each of coordinate axes as illustrated in figure \ref{figminconf3}. The asymptotics Young diagram in $x_1$ direction is in the plane parallel to the $x_2-x_3$ plane, and we will denote it by either by $\Lambda_{23}$ or $\Lambda_{32} \equiv (\Lambda_{32})^T$. The first index labels the direction of an increasing column and the second index labels the direction of an increasing row. In the example of figure \ref{figminconf3}, the asymptotics in $x_1$ direction can be denoted either by $\Lambda_{23} = \ydiagram{2,1,1}$ or by $\Lambda_{32} = \ydiagram{3,1}$. Exchanging the order of two labels just transposes the corresponding Young diagram. Analogously, we use the label $\Lambda_{13} = \ydiagram{3,2}$ or $\Lambda_{31} = \ydiagram{2,2,1}$ to describe the asymptotic Young diagram in the $x_2$ direction.

\subsection{Representations of $\mathcal{Y}$ on the plane partitions}
\label{secplanepart}

The most interesting class of representations of $\mathcal{Y}$ found so far are the MacMahon representations on the plane partitions. In \cite{tsymbaliukrev} an explicit form of the Fock representation of $\mathcal{Y}$ on the two-dimensional partitions was given, but the representations on the plane partitions are in principle easy to write down starting from the $q$-analogue of $\mathcal{Y}$ \cite{feigin2012quantum} and following \cite{tsymbaliukrev}. In this section we will study general properties of this class of representations by extracting the general structure of the representations from \cite{feigin2012quantum,tsymbaliukrev} and using the defining relations of $\mathcal{Y}$.

\begin{figure}
\begin{center}
\includegraphics[scale=0.4]{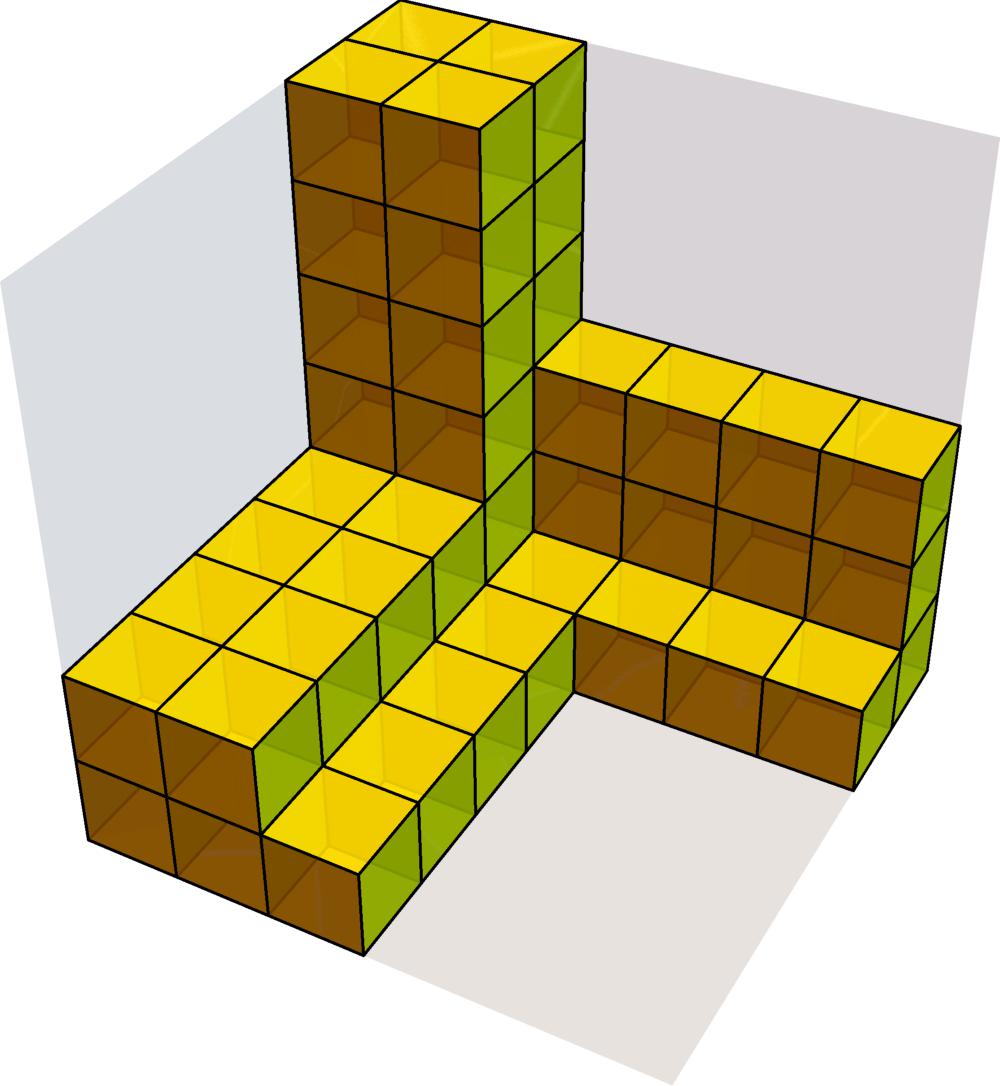}
\end{center}
\caption{An example of a minimal configuration of boxes with asymptotic Young diagrams $(\Lambda_{13},\Lambda_{23},\Lambda_{12}) = \left(\ydiagram{3,2},\ydiagram{2,1,1},\ydiagram{2,2}\right)$. There is an infinite number of boxes along each of the coordinate axes and no box in this configuration can be removed without violating the plane partition condition. This configuration corresponds to a highest weight (primary) state.}
\label{figminconf3}
\end{figure}

The representation space has a basis corresponding to the partitions or to the generalized plane partitions (possibly allowing for some geometric restrictions or certain asymptotics). The precise class of the combinatorial objects will depend in particular on the values of $h_j$ parameters and on the values of the charges of the highest weight state.

In the simplest case (generic values of $h_j$ and the values of $\psi_j$ charges corresponding to the vacuum representation discussed in the previous section) the highest weight state is given by an empty configuration, but other possible highest weight states corresponding to an infinite configuration of boxes with given Young diagram asymptotics as illustrated in figure \ref{figminconf3} are also allowed. The configuration in figure \ref{figminconf3} corresponds to a highest weight state, because we cannot remove any box from this configuration without violating the plane partition condition. Other states in the representation are plane partitions obtained by adding a finite number of boxes to the highest weight configuration, such that at every step the configuration satisfies the plane partition restrictions.

The generating field $\psi(z)$ acts diagonally in this basis,
\begin{empheq}[box=\fbox]{equation}
\psi(u) \ket{\Lambda} = \psi_{\Lambda}(u) \ket{\Lambda}
\end{empheq}
while the operators $e_j$ act by adding a box to the diagram,
\begin{equation}
\label{eansatz}
e_j \ket{\Lambda} = \sum_{\Box \in \Lambda^+} (q+h_{\Box})^j E(\Lambda \to \Lambda + \Box) \ket{\Lambda+\Box}.
\end{equation}
Here the sum is over all the diagrams which have a box added at geometrically allowed position and $E(\Lambda \to \Lambda+\Box)$ is the amplitude of this process, yet to be determined. The coefficient $h_{\Box}$ is the weighted position of a given box
\begin{empheq}[box=\fbox]{equation}
h_{\Box} = h_1 x_1(\Box) + h_2 x_2(\Box) + h_3 x_3(\Box).
\end{empheq}
The coordinates $x_j$ are positive integers (in fact the coordinates of the origin can be shifted by an arbitrary constant $x_j \to x_j + a$ without any effect because of the constraint $h_1 + h_2 + h_3 = 0$). Similarly, the operator $f_j$ removes a box from the diagram,
\begin{equation}
f_j \ket{\Lambda} = \sum_{\Box \in \Lambda^-} (q+h_{\Box})^j F(\Lambda \to \Lambda - \Box) \ket{\Lambda-\Box},
\end{equation}
where the sum runs over all possible boxes that can be removed such that the resulting diagram is still compatible with the plane partition rules. In terms of the generating functions, we can write these compactly as
\begin{align}
e(u) \ket{\Lambda} & = \sum_{\Box \in \Lambda^+} \frac{E(\Lambda\to\Lambda+\Box)}{u-q-h_{\Box}} \ket{\Lambda+\Box} \\
f(u) \ket{\Lambda} & = \sum_{\Box \in \Lambda^-} \frac{F(\Lambda\to\Lambda-\Box)}{u-q-h_{\Box}} \ket{\Lambda-\Box}.
\end{align}
Let us now verify the commutation relations. We can start with (Y4). Multiplying this relation by $\sigma_3 u^{-j-1}$ and summing over $j$, we find
\begin{equation}
0 = \left[ \psi(u), u^3 e_k - 3u^2 e_{k+1} + 3u e_{k+2} - e_{k+3} + \sigma_2 u e_k - \sigma_2 e_{k+1} \right] - \sigma_3 \left\{ \psi(u), e_k \right\}.
\end{equation}
Note that there was a non-trivial cancellation between $-\sigma_3 \left[ \psi_2, e_k \right]$ and $\sigma_3 \left\{ \mathbbm{1}, e_k \right\}$ which follows from (Y4'). Acting now on $\ket{\Lambda}$ and using the ansatz (\ref{eansatz}), we find
\begin{equation}
\frac{\psi_{\Lambda+\Box}(u)}{\psi_{\Lambda}(u)} = \varphi(u-q-h_{\Box})
\end{equation}
which can be used to iteratively determine the eigenvalues of $\psi(u)$ on any descendant state in terms of eigenvalues $\psi_0(u)$ of the highest weight state,
\begin{empheq}[box=\fbox]{equation}
\label{psiproduct}
\psi_{\Lambda}(u) = \psi_0(u) \prod_{\Box \in \Lambda} \varphi(u-q-h_{\Box}).
\end{empheq}
Here $\varphi(u)$ is again the rational function (\ref{magicfunction}). The relations (Y5) and (Y5') lead to the same conclusion. The value of $\psi_0(u)$ for the vacuum configuration (no boxes) was determined in the previous section, formula (\ref{vacuumcharges}). It is consistent to assume that the value of $\psi(u)$ charges of the highest weight state in the case of the highest weight state with the non-trivial Young diagram asymptotics is equal to the product of the vacuum $\psi_0(u)$ times the product of $\varphi(u-q-h_{\Box})$ over all boxes of the minimal configuration of boxes. This product converges, as will be discussed in greater detail and illustrated on examples in section \ref{seccharges}.

Let us now turn to (Y3). Given our ansatz, it implies
\begin{equation}
E(\Lambda+B\to\Lambda+A+B)F(\Lambda+A+B\to\Lambda+A) = F(\Lambda+B\to\Lambda)E(\Lambda\to\Lambda+A)
\end{equation}
for independent boxes $A$ and $B$ (so that the addition and removal of different boxes commute) as well as the relation
\begin{equation}
\psi_{\Lambda}(u) = 1 + \sigma_3 \sum_{\Box \in \Lambda^-} \frac{E(\Lambda-\Box\to\Lambda)F(\Lambda\to\Lambda-\Box)}{u-q-h_{\Box}} - \sigma_3 \sum_{\Box \in \Lambda^+} \frac{E(\Lambda\to\Lambda+\Box)F(\Lambda+\Box\to\Lambda)}{u-q-h_{\Box}}
\end{equation}
which relates the residue of $\psi_{\Lambda}(u)$ to the amplitudes of addition and removal of a box. We will see later in section \ref{secrhombic} that $\psi_{\Lambda}(u)$ has only simple poles in $u$-space and that these poles can appear only at the weighted positions of the boxes in $\Lambda^{\pm}$. Since we cannot have both $\Box \in \Lambda^+$ or $\Box \in \Lambda^-$ at the same time, in the case that $\Box \in \Lambda^+$ this tells us that	
\begin{equation}
\label{efres1}
\sigma_3 E(\Lambda\to\Lambda+\Box)F(\Lambda+\Box\to\Lambda) = - \res_{u \to q+h_{\Box}} \psi_{\Lambda}(u)
\end{equation}
and similarly for $\Box \in \Lambda^-$
\begin{equation}
\label{efres2}
\sigma_3 E(\Lambda-\Box\to\Lambda)F(\Lambda\to\Lambda-\Box) = + \res_{u \to q+h_{\Box}} \psi_{\Lambda}(u).
\end{equation}
The commutation relation (Y1) relates the amplitudes of adding two boxes in a different order,
\begin{equation}
\label{eeholo}
\frac{E(\Lambda\to\Lambda+B)E(\Lambda+B\to\Lambda+A+B)}{E(\Lambda\to\Lambda+A)E(\Lambda+A\to\Lambda+A+B)} = \varphi (h_A-h_B).
\end{equation}
Here $A$ and $B$ represent the two boxes that can be added to a diagram in different order (such that both amplitudes $E(\Lambda\to\Lambda+A)$ and $E(\Lambda\to\Lambda+B)$ are nonzero). Similarly, the relation (Y2) is imposes the constraint
\begin{equation}
\frac{F(\Lambda+A+B\to\Lambda+A)F(\Lambda+A\to\Lambda)}{F(\Lambda+A+B\to\Lambda+B)F(\Lambda+B\to\Lambda)} = \varphi (h_B-h_A).
\end{equation}
Finally, the relation (Y6) is a cubic relation on amplitudes $E(\Lambda\to\Lambda+A)$,
\begin{multline}
\sum_{\pi \in S_3} \left(h_{A_{\pi(1)}} - 2h_{A_{\pi(2)}} + h_{A_{\pi(3)}} \right) E(\Lambda\to\Lambda+A_{\pi(1)}) E(\Lambda+A_{\pi(1)}\to\Lambda+A_{\pi(1)}+A_{\pi(2)}) \times \\
\times E(\Lambda+A_{\pi(1)}+A_{\pi(2)}\to\Lambda+A_{\pi(1)}+A_{\pi(2)}+A_{\pi(3)}) = 0
\end{multline}
and similarly for (Y7).

\paragraph{Young's lattice}
We can picture the previous discussion diagrammatically in terms of a generalization of the Young's lattice to the plane partitions. We associate to a representation an oriented graph with nodes labelled by the plane partitions and with oriented edges connecting nodes which differ by an addition of a box, the arrow pointing in the direction of the addition of the box. The basis vectors (corresponding to the vertices of the graph) are uniquely determined up to an overall rescaling by a $\mathbbm{C}^{\times}$ factor. The box addition amplitudes $E(\Lambda\to\Lambda+\Box)$ are associated to the edges and we can think of them as a discretized $\mathbbm{C}^{\times}$ connection - under the rescaling of the basis vectors they transform as a holonomy from one vertex to another.

For every plane partition to which we can add two different boxes we have a square in the lattice and the quadratic relation (\ref{eeholo}) expresses that the holonomy around this square is given by $\varphi(h_A-h_B)$. In this way $\varphi(h_A-h_B)$ associated to squares behaves as an integrated curvature (holonomy around the closed loop).

We may now count the number of unknown $E(\Lambda\to\Lambda+\Box)$ coefficients and determine to which extent the relation (\ref{eeholo}) determines them uniquely. Let us first introduce some notation. We will denote by $c_{nk}$ the number of partitions with $n$ boxes to which we can add another box in $k$ possible ways (so there are $k$ arrows starting at this partition). For example, if we specialize to the usual 2 dimensional partitions, the generating function of these integers is
\begin{equation}
\prod_{j=1}^{\infty} \left( 1 + \frac{tq^j}{1-q^j} \right) = \sum_{n=0}^{\infty} \sum_{k>0} c_{nk} q^n t^{k-1} = \sum_{\Lambda} q^{|\Lambda|} t^{|\Lambda^+|-1}
\end{equation}
For a given partition $\Lambda$ we denoted $|\Lambda^+|$ the number of ways to add a box to $\Lambda$. Using these integers, we can count the number of $d$-dimensional hypercubes starting at $\Lambda$ - it is simply given by
\begin{equation}
{|\Lambda^+| \choose d}.
\end{equation}

The amount of gauge freedom (basis vector rescaling freedom) at level $n$ of the Young's lattice is given by the number of partitions with a given number of boxes,
\begin{equation}
\sum_{|\Lambda|=n} 1 = \sum_{|\Lambda|=n} {|\Lambda^+| \choose 0}.
\end{equation}
The number of unknown $E(\Lambda\to\Lambda+\Box)$ amplitudes that end at level $n$ is equal to the number of edges connecting levels $n-1$ and $n$,
\begin{equation}
\sum_{|\Lambda|=n-1} |\Lambda^+| = \sum_{|\Lambda|=n-1} {|\Lambda^+| \choose 1}.
\end{equation}
Finally, the number of curvature constraints (\ref{eeholo}) coming from squares between levels $n-2$ and $n$ is
\begin{equation}
\sum_{|\Lambda|=n-2} {|\Lambda^+| \choose 2}.
\end{equation}
We see that the conditions (\ref{eeholo}) would determine amplitudes $E(\Lambda\to\Lambda+\Box)$ uniquely up to the rescaling freedom if
\begin{equation}
\sum_{j=0}^2 (-1)^j \sum_{|\Lambda|=n-j} {|\Lambda^+| \choose j} \stackrel{?}{=} 0
\end{equation}
This relation is only satisfied at lower levels. But it looks as a beginning of a graded generalization of the Euler characteristic. We can guess the full formula to be
\begin{equation}
\sum_{j=0}^n (-1)^j \sum_{|\Lambda|=n-j} {|\Lambda^+| \choose j} \stackrel{?}{=} 0
\end{equation}
and in fact it is satisfied both for the usual two-dimensional plane partitions as well as for the plane partitions at lower levels and plane partitions with non-trivial asymptotics. This is illustrated for the vacuum MacMahon representation in the following table.

\begin{center}
\begin{tabular}{|c|c|ccccccc|}
\hline
level $n$ & GF of starting edges & 0d & 1d & 2d & 3d & 4d & 5d & 6d \\
\hline
0 & $1$ & $1$ & $1$ & $0$ & $0$ & $0$ & $0$ & $0$ \\
1 & $z^2$ & $1$ & $3$ & $3$ & $1$ & $0$ & $0$ & $0$ \\
2 & $3z^2$ & $3$ & $9$ & $9$ & $3$ & $0$ & $0$ & $0$ \\
3 & $3z^2 + 3z^3$ & $6$ & $21$ & $27$ & $15$ & $3$ & $0$ & $0$ \\
4 & $6z^2 + 6z^3 + z^5$ & $13$ & $48$ & $69$ & $50$ & $21$ & $6$ & $1$ \\
5 & $3z^2 + 21 z^3 + 3z^4 + 3z^5$ & $24$ & $102$ & $174$ & $153$ & $75$ & $21$ & $3$ \\
6 & $9z^2 + 21 z^3 + 6z^4 + 12 z^5$ & $48$ & $213$ & $393$ & $393$ & $231$ & $78$ & $12$ \\
\hline
\end{tabular}
\end{center}

We can immediately see the two types of sequences giving us vanishing Euler characteristic: one is along the rows (which is the Euler characteristic for hypercubes starting at a given level), for example at level $6$
\begin{equation}
48 - 213 + 393 - 393 + 231 - 78 + 12 = 0.
\end{equation}
Second vanishing Euler characteristic are associated to antidiagonals of the table. This is the Euler characteristic for hypercubes ending at a given level. For instance for the hypercubes ending at level $6$ we have
\begin{equation}
48 - 102 + 69 - 15 = 0.
\end{equation}

\paragraph{Consistent solution}
Although the combinatorial analysis of the previous section showed us that relation (\ref{eeholo}) itself cannot uniquely fix the amplitudes $E(\Lambda\to\Lambda+\Box)$, we also have other relations at our disposal. We can for example require $f_j$ to be conjugate to $e_j$ which fixes the normalization of the basis vectors:
\begin{equation}
\label{efconnection}
E(\Lambda\to\Lambda+\Box) = F(\Lambda+\Box\to\Lambda)
\end{equation}
Using this additional condition together with (\ref{efres1}), we can solve for $E(\Lambda\to\Lambda+\Box)$ and find \footnote{See also equations (2.18-2.21) of \cite{Bourgine:2015szm} where authors arrive at analogous conclusions in the context of $SH^c$ algebra.}
\begin{empheq}[box=\fbox]{align}
\label{esol}
E(\Lambda\to\Lambda+\Box) = \sqrt{-\frac{1}{\sigma_3} \res_{u \to q + h_{\Box}} \psi_{\Lambda}(u)}
\end{empheq}
We should check that this expression for $E(\Lambda\to\Lambda+\Box)$ coefficients solves all the relations of $\mathcal{Y}$. The only ambiguity would remain in the choice of signs of square roots in (\ref{esol}). We don't give here a proof that (\ref{esol}) solves all the relations of $\mathcal{Y}$, but in all examples checked there was a consistent choice of signs such that the quadratic and cubic identities in $\mathcal{Y}$ were satisfied. If this is true in general, the form of (\ref{esol}) implies in particular that everything can determined in terms of the function $\varphi(u)$ (\ref{magicfunction}) and the geometrical shape of the corresponding plane partition.

In the following, we will only assume that there exists a consistent solution satisfying the relations of section \ref{secplanepart}. For the usual partitions, such a solution is given in \cite{tsymbaliukrev} while for plane partitions such a solution is known to exist for $q$-deformed versions of the algebra \cite{feigin2012quantum}, so in particular is expected to exist also in $\mathcal{Y}$.

\subsection{Rhombic tilings and shell formula}
\label{secrhombic}

Although the formula (\ref{psiproduct}) for the value of the generating function of $\psi$-charges is rather explicit, there are many cancellations between the contributions from the neighbouring boxes. There is another formula by \cite{feigin2012quantum}, the shell formula, which does not involve the boxes in the bulk part of the plane partition, but only those that lie on the surface.

\begin{figure}
\centering
\begin{tikzpicture}[scale=0.6]
\definecolor{mbr}{RGB}{128,71,0}
\definecolor{mgr}{RGB}{148,166,0}
\definecolor{mor}{RGB}{242,198,0}
\filldraw[fill=mbr,draw=black] (0,0) -- ++(330:1) -- ++(90:1) -- ++(150:1) -- ++(270:1);
\filldraw[fill=mbr,draw=black] ++(330:1) -- ++(330:1) -- ++(90:1) -- ++(150:1) -- ++(270:1);
\filldraw[fill=mbr,draw=black] ++(330:2) -- ++(330:1) -- ++(90:1) -- ++(150:1) -- ++(270:1);
\filldraw[fill=mbr,draw=black] ++(330:3) -- ++(330:1) -- ++(90:1) -- ++(150:1) -- ++(270:1);
\filldraw[fill=mbr,draw=black] (90:1) -- ++(330:1) -- ++(90:1) -- ++(150:1) -- ++(270:1);
\filldraw[fill=mbr,draw=black] (90:1) ++(330:1) -- ++(330:1) -- ++(90:1) -- ++(150:1) -- ++(270:1);
\filldraw[fill=mbr,draw=black] (90:1) ++(330:2) -- ++(330:1) -- ++(90:1) -- ++(150:1) -- ++(270:1);
\filldraw[fill=mbr,draw=black] (90:1) ++(330:3) -- ++(330:1) -- ++(90:1) -- ++(150:1) -- ++(270:1);
\filldraw[fill=mbr,draw=black] (90:2) -- ++(330:1) -- ++(90:1) -- ++(150:1) -- ++(270:1);
\filldraw[fill=mbr,draw=black] (90:2) ++(330:1) -- ++(330:1) -- ++(90:1) -- ++(150:1) -- ++(270:1);
\filldraw[fill=mbr,draw=black] (90:2) ++(330:2) -- ++(330:1) -- ++(90:1) -- ++(150:1) -- ++(270:1);
\filldraw[fill=mbr,draw=black] (90:2) ++(330:3) -- ++(330:1) -- ++(90:1) -- ++(150:1) -- ++(270:1);
\filldraw[fill=mbr,draw=black] (90:3) -- ++(330:1) -- ++(90:1) -- ++(150:1) -- ++(270:1);
\filldraw[fill=mbr,draw=black] (90:3) ++(330:1) -- ++(330:1) -- ++(90:1) -- ++(150:1) -- ++(270:1);
\filldraw[fill=mbr,draw=black] (90:3) ++(330:2) -- ++(330:1) -- ++(90:1) -- ++(150:1) -- ++(270:1);
\filldraw[fill=mbr,draw=black] (90:3) ++(330:3) -- ++(330:1) -- ++(90:1) -- ++(150:1) -- ++(270:1);
\filldraw[fill=mgr,draw=black] (0,0) -- ++(90:1) -- ++(210:1) -- ++(270:1) -- ++(30:1);
\filldraw[fill=mgr,draw=black] ++(90:1) -- ++(90:1) -- ++(210:1) -- ++(270:1) -- ++(30:1);
\filldraw[fill=mgr,draw=black] ++(90:2) -- ++(90:1) -- ++(210:1) -- ++(270:1) -- ++(30:1);
\filldraw[fill=mgr,draw=black] ++(90:3) -- ++(90:1) -- ++(210:1) -- ++(270:1) -- ++(30:1);
\filldraw[fill=mgr,draw=black] (210:1) -- ++(90:1) -- ++(210:1) -- ++(270:1) -- ++(30:1);
\filldraw[fill=mgr,draw=black] (210:1) ++(90:1) -- ++(90:1) -- ++(210:1) -- ++(270:1) -- ++(30:1);
\filldraw[fill=mgr,draw=black] (210:1) ++(90:2) -- ++(90:1) -- ++(210:1) -- ++(270:1) -- ++(30:1);
\filldraw[fill=mgr,draw=black] (210:1) ++(90:3) -- ++(90:1) -- ++(210:1) -- ++(270:1) -- ++(30:1);
\filldraw[fill=mgr,draw=black] (210:2) -- ++(90:1) -- ++(210:1) -- ++(270:1) -- ++(30:1);
\filldraw[fill=mgr,draw=black] (210:2) ++(90:1) -- ++(90:1) -- ++(210:1) -- ++(270:1) -- ++(30:1);
\filldraw[fill=mgr,draw=black] (210:2) ++(90:2) -- ++(90:1) -- ++(210:1) -- ++(270:1) -- ++(30:1);
\filldraw[fill=mgr,draw=black] (210:2) ++(90:3) -- ++(90:1) -- ++(210:1) -- ++(270:1) -- ++(30:1);
\filldraw[fill=mgr,draw=black] (210:3) -- ++(90:1) -- ++(210:1) -- ++(270:1) -- ++(30:1);
\filldraw[fill=mgr,draw=black] (210:3) ++(90:1) -- ++(90:1) -- ++(210:1) -- ++(270:1) -- ++(30:1);
\filldraw[fill=mgr,draw=black] (210:3) ++(90:2) -- ++(90:1) -- ++(210:1) -- ++(270:1) -- ++(30:1);
\filldraw[fill=mgr,draw=black] (210:3) ++(90:3) -- ++(90:1) -- ++(210:1) -- ++(270:1) -- ++(30:1);
\filldraw[fill=mor,draw=black] (0,0) -- ++(210:1) -- ++(330:1) -- ++(30:1) -- ++(150:1);
\filldraw[fill=mor,draw=black] ++(330:1) -- ++(210:1) -- ++(330:1) -- ++(30:1) -- ++(150:1);
\filldraw[fill=mor,draw=black] ++(330:2) -- ++(210:1) -- ++(330:1) -- ++(30:1) -- ++(150:1);
\filldraw[fill=mor,draw=black] ++(330:3) -- ++(210:1) -- ++(330:1) -- ++(30:1) -- ++(150:1);
\filldraw[fill=mor,draw=black] ++(210:1) -- ++(210:1) -- ++(330:1) -- ++(30:1) -- ++(150:1);
\filldraw[fill=mor,draw=black] ++(210:1) ++(330:1) -- ++(210:1) -- ++(330:1) -- ++(30:1) -- ++(150:1);
\filldraw[fill=mor,draw=black] ++(210:1) ++(330:2) -- ++(210:1) -- ++(330:1) -- ++(30:1) -- ++(150:1);
\filldraw[fill=mor,draw=black] ++(210:1) ++(330:3) -- ++(210:1) -- ++(330:1) -- ++(30:1) -- ++(150:1);
\filldraw[fill=mor,draw=black] ++(210:2) -- ++(210:1) -- ++(330:1) -- ++(30:1) -- ++(150:1);
\filldraw[fill=mor,draw=black] ++(210:2) ++(330:1) -- ++(210:1) -- ++(330:1) -- ++(30:1) -- ++(150:1);
\filldraw[fill=mor,draw=black] ++(210:2) ++(330:2) -- ++(210:1) -- ++(330:1) -- ++(30:1) -- ++(150:1);
\filldraw[fill=mor,draw=black] ++(210:2) ++(330:3) -- ++(210:1) -- ++(330:1) -- ++(30:1) -- ++(150:1);
\filldraw[fill=mor,draw=black] ++(210:3) -- ++(210:1) -- ++(330:1) -- ++(30:1) -- ++(150:1);
\filldraw[fill=mor,draw=black] ++(210:3) ++(330:1) -- ++(210:1) -- ++(330:1) -- ++(30:1) -- ++(150:1);
\filldraw[fill=mor,draw=black] ++(210:3) ++(330:2) -- ++(210:1) -- ++(330:1) -- ++(30:1) -- ++(150:1);
\filldraw[fill=mor,draw=black] ++(210:3) ++(330:3) -- ++(210:1) -- ++(330:1) -- ++(30:1) -- ++(150:1);
\end{tikzpicture}
\caption{A representation of the highest weight state of the vacuum representation of $\mathcal{Y}$ as a rhombic tiling of the plane. There is one trivalent vertex (point defect) at the centre and three half-lines (line defects) extending from the centre to infinity. We can think of these half-lines as of defect lines which bound the domains of tiles of the same orientation. All the vertices except for the central one are four-valent.}
\label{vacuumtiling}
\end{figure}

In order to explain the shell formula combinatorially, we will use the known map from the plane partition to the dual picture of rhombic tilings of the plane. Let us focus for simplicity on the plane partitions which have a trivial Young diagram asymptotics (these plane partitions are in 1-1 correspondence with states in the vacuum representation) and furthermore let us consider the case of the uncharged representations ($\psi_1 = 0$). It is useful to consider $\mathbbm{R}^3$ with all the octants filled with boxes except for the positive one (until now we didn't consider these octants at all and we were only building plane partitions inside of the positive octant). When viewed from a distant point in $(1,1,1)$-direction towards the origin $(0,0,0)$, the vacuum configuration looks as illustrated in figure \ref{vacuumtiling} - it is a rhombic plane tiling with three sectors. In each of these three sectors the tiles have the same orientation and the sectors are separated by line defects. These line defects in turn all end at a point defect which in this case is just the origin. Any other plane partition will correspond to a unique rhombic tiling of the plane that is asymptotically same as the vacuum tiling. From the definition of the plane partition we know that if there is a box at position $(x_1,x_2,x_3)$, all the boxes $(x_1-l,x_2-l,x_3-l)$, $l \in \mathbbm{N}$ are also in $\Lambda$ (or in the filled octants). This lets us uniquely reconstruct the plane partition from its associated rhombic tiling. Addition or removal of a box in the plane partition picture corresponds to the operation
\begin{equation}
\label{trimove}
\begin{tikzpicture}[scale=0.6]
\definecolor{mbr}{RGB}{128,71,0}
\definecolor{mgr}{RGB}{148,166,0}
\definecolor{mor}{RGB}{242,198,0}
\filldraw[fill=mor] (0,0) ++(270:1) -- ++(30:1) -- ++(150:1) -- ++(210:1) -- ++(330:1);
\filldraw[fill=mgr] (0,0) ++(150:1) -- ++(270:1) -- ++(30:1) -- ++(90:1) -- ++(210:1);
\filldraw[fill=mbr] (0,0) ++(30:1) -- ++(150:1) -- ++(270:1) -- ++(330:1) -- ++(90:1);
\filldraw[fill=mor] (5,0) -- ++(30:1) -- ++(150:1) -- ++(210:1) -- ++(330:1);
\filldraw[fill=mgr] (5,0) -- ++(270:1) -- ++(30:1) -- ++(90:1) -- ++(210:1);
\filldraw[fill=mbr] (5,0) -- ++(150:1) -- ++(270:1) -- ++(330:1) -- ++(90:1);
\draw[thick,<->] (1.5,0) -- (3.5,0);
\end{tikzpicture}\end{equation}
on the tiling.

We can now turn to the generating function of the conserved charges $\psi_\Lambda(u)$ associated to plane partition $\Lambda$. From (\ref{psiproduct}) we know that it is a rational function with zeros and poles at $u = x_1 h_1 + x_2 h_2 + x_3 h_3$ where $x_j \in \mathbbm{Z}$. Since the overall normalization of $\psi_\Lambda(u)$ is fixed by its definition (\ref{genfunc}), we only need to understand the order of zero or pole at each lattice point. Let us fix a point with coordinates $(x_1,x_2,x_3)$ such that the points
\begin{equation}
(x_1,x_2,x_3)+l(1,1,1)
\end{equation}
with $l<0$ are in $\Lambda$ while points with $l>0$ are not in $\Lambda$. The order of zero or pole at $u = \sum_j x_j h_j$ depends only on the boxes
\begin{equation}
(x_1 \pm 1,x_2,x_3), \quad \quad (x_1, x_2 \pm 1, x_3) \quad \quad \text{and} \quad \quad (x_1, x_2, x_3 \pm 1) \quad \quad mod \; (1,1,1),
\end{equation}
see (\ref{magicfunction}). Because of the plane partition property, the contribution of boxes deeper in $(1,1,1)$ direction exactly cancels out. This reduces the analysis only to the neighbouring boxes. There are $18$ possible neighbouring configurations:
\begin{align}
\nonumber
\text{simple pole:}\quad & \begin{ytableau} \scriptstyle 1 & \scriptstyle 0 \\ \scriptstyle 0 & \scriptstyle 0 \end{ytableau} \quad \begin{ytableau} \scriptstyle 2 & \scriptstyle 2 \\ \scriptstyle 2 & \scriptstyle 1 \end{ytableau}\\
\nonumber
\text{no contribution:}\quad & \begin{ytableau} \scriptstyle 1 & \scriptstyle 1 \\ \scriptstyle 0 & \scriptstyle 0 \end{ytableau} \quad \begin{ytableau} \scriptstyle 1 & \scriptstyle 0 \\ \scriptstyle 1 & \scriptstyle 0 \end{ytableau} \quad \begin{ytableau} \scriptstyle 2 & \scriptstyle 0 \\ \scriptstyle 0 & \scriptstyle 0 \end{ytableau} \quad \begin{ytableau} \scriptstyle 1 & \scriptstyle 1 \\ \scriptstyle 1 & \scriptstyle 1 \end{ytableau} \quad \begin{ytableau} \scriptstyle 2 & \scriptstyle 2 \\ \scriptstyle 0 & \scriptstyle 0 \end{ytableau} \quad \begin{ytableau} \scriptstyle 2 & \scriptstyle 0 \\ \scriptstyle 2 & \scriptstyle 0 \end{ytableau} \quad \begin{ytableau} \scriptstyle 2 & \scriptstyle 2 \\ \scriptstyle 1 & \scriptstyle 1 \end{ytableau} \quad \begin{ytableau} \scriptstyle 2 & \scriptstyle 1 \\ \scriptstyle 2 & \scriptstyle 1 \end{ytableau} \quad \begin{ytableau} \scriptstyle 2 & \scriptstyle 2 \\ \scriptstyle 2 & \scriptstyle 0 \end{ytableau} \\
\text{simple zero:}\quad & \begin{ytableau} \scriptstyle 1 & \scriptstyle 1 \\ \scriptstyle 1 & \scriptstyle 0 \end{ytableau} \quad \begin{ytableau} \scriptstyle 2 & \scriptstyle 1 \\ \scriptstyle 0 & \scriptstyle 0 \end{ytableau} \quad \begin{ytableau} \scriptstyle 2 & \scriptstyle 0 \\ \scriptstyle 1 & \scriptstyle 0 \end{ytableau} \quad \begin{ytableau} \scriptstyle 2 & \scriptstyle 1 \\ \scriptstyle 1 & \scriptstyle 1 \end{ytableau} \quad \begin{ytableau} \scriptstyle 2 & \scriptstyle 1 \\ \scriptstyle 2 & \scriptstyle 0 \end{ytableau} \quad \begin{ytableau} \scriptstyle 2 & \scriptstyle 2 \\ \scriptstyle 1 & \scriptstyle 0 \end{ytableau} \\
\nonumber
\text{double zero:}\quad & \begin{ytableau} \scriptstyle 2 & \scriptstyle 1 \\ \scriptstyle 1 & \scriptstyle 0 \end{ytableau}
\end{align}
For example the configuration of boxes
\begin{equation}
\begin{ytableau} \scriptstyle 1 & \scriptstyle 1 \\ \scriptstyle 1 & \scriptstyle 0 \end{ytableau} \; \leftrightarrow \parbox[c]{1.8cm}{\includegraphics[width=3cm]{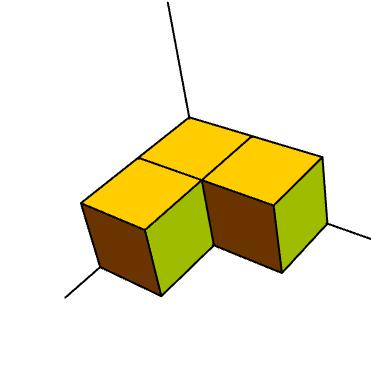}}
\end{equation}
contributes to $\psi_\Lambda$ by a simple zero at position of the vertex at $(1,1,1)$ (we don't consider other zeros and poles at the moment). In the rhombic tiling picture, we find $5$ different types of local configurations as shown in the following table:
\begin{center}
\begin{tabular}{|c|c|c|c|c|c|}
\hline
local configuration & 
\begin{tikzpicture}[scale=0.6]
\definecolor{mbr}{RGB}{128,71,0}
\definecolor{mgr}{RGB}{148,166,0}
\definecolor{mor}{RGB}{242,198,0}
\filldraw[fill=mgr,draw=black] (0,0) -- ++(30:1) -- ++(90:1) -- ++(210:1) -- ++(270:1);
\filldraw[fill=mbr,draw=black] (0,0) -- ++(90:1) -- ++(150:1) -- ++(270:1) -- ++(330:1);
\filldraw[fill=mor,draw=black] (0,0) -- ++(150:1) -- ++(210:1) -- ++(330:1) -- ++(30:1);
\filldraw[fill=mgr,draw=black] (0,0) -- ++(210:1) -- ++(270:1) -- ++(30:1) -- ++(90:1);
\filldraw[fill=mbr,draw=black] (0,0) -- ++(270:1) -- ++(330:1) -- ++(90:1) -- ++(150:1);
\filldraw[fill=mor,draw=black] (0,0) -- ++(330:1) -- ++(30:1) -- ++(150:1) -- ++(210:1);
\end{tikzpicture} &
\begin{tikzpicture}[scale=0.6]
\definecolor{mbr}{RGB}{128,71,0}
\definecolor{mgr}{RGB}{148,166,0}
\definecolor{mor}{RGB}{242,198,0}
\filldraw[fill=mor,draw=black] (0,0) -- ++(150:1) -- ++(210:1) -- ++(330:1) -- ++(30:1);
\filldraw[fill=mor,draw=black] (0,0) -- ++(330:1) -- ++(30:1) -- ++(150:1) -- ++(210:1);
\filldraw[fill=mor,draw=black] (0,0) -- ++(30:1) -- ++(150:1) -- ++(210:1) -- ++(330:1);
\filldraw[fill=mgr,draw=black] (0,0) -- ++(210:1) -- ++(270:1) -- ++(30:1) -- ++(90:1);
\filldraw[fill=mbr,draw=black] (0,0) -- ++(270:1) -- ++(330:1) -- ++(90:1) -- ++(150:1);
\end{tikzpicture} &
\begin{tikzpicture}[scale=0.6]
\definecolor{mbr}{RGB}{128,71,0}
\definecolor{mgr}{RGB}{148,166,0}
\definecolor{mor}{RGB}{242,198,0}
\filldraw[fill=mor,draw=black] (0,0) -- ++(30:1) -- ++(150:1) -- ++(210:1) -- ++(330:1);
\filldraw[fill=mor,draw=black] (0,0) -- ++(330:1) -- ++(30:1) -- ++(150:1) -- ++(210:1);
\filldraw[fill=mbr,draw=black] (0,0) -- ++(150:1) -- ++(270:1) -- ++(330:1) -- ++(90:1);
\filldraw[fill=mbr,draw=black] (0,0) -- ++(270:1) -- ++(330:1) -- ++(90:1) -- ++(150:1);
\end{tikzpicture} &
\begin{tikzpicture}[scale=0.6]
\definecolor{mbr}{RGB}{128,71,0}
\definecolor{mgr}{RGB}{148,166,0}
\definecolor{mor}{RGB}{242,198,0}
\filldraw[fill=mor,draw=black] (0,0) -- ++(30:1) -- ++(150:1) -- ++(210:1) -- ++(330:1);
\filldraw[fill=mor,draw=black] (0,0) -- ++(150:1) -- ++(210:1) -- ++(330:1) -- ++(30:1);
\filldraw[fill=mor,draw=black] (0,0) -- ++(210:1) -- ++(330:1) -- ++(30:1) -- ++(150:1);
\filldraw[fill=mor,draw=black] (0,0) -- ++(330:1) -- ++(30:1) -- ++(150:1) -- ++(210:1);
\end{tikzpicture} &
\begin{tikzpicture}[scale=0.6]
\definecolor{mbr}{RGB}{128,71,0}
\definecolor{mgr}{RGB}{148,166,0}
\definecolor{mor}{RGB}{242,198,0}
\filldraw[fill=mor,draw=black] (0,0) -- ++(30:1) -- ++(150:1) -- ++(210:1) -- ++(330:1);
\filldraw[fill=mgr,draw=black] (0,0) -- ++(270:1) -- ++(30:1) -- ++(90:1) -- ++(210:1);
\filldraw[fill=mbr,draw=black] (0,0) -- ++(150:1) -- ++(270:1) -- ++(330:1) -- ++(90:1);
\end{tikzpicture} \\
\hline
example partition & $\begin{ytableau} \scriptstyle 2 & \scriptstyle 1 \\ \scriptstyle 1 & \scriptstyle 0 \end{ytableau}$ & $\begin{ytableau} \scriptstyle 1 & \scriptstyle 1 \\ \scriptstyle 1 & \scriptstyle 0 \end{ytableau}$ & $\begin{ytableau} \scriptstyle 1 & \scriptstyle 1 \\ \scriptstyle 0 & \scriptstyle 0 \end{ytableau}$ & $\begin{ytableau} \scriptstyle 1 & \scriptstyle 1 \\ \scriptstyle 1 & \scriptstyle 1 \end{ytableau}$ & $\begin{ytableau} \scriptstyle 1 & \scriptstyle 0 \\ \scriptstyle 0 & \scriptstyle 0 \end{ytableau}$\\
\hline
valence of the vertex & 6 & 5 & 4 & 4 & 3 \\
\hline
order of zero in $\psi_\Lambda(u)$ & 2 & 1 & 0 & 0 & -1 \\
\hline
\end{tabular}
\end{center}
It is clear from the table that the quantity that ultimately determines the order of pole or zero at any given point is the number of edges entering the given vertex. Looking back at the vacuum charges (\ref{vacuumcharges}), we see that the pole at $u = 0$ comes from the trivalent vertex at the origin (see figure \ref{vacuumtiling}) and there is an overall factor of $u+\psi_0 \sigma_3$ which is not related to the tiling (for generic $\psi_0$; for special values of $\psi_0$ it actually corresponds to quintuple vertex consistently with the table above). To summarize, in the rhombic tiling picture the generating function of the charges $\psi_\Lambda(u)$ of a given configuration $\Lambda$ is given by
\begin{empheq}[box=\fbox]{equation}
\psi_\Lambda(u) = (u+\psi_0 \sigma_3) \prod_{v \in \text{vert}(\Lambda)} (u - h_v)^{\text{val}(v)-4}
\end{empheq}
Here the product is over all vertices $v$ of the tiling and for each vertex $h_v$ is the weighted coordinate of the vertex $x_1 h_1 + x_2 h_2 + x_3 h_3$ and $\text{val}(v)$ is its valence (number of edges that end at this vertex). The constraint (\ref{sigma1zero}) has also a natural meaning in the tiling picture: the direction $(1,1,1)$ is projected out and we don't see it,
\begin{center}
\begin{tikzpicture}[scale=1]
\draw[thick,->] (0,0) -- (330:2);
\draw[thick,->] (0,0) -- (90:2);
\draw[thick,->] (0,0) -- (210:2);
\node at (1,-0.3) {$h_1$};
\node at (-1.1,-0.3) {$h_2$};
\node at (0.3,1) {$h_3$};
\end{tikzpicture}
\end{center}
The points whose $x_j$ coordinates differ by a multiple of $(1,1,1)$ are identified in the tiling picture.

Now it is clear why the residue formulas (\ref{efres1}) and (\ref{efres2}) require a simple pole at position where we can add or remove a box: the basic move (\ref{trimove}) only affects the three faces around a trivalent vertex (simple pole) and these are the only places where $e_j$ and $f_j$ can act.

\begin{figure}
\centering
\begin{tikzpicture}[scale=0.6]
\definecolor{mbr}{RGB}{128,71,0}
\definecolor{mgr}{RGB}{148,166,0}
\definecolor{mor}{RGB}{242,198,0}
\filldraw[fill=mbr,draw=black] ++(330:2) -- ++(330:1) -- ++(90:1) -- ++(150:1) -- ++(270:1);
\filldraw[fill=mbr,draw=black] ++(330:3) -- ++(330:1) -- ++(90:1) -- ++(150:1) -- ++(270:1);
\filldraw[fill=mbr,draw=black] (90:1) -- ++(330:1) -- ++(90:1) -- ++(150:1) -- ++(270:1);
\filldraw[fill=mbr,draw=black] (90:1) ++(330:1) -- ++(330:1) -- ++(90:1) -- ++(150:1) -- ++(270:1);
\filldraw[fill=mbr,draw=black] (90:1) ++(330:2) -- ++(330:1) -- ++(90:1) -- ++(150:1) -- ++(270:1);
\filldraw[fill=mbr,draw=black] (90:1) ++(330:3) -- ++(330:1) -- ++(90:1) -- ++(150:1) -- ++(270:1);
\filldraw[fill=mbr,draw=black] (90:2) -- ++(330:1) -- ++(90:1) -- ++(150:1) -- ++(270:1);
\filldraw[fill=mbr,draw=black] (90:2) ++(330:1) -- ++(330:1) -- ++(90:1) -- ++(150:1) -- ++(270:1);
\filldraw[fill=mbr,draw=black] (90:2) ++(330:2) -- ++(330:1) -- ++(90:1) -- ++(150:1) -- ++(270:1);
\filldraw[fill=mbr,draw=black] (90:2) ++(330:3) -- ++(330:1) -- ++(90:1) -- ++(150:1) -- ++(270:1);
\filldraw[fill=mbr,draw=black] (90:3) -- ++(330:1) -- ++(90:1) -- ++(150:1) -- ++(270:1);
\filldraw[fill=mbr,draw=black] (90:3) ++(330:1) -- ++(330:1) -- ++(90:1) -- ++(150:1) -- ++(270:1);
\filldraw[fill=mbr,draw=black] (90:3) ++(330:2) -- ++(330:1) -- ++(90:1) -- ++(150:1) -- ++(270:1);
\filldraw[fill=mbr,draw=black] (90:3) ++(330:3) -- ++(330:1) -- ++(90:1) -- ++(150:1) -- ++(270:1);
\filldraw[fill=mgr,draw=black] ++(90:1) -- ++(90:1) -- ++(210:1) -- ++(270:1) -- ++(30:1);
\filldraw[fill=mgr,draw=black] ++(90:2) -- ++(90:1) -- ++(210:1) -- ++(270:1) -- ++(30:1);
\filldraw[fill=mgr,draw=black] ++(90:3) -- ++(90:1) -- ++(210:1) -- ++(270:1) -- ++(30:1);
\filldraw[fill=mgr,draw=black] (210:1) ++(90:1) -- ++(90:1) -- ++(210:1) -- ++(270:1) -- ++(30:1);
\filldraw[fill=mgr,draw=black] (210:1) ++(90:2) -- ++(90:1) -- ++(210:1) -- ++(270:1) -- ++(30:1);
\filldraw[fill=mgr,draw=black] (210:1) ++(90:3) -- ++(90:1) -- ++(210:1) -- ++(270:1) -- ++(30:1);
\filldraw[fill=mgr,draw=black] (210:2) -- ++(90:1) -- ++(210:1) -- ++(270:1) -- ++(30:1);
\filldraw[fill=mgr,draw=black] (210:2) ++(90:1) -- ++(90:1) -- ++(210:1) -- ++(270:1) -- ++(30:1);
\filldraw[fill=mgr,draw=black] (210:2) ++(90:2) -- ++(90:1) -- ++(210:1) -- ++(270:1) -- ++(30:1);
\filldraw[fill=mgr,draw=black] (210:2) ++(90:3) -- ++(90:1) -- ++(210:1) -- ++(270:1) -- ++(30:1);
\filldraw[fill=mgr,draw=black] (210:3) -- ++(90:1) -- ++(210:1) -- ++(270:1) -- ++(30:1);
\filldraw[fill=mgr,draw=black] (210:3) ++(90:1) -- ++(90:1) -- ++(210:1) -- ++(270:1) -- ++(30:1);
\filldraw[fill=mgr,draw=black] (210:3) ++(90:2) -- ++(90:1) -- ++(210:1) -- ++(270:1) -- ++(30:1);
\filldraw[fill=mgr,draw=black] (210:3) ++(90:3) -- ++(90:1) -- ++(210:1) -- ++(270:1) -- ++(30:1);
\filldraw[fill=mor,draw=black] ++(330:2) -- ++(210:1) -- ++(330:1) -- ++(30:1) -- ++(150:1);
\filldraw[fill=mor,draw=black] ++(330:3) -- ++(210:1) -- ++(330:1) -- ++(30:1) -- ++(150:1);
\filldraw[fill=mor,draw=black] ++(210:1) ++(330:1) -- ++(210:1) -- ++(330:1) -- ++(30:1) -- ++(150:1);
\filldraw[fill=mor,draw=black] ++(210:1) ++(330:2) -- ++(210:1) -- ++(330:1) -- ++(30:1) -- ++(150:1);
\filldraw[fill=mor,draw=black] ++(210:1) ++(330:3) -- ++(210:1) -- ++(330:1) -- ++(30:1) -- ++(150:1);
\filldraw[fill=mor,draw=black] ++(210:2) -- ++(210:1) -- ++(330:1) -- ++(30:1) -- ++(150:1);
\filldraw[fill=mor,draw=black] ++(210:2) ++(330:1) -- ++(210:1) -- ++(330:1) -- ++(30:1) -- ++(150:1);
\filldraw[fill=mor,draw=black] ++(210:2) ++(330:2) -- ++(210:1) -- ++(330:1) -- ++(30:1) -- ++(150:1);
\filldraw[fill=mor,draw=black] ++(210:2) ++(330:3) -- ++(210:1) -- ++(330:1) -- ++(30:1) -- ++(150:1);
\filldraw[fill=mor,draw=black] ++(210:3) -- ++(210:1) -- ++(330:1) -- ++(30:1) -- ++(150:1);
\filldraw[fill=mor,draw=black] ++(210:3) ++(330:1) -- ++(210:1) -- ++(330:1) -- ++(30:1) -- ++(150:1);
\filldraw[fill=mor,draw=black] ++(210:3) ++(330:2) -- ++(210:1) -- ++(330:1) -- ++(30:1) -- ++(150:1);
\filldraw[fill=mor,draw=black] ++(210:3) ++(330:3) -- ++(210:1) -- ++(330:1) -- ++(30:1) -- ++(150:1);
\filldraw[fill=mor,draw=black] ++(90:1) -- ++(210:1) -- ++(330:1) -- ++(30:1) -- ++(150:1);
\filldraw[fill=mor,draw=black] ++(150:1) -- ++(210:1) -- ++(330:1) -- ++(30:1) -- ++(150:1);
\filldraw[fill=mor,draw=black] ++(30:1) -- ++(210:1) -- ++(330:1) -- ++(30:1) -- ++(150:1);
\filldraw[fill=mgr,draw=black] (270:1) -- ++(90:1) -- ++(210:1) -- ++(270:1) -- ++(30:1);
\filldraw[fill=mgr,draw=black] (270:1) ++(30:1) ++(330:1) -- ++(90:1) -- ++(210:1) -- ++(270:1) -- ++(30:1);
\filldraw[fill=mbr,draw=black] (270:1) -- ++(330:1) -- ++(90:1) -- ++(150:1) -- ++(270:1);
\filldraw[fill=mbr,draw=black] (210:2) -- ++(330:1) -- ++(90:1) -- ++(150:1) -- ++(270:1);
\end{tikzpicture}
\caption{An example of a rhombic tiling of plane which corresponds to a plane partition with three boxes. Apart from 4-valent vertices, there are six trivalent vertices and five 5-valent vertices.}
\label{extiling}
\end{figure}

To illustrate these results on an example, let us consider the plane partition
\begin{equation}
\begin{ytableau} \scriptstyle 1 & \scriptstyle 1 \\ \scriptstyle 1 & \scriptstyle 0 \end{ytableau}
\end{equation}
The corresponding tiling is shown in figure \ref{extiling}. Apart from four-valent vertices which give no contribution to $\psi_\Lambda(u)$, we have $6$ trivalent vertices at values of $u$
\begin{equation}
h_1, \quad h_2, \quad h_3, \quad h_1+h_2, \quad 2h_1, \quad 2h_2
\end{equation}
and five $5$-valent vertices at $u$ equal to
\begin{equation}
0, \quad h_1+2h_2, \quad 2h_1+h_2, \quad 2h_1+h_3, \quad 2h_2+h_3
\end{equation}
We can thus immediately write the generating function of the charges for this configuration
\begin{equation}
\frac{(u+\psi_0\sigma_3)u(u-h_1-2h_2)(u-2h_1-h_2)(u-2h_1-h_3)(u-2h_2-h_3)}{(u-h_1)(u-h_2)(u-h_3)(u-2h_1)(u-2h_2)(u-h_1-h_2)}
\end{equation}
which is the same as what (\ref{psiproduct}) gives in this case.

\subsection{Free boson representation}
\label{freebosonrep}
If we want to relate $\mathcal{Y}$ to $\mathcal{W}_{1+\infty}$, we should be able to find a free-boson representation of $\mathcal{Y}$ which is just the usual Fock representation of $\widehat{\mathfrak{u}(1)}$ factor of $\mathcal{W}_{1+\infty}$. We will start with the free boson representation of the linear $\mathcal{W}_{1+\infty}$ and we will deform it to be a representation of $\mathcal{Y}$ even for $\sigma_3 \neq 0$.

Let us first fix the parameters of algebra. First, let us fix the rescaling symmetry $\alpha$ by requiring
\begin{equation}
\psi_0 = 1.
\end{equation}
We expect one of $\mathcal{W}_{\infty}$ $\lambda$-parameters to be equal to $1$ in order to have a representation of $\mathcal{Y}$ on one free boson (otherwise the Virasoro subalgebra of $\mathcal{W}_{\infty}$ factor would not act trivially and we would not have a representation of $\hat{\mathfrak{u}}(1)$ factor only). The representations in this case will be on (two-dimensional) partitions instead of the full plane partitions. In other words the condition $\lambda = 1$ restricts to partitions with one layer only. In section \ref{specializations} we will see how this is related to appearance of a null state in the vacuum representation at level $2$. As a result of existence of this null state, the box creation operators in the irreducible representation will not produce any boxes outside of the first layer of boxes.

With no loss of generality, we can choose $\lambda_3 = 1$, so that
\begin{align}
\label{bosonparams}
\nonumber
h_1 & = h \\
\nonumber
h_2 & = -h^{-1} \\
h_3 & = -h + h^{-1} \\
\nonumber
\sigma_2 & = 1 - h^2 - h^{-2} \\
\nonumber
\sigma_3 & = h - h^{-1}
\end{align}
Assuming the same mode grading and spin filtration as in the linear $\mathcal{W}_{1+\infty}$, we have to identify
\begin{align}
\nonumber
e_0 & \leftrightarrow a_{-1} \\
f_0 & \leftrightarrow -a_{+1}
\end{align}
as before. So all that remains is to fix the remaining $\psi_3$ generator.

To find $\psi_3$, we will proceed indirectly. It is well known that $\widehat{\mathfrak{u}(1)}$ is naturally represented on Young diagrams. In fact, the usual representation has a natural basis where the Young diagrams correspond to Schur functions (and through boson-fermion correspondence this basis can be mapped to similarly simple basis of states created by the fermion creation operators). We know a representation of $\mathcal{Y}$ on Young diagrams, the representation of section \ref{secplanepart}. Specializing (\ref{vacuumcharges}) to the choice of parameters (\ref{bosonparams}), we see that the vacuum should have charges
\begin{equation}
\psi(u) \ket{0} = \frac{u+h-h^{-1}}{u} \ket{0}
\end{equation}
and the state corresponding to Young diagram $\lambda = (\lambda_1,\lambda_2,\ldots)$ should have the charges (\ref{psiproduct})
\begin{align}
\nonumber
\psi_\lambda(u) & = \frac{u+h-h^{-1}}{u} \prod_{j > 0} \frac{(u-\lambda_j h_1-j h_2)(u-\lambda_j h_1-j h_2 + h_1 + 2h_2)}{(u-\lambda_j h_1 - j h_2 + h_1 + h_2)(u-\lambda_j h_1 - j h_2 + h_2)} \times \\
& \quad \times \frac{(u - j h_2 + h_1 + h_2)(u-j h_2 + h_2)}{(u-j h_2)(u-j h_2 + h_1 + 2h_2)}.
\end{align}
From this we can easily extract the individual charges up to spin $3$,
\begin{align}
\nonumber
\psi_1 \ket{\lambda} & = 0 \\
\psi_2 \ket{\lambda} & = 2 \sum_j \lambda_j
\end{align}
and
\begin{equation}
\psi_3 \ket{\lambda} = \sum_j \left[ 3 \lambda_j (\lambda_j-1) h_1 + 6 (j-1) \lambda_j h_2 + 2 \lambda_j h_1 h_2 h_3 \right] \ket{\lambda}.
\end{equation}
These relations are clearly symmetric under exchange $h_1 \leftrightarrow h_2$ with simultaneous transposition of $\lambda$, since for any Young diagram
\begin{equation}
\sum_j \lambda_j (\lambda_j-1) = \sum_j 2(j-1) \lambda^T_j.
\end{equation}
The main difference between the deformed case, $h \neq \pm 1$ and the symmetric case where we have Schur functions (and linear $\mathcal{W}_{1+\infty}$) is that for $h \neq \pm 1$ the rows and columns of the Young diagrams are weighted by a different factor. The corresponding generalization of Schur polynomials with similar row-column asymmetry are the well-known Jack polynomials. We want to find an operator whose action on Jack polynomials has same eigenvalues as $\psi_3$. First of all, the Laplace-Beltrami operator \cite{stanleyjack,macdonaldsym}
\begin{equation}
D(\alpha) = \frac{\alpha}{2} \sum_j x_j^2 \partial_j^2 + \sum_{j \neq k} \frac{x_j^2}{x_j - x_k} \partial_j
\end{equation}
is diagonal in the basis of Jack polynomials with eigenvalue
\begin{equation}
D(\alpha) J_{\lambda}^{(\alpha)}(x) = \sum_j \left[ \frac{\alpha}{2} \lambda_j(\lambda_j-1) - (j-1)\lambda_j + (n-1)\lambda_j \right] J_{\lambda}^{(\alpha)}(x).
\end{equation}
Comparing this to values of $\psi_3$ above, we can identify
\begin{equation}
\psi_3 = 6h^{-1} D(\alpha=h^2) - 6h^{-1} (n-1)E + 2(h-h^{-1}) E
\end{equation}
where $E$ is the Euler operator. To express this in terms of the bosonic oscillators, we need to rewrite the action of these operators on the Newton power sum symmetric polynomials (which are the objects usually corresponding to bosonic mode operators). We find
\begin{align}
\nonumber
D(\alpha) & = \frac{\alpha}{2} \sum_{l>0} l(l-1) p_l \frac{\partial}{\partial p_l} + \frac{\alpha}{2} \sum_{k,l>0} k l p_{k+l} \frac{\partial^2}{\partial p_k \partial p_l} + \frac{1}{2} \sum_{k,l>0} p_k p_l (k+l) \frac{\partial}{\partial p_{k+l}} \\
& - \frac{1}{2} \sum_{l>0} l(l+1) p_l \frac{\partial}{\partial p_l} + n \sum_{l>0} l p_l \frac{\partial}{\partial p_l}.
\end{align}
Using this, we can write $\psi_3$ as differential operator acting on symmetric polynomials expressed as functions of $p_j$
\begin{equation}
\psi_3 = 3h \sum_{k,l>0} k l p_{k+l} \frac{\partial^2}{\partial p_k \partial p_l} + 3h^{-1} \sum_{k,l>0} p_k p_l (k+l) \frac{\partial}{\partial p_{k+l}} + (h-h^{-1}) \sum_{l>0} (3l-1) l p_l \frac{\partial}{\partial p_l}.
\end{equation}
Identifying the bosonic creation and annihilation operators
\begin{align}
\nonumber
h^{-1} p_l & \leftrightarrow a_{-l} \\
h l \frac{\partial}{\partial p_l} & \leftrightarrow a_{+l}
\end{align}
(which differs from the usual identification only by an irrelevant rescaling $\beta$) we finally find
\begin{equation}
\psi_3 = 3 \sum_{k,l>0} ( a_{-k-l} a_k a_l + a_{-k} a_{-l} a_{k+l} ) + (h-h^{-1}) \sum_{l>0} (3l-1) a_{-l} a_l.
\end{equation}
This is our candidate for $\psi_3$ in the bosonic representation of $\mathcal{Y}$. By construction, this operator acting on bosonic Fock space has exactly the same eigenvalues that we expect from $\psi_3$.

\paragraph{Verification of the commutation relations}
Let us verify that the first of relations (Y1), for $j=k=0$ is satisfied. First, we compute $e_j$ and $f_j$ using (Y4),
\begin{align}
\nonumber
e_1 & = \sum_{j>0} a_{-j-1} a_j \\
\nonumber
f_1 & = -\sum_{j>0} a_{-j} a_{j+1} \\
\nonumber
e_2 & = \sum_{j,k>0} \left[ a_{-j-k-1} a_j a_k + a_{-j} a_{-k} a_{j+k-1} \right] + (h-h^{-1}) \sum_j j a_{-j-1} a_j \\
f_2 & = -\sum_{j,k>0} \left[ a_{-j-k+1} a_j a_k + a_{-j} a_{-k} a_{j+k+1} \right] - (h-h^{-1}) \sum_j j a_{-j} a_{j+1} \\
\nonumber
e_3 & = \sum_{j,k,l>0} \left[ a_{-j-k-l-1} a_j a_k a_l + a_{-j} a_{-k} a_{-l} a_{j+k+l-1} \right] \\
\nonumber
& \quad + \frac{3}{2} \sum_{j+k=l+m+1} a_{-j} a_{-k} a_l a_m + \frac{1}{2} \sum_j j(j+1) a_{-j-1} a_j \\
\nonumber
& \quad + \frac{\sigma_3}{2} \sum_{j,k>0} (3j+3k) a_{-j-k-1} a_j a_k + \frac{\sigma_3}{2} \sum_{j,k>0} (3j+3k-4) a_{-j} a_{-k} a_{j+k-1} \\
\nonumber
& \quad + \sigma_3^2 \sum_{j>0} j^2 a_{-j-1} a_j
\end{align}
and from these we easily see that
\begin{align}
\nonumber
\psi_1 & = 0 \\
\psi_2 & = 2 \sum_{j>0} a_{-j} a_j.
\end{align}
Now we can verify that the first quadratic relation (Y1) is satisfied: we have
\begin{align}
\nonumber
\left[ e_3, e_0 \right] & = 3 \sum_{j,k>0,j+k>2} a_{-j} a_{-k} a_{j+k-2} + 3 \sum_{j,k>0} a_{-j-k-2} a_j a_k \\
\nonumber
& \quad + 3 \sigma_3 \sum_{j>0} (j+1) a_{-j-2} a_j + (1+\sigma_3^2) a_{-2} + \sigma_3 a_{-1}^2 \\
\nonumber
\left[ e_2, e_1 \right] & = \sum_{j,k>0,j+k>2} a_{-j} a_{-k} a_{j+k-2} + \sum_{j,k>0} a_{-j-k-2} a_j a_k + \sigma_3 \sum_{j>0} (j+1) a_{-j-2} a_j \\
\left[ e_1, e_0 \right] & = a_{-2} \\
\nonumber
\left\{ e_0, e_0 \right\} & = 2 a_{-1}^2
\end{align}
which satisfy
\begin{equation}
2\left[ e_3, e_0 \right] - 6 \left[ e_2, e_1 \right] + 2\sigma_2 \left[ e_1, e_0 \right] - \sigma_3 \left\{ e_0, e_0 \right\} = 0
\end{equation}
as we need. This does not prove that all the relations are satisfied, but is a non-trivial check of the identification of $\psi_3$.

\paragraph{Zero modes}
Our choice of $\psi_3$ did not contain any zero mode operators $a_0$ (which slightly simplified the computations). From point of view of $\widehat{\mathfrak{u}(1)}$ this element is central (just like $\psi_1$). It can be easily reintroduced by the spectral shift transformation of section \ref{spectralshift}. Since it will be useful to have an expression for shifted $\psi_j$ in the following, here they are:
\begin{align}
\nonumber
\psi_0 & = 1 \\
\psi_1 & = a_0 \\
\nonumber
\psi_2 & = a_0^2 + 2\sum_{j>0} a_{-j} a_j
\end{align}
and most importantly
\begin{empheq}[box=\fbox]{align}
\nonumber
\psi_3 & = a_0^3 + 6 a_0 \sum_{j>0} a_{-j} a_j + 3 \sum_{j,k>0} ( a_{-j-k} a_j a_k + a_{-j} a_{-k} a_{j+k} ) \\
& \quad + \sigma_3 \sum_{j>0} (3j-1) a_{-j} a_j
\end{empheq}
Except for the last term in $\psi_3$, these can be written as zero modes of normal-ordered powers of $\widehat{\mathfrak{u}(1)}$ current which are local operators. The last term, however, contains
\begin{equation}
\sum_{j>0} j a_{-j} a_j = \frac{1}{2} \sum_{j \in \mathbbm{Z}} |j| \normord{a_{-j} a_j}
\end{equation}
which is not of this form. One could use the Hilbert transform on a circle as in the discussions of the Benjamin-Ono equation \cite{nazarovskl} to rewrite this term as zero mode of some non-local operator, but we won't need this in the following. The necessity of using this Hilbert transform is manifestation of the non-local nature of the transformation between $\mathcal{W}_{1+\infty}$ and Yangian generators. Notice that this non-local term is related to the form of the `improvement' terms in the coproduct (\ref{newcoproduct}).

\section{$\mathcal{W}_{1+\infty}$ representations}

In this last part, we apply what we learned about $\mathcal{Y}$ to $\mathcal{W}_{1+\infty}$ representations. First we study the large $N$ behaviour of the characters of the completely degenerate representations of $\mathcal{W}_N$. For generic values of the parameters $\lambda_j$, these characters turn out to be given by a specialization of the topological vertex from the topological strings to two non-trivial Young diagram labels (the third label is in this case the empty Young diagram). Application of the triality symmetry shows that the set of completely degenerate representations of $\mathcal{W}_{1+\infty}$ parametrized by two Young diagrams as in \cite{Gaberdiel:2012ku} is not closed under this symmetry, so even without the knowledge of the topological vertex one is tempted to include the third label. The equality with the topological vertex reinforces this expectation and one is lead to conjecture that the set of completely degenerate representations of $\mathcal{W}_{1+\infty}$ that we are considering should be labelled by three Young diagram instead of just two. The reason why the third label did not appear in the study of $\mathcal{W}_N$ characters will be clear later from the discussion of the degenerate states: truncation from $\mathcal{W}_{1+\infty}$ to $\mathcal{W}_{1+N}$ corresponds combinatorially to the restriction to the plane partitions which have height in one direction bounded by $N$. This geometric condition, which is easily verified for the vacuum character, does not for the plane partitions to have a non-trivial asymptotics in the truncated direction (see also figure \ref{h3null} for the illustration of this truncation).

In the next section, we find the translation between the $U$-charges of $\mathcal{W}_{1+\infty}$ of the primary states considered in \cite{Prochazka:2014gqa} and $\psi(u)$-charges which can be read of combinatorially from the plane partition or the plane tiling picture. The identification is quite nice if we introduce a new generating function for $U$-charges motivated by the study of shifted Schur polynomials by Okounkov and Olshanski \cite{ooshifted}.

The following short section discusses the Schiffmann-Vasserot linearization - the construction of a different basis of the Baxter subalgebra which has simpler commutation relations with $e_j$ and $f_j$ creation and annihilation operators. Using this map, we can see that $\mathcal{Y}$ and the Schiffmann-Vasserot algebra constructed as a limit of the spherical degenerate double affine Hecke algebras of $GL(N)$ are in fact the same algebra.

In the next section we look for an explicit map between $\mathcal{Y}$ and $\mathcal{W}_{1+\infty}$ using the free boson representations and the coproduct. Using the coproduct (\ref{newcoproduct}) is necessary for this identification to go through - we would not get the correct match if we instead used the simpler (\ref{delta0}). As a result, we find an expression for $\psi_3$ in terms of $U$-basis mode operators in $\mathcal{W}_{1+\infty}$. This formula is one of the main results of this article and allows us to map between $\mathcal{Y}$ and $\mathcal{W}_{1+\infty}$. Although it relies on (\ref{newcoproduct}) being the coproduct which we didn't prove, every test of the identification between $\mathcal{Y}$ and $\mathcal{W}_{1+\infty}$ that we were able to do matches nicely. The most non-trivial of this is the exact match of $\psi_3$ eigenvalues in the vacuum representation up to level $6$, which means that all $95$ eigenvalues that we computed are exactly the same for all values of $h_j$.

The following section focuses on one example of completely degenerate representations of $\mathcal{Y}$, the representation with Young diagram asymptotics $(\Box,\Box,\Box)$. We compute the character of this representation in two different ways, verifying most of the results found so far. First, the topological vertex gives us directly a prediction for the character of this representation. On the other hand, from the plane partition picture we can determine the $\psi(u)$ charges of the highest weight state, translate these charges to $U$-charges of $\mathcal{W}_{1+\infty}$ and use the commutation relations of $\mathcal{W}_{1+\infty}$ from \cite{Prochazka:2014gqa} to compute the rank of the Shapovalov form. The fact that these two computations match is a non-trivial verification of all the steps involved in this check. It is also an independent check of the commutation relations of $\mathcal{W}_{1+\infty}$ found in \cite{Prochazka:2014gqa}.

In the last part we discuss what happens if $h_j$ parameters are non-generic. First of all, we translate the equations for null states in the vacuum representation found in \cite{Prochazka:2014gqa} from $\mathcal{W}_{1+\infty}$ to $\mathcal{Y}$ and show on examples that in $\mathcal{Y}$ these states are eigenstates of the Baxter subalgebra. Next we consider the minimal model describing the Lee-Yang singularity in detail. The characters of irreducible representations studied in detail in \cite{Fukuda:2015ura} can be described in terms of the box counting.

\subsection{Characters and topological vertex}

In this section we will study the characters of the irreducible representations of $\mathcal{W}_{1+\infty}$ and find a correspondence between these and the topological vertex of the topological string theory. Let us first review what is known from the representation theory of $\mathcal{W}_N$ \cite{Bouwknegt:1992wg}, which is in many aspects quite similar to the representation theory of the Virasoro algebra. The highest weight representations of $\mathcal{W}_N$ are labelled by $N-1$ complex numbers, one of them being the $L_0$ eigenvalue (the conformal dimension). The precise choice of these charges depends on the choice of the generating fields of $\mathcal{W}_N$. For generic values of the highest weights and the central charge $c$ the whole Verma module is irreducible and its character is simply
\begin{equation}
\chi_V(q) = \Tr_V q^{L_0 - c/24} = q^{h-c/24} \prod_{n=1}^{\infty} \frac{1}{(1-q^n)^{N-1}}.
\end{equation}
If we tune the values of the highest weights to non-generic values (still keeping the generic values of the central charge), the Verma module can develop null states and it is no longer irreducible. If we in addition tune the value of the central charge, more null states can appear for special values of the central charge (this is what happens in the minimal models) and to get an irreducible representation, we must remove also these states.

In the case of the Virasoro algebra we have for a generic value of $c$ a class of special degenerate representations with highest weight $h_{r,s}$ parametrized by two positive integers $r$ and $s$. These representations have their first null state at level $rs$. For $\mathcal{W}_N$ (which is rank $N-1$) there can be up to $N-1$ independent null states of this kind, each of them being associated to an algebraic condition on the highest weights. This means that the highest weight representations which have maximal number of the null states (still for generic $c$) have highest weights that are a discrete (codimension $N-1$) subset of the full $N-1$-dimensional space of the highest weights. In the following, we will be interested in representations with these special highest weights and we will call them (following \cite{Bouwknegt:1992wg}, section 6.4.2) the completely degenerate representations. For $\mathcal{W}_N$, these completely degenerate representations are parametrized by a pair of Young diagrams with at most $N-1$ rows (this is the usual condition for representations of $\mathfrak{su}(N)$). In contrast to the Virasoro case, where the representation is either generic or completely degenerate (since there is just one highest weight), in $\mathcal{W}_N$ case with $N \geq 3$ we have also intermediate classes of representations, where the number of null states is neither zero nor maximal. We will not consider these representations here.

Our goal is to find the character of completely degenerate representations of $\mathcal{W}_{1+\infty}$ at generic values of $\lambda_j$. In order to do this, we will first consider the characters of completely degenerate representations of $\mathcal{W}_N$ for a generic value of the central charge and argue that taking formally the limit $N \to \infty$ results in character of $\mathcal{W}_{\infty}$ that we want to find. The reason why this works is that the $q$-expansion of characters of completely degenerate representations stabilizes as we increase $N$. From $\mathcal{W}_{\infty}$ point of view, the lowest lying null state that appears as a result of specialization to $\lambda=N$ is at level $N+1$, so as $N$ increases, the characters of $\mathcal{W}_N$ approach those of $\mathcal{W}_{\infty}$ (the null states associated to $\mathcal{W}_N$ truncation move to infinity).

We can see a simple example of this phenomenon in the vacuum representation of $\mathcal{W}_N$ for generic value of the central charge. Here the character is
\begin{equation}
\chi_{\mathrm{vac},\mathcal{W}_N} \sim \prod_{s=2}^N \prod_{j=0}^{\infty} \frac{1}{1-q^{s+j}}
\end{equation}
which has a $q$-expansion
\begin{align}
\nonumber
\chi_{\mathrm{vac},\mathfrak{Vir}} & \sim 1 + q^2 + q^3 + 2q^4 + 2q^5 + \ldots \\
\nonumber
\chi_{\mathrm{vac},\mathcal{W}_3} & \sim 1 + q^2 + 2q^3 + 3q^4 + 4q^5 + \ldots \\
\chi_{\mathrm{vac},\mathcal{W}_4} & \sim 1 + q^2 + 2q^3 + 4q^4 + 5q^5 + \ldots \\
\nonumber
\chi_{\mathrm{vac},\mathcal{W}_5} & \sim 1 + q^2 + 2q^3 + 4q^4 + 6q^5 + \ldots \\
\nonumber
\chi_{\mathrm{vac},\mathcal{W}_{\infty}} & \sim 1 + q^2 + 2q^3 + 4q^4 + 6q^5 + \ldots.
\end{align}
We see that the series expansion of the character of the vacuum representation for $\mathcal{W}_N$ is equal to that of $\mathcal{W}_{\infty}$ for all levels up to level $N$ (while at the next level the null state appears). Analogous results hold for characters of other completely degenerate representations. Note however, that this is not necessarily true for other classes of representations. For example, the Verma modules of $\mathcal{W}_N$ (which are irreducible for generic choice of the highest weights and so stand at the opposite end than the completely degenerate representations) have $N-1$ states at level $1$, so there is no analogous stabilization as $N \to \infty$. In fact the Verma module of $\mathcal{W}_{1+\infty}$ has an infinite number of states at level $1$ so the usual notion of (unrefined) character does not make sense in this case. Fixing the Young diagram labels of the completely degenerate representation when sending $N \to \infty$ is a natural choice for the $N$-dependence of the weights as we take the large $N$ limit.

Let us now proceed to computation of the characters. The starting point is the character formula for the irreducible completely degenerate representation $(\Lambda_+,\Lambda_-)$ of $\mathcal{W}_N$ minimal model $(p^\prime,p)$ (see \cite{Bouwknegt:1992wg,Gaberdiel:2011zw}),
\begin{equation}
\label{wncharacter}
\chi(\Lambda_+,\Lambda_-) \equiv q^{-c/24} \Tr q^{L_0} = \frac{1}{\eta^{N-1}} \sum_{\alpha} \sum_{w \in \mathcal{W}} \epsilon(w) q^{\frac{1}{2pp^\prime}\left[ p^\prime w(\Lambda_{+}+\rho)-p(\Lambda_{-}+\rho)+pp^\prime\alpha \right]^2}.
\end{equation}
Here $(\Lambda_+,\Lambda_-)$ are the two Young diagrams parametrizing the highest weight representation and $\eta(q)$ is the Dedekind eta function
\begin{equation}
\eta(q) = q^{1/24} \prod_{n=1}^{\infty} (1-q^n).
\end{equation}
The sum over $\alpha$ in (\ref{wncharacter}) goes over the roots of $\mathfrak{sl}(N)$ root lattice (which we represent as elements of $\mathbbm{Z}^N$ constrained to have the total sum of the entries equal to zero), $\mathcal{W}$ is the Weyl group of $\mathfrak{sl}(N)$ which is the symmetric group $\mathcal{S}_N$. $\epsilon(w)$ is the sign of the permutation $w$, $\rho$ is the Weyl vector of $\mathfrak{sl}(N)$ with the components
\begin{equation}
\label{weylrho}
\rho = \left(\frac{N-1}{2}, \frac{N-3}{2}, \ldots, -\frac{N-1}{2} \right)
\end{equation}
and the square in the exponent is evaluated in the weight space of $\mathfrak{sl}(N)$ with the usual normalizations, for example
\begin{equation}
\rho^2 = \frac{1}{4} \sum_{j=1}^N (N+1-2j)^2 = \frac{(N-1)N(N+1)}{12}.
\end{equation}

In the case of $\mathcal{W}_N$ minimal models, $p^\prime$ and $p$ are two coprime integers which parametrize the central charge. If we move the central charge away from these special minimal model values, the null states that are taken into account by summation over the roots $\alpha$ of the root lattice do not contribute. The correct character formula in this case is simpler,
\begin{equation}
\chi(\Lambda_+,\Lambda_-) = \frac{1}{\eta^{N-1}} \sum_{w \in \mathcal{W}} \epsilon(w) q^{\frac{1}{2pp^\prime}\left[ p^\prime w(\Lambda_{+}+\rho)-p(\Lambda_{-}+\rho)\right]^2}.
\end{equation}
One can argue similarly as above using the stabilization of characters as we increase $p^\prime$ and $p$. Notice also that for $p^\prime / p$ irrational the powers of $q$ would not differ by integers which is clearly impossible in an irreducible representation.

Before taking the limit $N \to \infty$, we should strip off the factor $q^{h-c/24}$ which is $N$-dependent:
\begin{equation}
\chi(\Lambda_+,\Lambda_-) = q^{h(\Lambda_+,\Lambda_-)-c/24} \frac{\sum_{w\in\mathcal{W}} \epsilon(w) q^{-(w(\Lambda_++\rho)-(\Lambda_++\rho),\Lambda_-+\rho)}}{\prod_{j=1}^{\infty} (1-q^j)^{N-1}}
\end{equation}
Here and in the following we use the notation $(\cdot,\cdot)$ for the inner product in the weight space and
\begin{equation}
\label{hwn}
h(\Lambda_+,\Lambda_-) = \frac{1}{2pp^\prime} \left[ p^\prime(\Lambda_++\rho) - p(\Lambda_-+\rho) \right]^2 - \frac{(p^\prime-p)^2}{2pp^\prime} \rho^2
\end{equation}
is the conformal dimension of the highest weight vector in representation $(\Lambda_+,\Lambda_-)$ and
\begin{equation}
c = (N-1) - \frac{12(p^\prime-p)^2}{pp^\prime} \rho^2 = (N-1) \left( 1 - \frac{N(N+1)(p^\prime-p)^2}{pp^\prime} \right)
\end{equation}
is the central charge. Because of the inequality
\begin{equation}
(w(\Lambda_++\rho),\Lambda_-+\rho) < (\Lambda_++\rho,\Lambda_-+\rho)
\end{equation}
which holds for any $\Lambda_{\pm}$ in (the closure of) the positive Weyl chamber, the sum over Weyl reflections in the numerator has an expansion $1 + \mathcal{O}(q)$. The same is clearly true for the denominator so we are thus lead to define
\begin{equation}
\tilde{\chi}(\Lambda_+,\Lambda_-) = \lim_{N \to \infty} \frac{\sum_{w\in\mathcal{S}_N} \epsilon(w) q^{-(w(\Lambda_++\rho)-(\Lambda_++\rho),\Lambda_-+\rho)}}{\prod_{n=1}^{\infty} (1-q^n)^N}
\end{equation}
as a (rescaled) character of the irreducible completely degenerate $\mathcal{W}_{1+\infty}$\footnote{A careful reader notices that we reintroduced the $\mathfrak{u}(1)$ factor of $\mathcal{W}_{1+\infty}$ compared to $\mathcal{W}_{\infty}$ considered previously by shifting $N-1 \to N$ in the exponent of the denominator.} representation $(\Lambda_+,\Lambda_-)$ for generic values of $\lambda$-parameters (in particular at generic value of the central charge). As discussed above, the limit above is well-defined (both algebraically and analytically). It is useful to factorize it into two parts,
\begin{equation}
\tilde{\chi}(\Lambda_+,\Lambda_-) = \lim_{N \to \infty} \frac{\sum_{w\in\mathcal{S}_N} \epsilon(w) q^{-(w(\Lambda_++\rho)-(\Lambda_++\rho),\Lambda_-+\rho)}}{\sum_{w\in\mathcal{S}_N} \epsilon(w) q^{-(w(\rho)-\rho,\rho)}} \frac{\sum_{w\in\mathcal{S}_N} \epsilon(w) q^{-(w(\rho)-\rho,\rho)}}{\prod_{j=1}^{\infty} (1-q^j)^N}.
\end{equation}
Using the Weyl denominator formula
\begin{equation}
\sum_{w\in\mathcal{W}} \epsilon(w) q^{-(w(\rho)-\rho,\Lambda)} = \prod_{\alpha\in\Delta^+} (1-q^{(\alpha,\Lambda)})
\end{equation}
(the sum runs over the positive roots) which in our case reduces to
\begin{equation}
\sum_{w\in\mathcal{W}} \epsilon(w) q^{-(w(\rho)-\rho,\rho)} = \prod_{\alpha\in\Delta^+} (1-q^{(\alpha,\rho)}) = \prod_{j=1}^N (1-q^j)^{N-j}
\end{equation}
we find that the second part (which is the vacuum character) goes to
\begin{equation}
\frac{\sum_{w\in\mathcal{S}_N} \epsilon(w) q^{-(w(\rho)-\rho,\rho)}}{\prod_{j=1}^{\infty} (1-q^j)^N} = \prod_{j=1}^N \prod_{k=j}^{\infty} \frac{1}{1-q^k} \to \tilde{\chi}_{\textrm{vac}} = \prod_{j=1}^{\infty} \frac{1}{(1-q^j)^j} \equiv M(q)
\end{equation}
 which is the famous MacMahon function counting the plane partitions. We are thus left with
\begin{equation}
\frac{\tilde{\chi}(\Lambda_+,\Lambda_-)}{\tilde{\chi}_{\mathrm{vac}}} \equiv \mathcal{P}(\Lambda_+,\Lambda_-) = \lim_{N \to \infty} \mathcal{P}_{N}(\Lambda_+,\Lambda_-)
\end{equation}
where
\begin{equation}
\label{cnfunctions}
\mathcal{P}_N(\Lambda_+,\Lambda_-) = \frac{\sum_{w\in\mathcal{S}_N} \epsilon(w) q^{-(w(\Lambda_++\rho)-(\Lambda_++\rho),\Lambda_-+\rho)}}{\sum_{w\in\mathcal{S}_N} \epsilon(w) q^{-(w(\rho)-\rho,\rho)}}.
\end{equation}
The functions $\mathcal{P}_N$ appear (up to an overall factor) as matrix elements of modular $S$-transformation of affine $\mathfrak{su}(N)$ characters or as Chern-Simons Hopf link invariants \cite{Marino:2005sj}. They are symmetric in Young diagrams $\Lambda_+$ and $\Lambda_-$ (because the Weyl reflections are orthogonal with respect to the Killing form) and by definition have normalization $1$ at $q=0$. For finite $N$, they are polynomials in $q$, but as $N\to\infty$ they greatly simplify and become simple rational functions of $q$. For example, the simplest one is
\begin{equation}
\mathcal{P}_N(\ydiagram{1},\cdot) = \sum_{j=0}^{N-1} q^j = \frac{1-q^N}{1-q} \to \frac{1}{1-q}.
\end{equation}
The Weyl character formula for $\mathfrak{su}(N)$ contains $\mathcal{P}_N$ so we can express $\mathcal{P}_N$ using the Schur polynomials. First of all, we have
\begin{equation}
\label{schurpoly}
S^{(N)}_\lambda(x_j=q^{-\mu_j-\rho_j}) = q^{-\frac{|\lambda||\mu|}{N}} \frac{\sum_{w\in\mathcal{S}_N} \epsilon(w) q^{-(w(\lambda+\rho),\mu+\rho)}}{\sum_{w\in\mathcal{S}_N} \epsilon(w) q^{-(w \rho,\mu+\rho)}}
\end{equation}
with $\rho$ as in (\ref{weylrho}). We can view this formula as a determinantal expression for Schur polynomials or as Weyl character formula for $\mathfrak{su}(N)$ knowing that the Schur polynomials are characters of irreducible $\mathfrak{su}(N)$ representations. A small subtlety is the extra power of $q$ which comes from the fact that on the right-hand side we are using the Killing form on the weight space of $\mathfrak{su}(N)$,
\begin{equation}
(\lambda,\mu) = \sum_j \lambda_j \mu_j - \frac{1}{N} \sum_j \lambda_j \sum_k \mu_k \equiv \sum_j \lambda_j \mu_j - \frac{|\lambda||\mu|}{N}.
\end{equation}
Combining expressions (\ref{cnfunctions}) and (\ref{schurpoly}) we find
\begin{align}
\label{cnlambdapm}
\nonumber
\mathcal{P}_N(\Lambda_+,\Lambda_-) & = q^{(\Lambda_++\rho,\Lambda_-+\rho)-(\rho,\rho)} q^{\frac{|\Lambda_+||\Lambda_-|}{N}} S^{(N)}_{\Lambda_+}(x = q^{-\Lambda_--\rho}) S^{(N)}_{\Lambda_-}(x=q^{-\rho}) \\
& = q^{\sum_j (\Lambda_{+j} \Lambda_{-j} + \rho_j \Lambda_{+j} + \rho_j \Lambda_{-j})} S^{(N)}_{\Lambda_+}(x_j=q^{-\Lambda_{-j}-\rho_j}) S^{(N)}_{\Lambda_-}(x_j=q^{-\rho_j})
\end{align}
and in particular
\begin{equation}
\mathcal{P}_N(\Lambda,\cdot) = q^{(\Lambda,\rho)} S^{(N)}_\Lambda(x=q^{-\rho})
\end{equation}
We can use this to invert the relations above and find expression for Schur polynomials in terms of $\mathcal{P}_N$ functions,
\begin{equation}
S^{(N)}_{\Lambda_+}(x=q^{-\Lambda_--\rho}) = q^{-\sum_j \Lambda_{+j}(\Lambda_{-j}+\rho_j)} \frac{\mathcal{P}_N(\Lambda_+,\Lambda_-)}{\mathcal{P}_N(\cdot,\Lambda_-)}
\end{equation}
Characters evaluated at $q^{\pm\rho}$ are often called the quantum dimension of a given representation. At finite $N$, there is an explicit formula for those in terms of the Young diagram $\Lambda$ \cite{Marino:2005sj, Marino:2008vja},
\begin{equation}
S^{(N)}_\Lambda(q^{\rho}) = \prod_{1 \leq j < k \leq \rows(\Lambda)} \frac{q^{\frac{\Lambda_j - \Lambda_k - j + k}{2}} - q^{-\frac{\Lambda_j - \Lambda_k - j + k}{2}}}{q^{\frac{k-j}{2}} - q^{-\frac{k-j}{2}}} \prod_{j=1}^{\rows(\Lambda)} \prod_{k=1}^{\Lambda_j} \frac{q^{\frac{N+k-j}{2}} - q^{-\frac{N+k-j}{2}}}{q^{\frac{k-j+\rows(\Lambda)}{2}} - q^{-\frac{k-j+\rows(\Lambda)}{2}}}.
\end{equation}
which simplifies in the $N\to\infty$ limit,
\begin{align}
\label{hookform}
\nonumber
\mathcal{P}(\Lambda,\cdot) & = \prod_{1 \leq j < k \leq \rows(\Lambda)} \frac{1-q^{\Lambda_j-\Lambda_k+k-j}}{1-q^{k-j}} \prod_{j=1}^{\rows(\Lambda)} \prod_{k=1}^{\Lambda_j} \frac{1}{1-q^{k-j+\rows(\Lambda)}} \\
& = \prod_{\Box \in \Lambda} \frac{1}{1-q^{\hook(\Box)}}.
\end{align}
This hook formula gives an easy combinatorial expression for the characters when one of the Young diagrams is empty.

\paragraph{Topological vertex}
The $N \to \infty$ limit of (\ref{cnlambdapm}) can be expressed using a specialization of the topological vertex \cite{Aganagic:2003db}, which is the partition function of open topological A model on $\mathbbm{C}^3$. Combinatorially, it is a counting function of the plane partitions with asymptotics given by a triplet of Young diagrams $(\lambda,\mu,\nu)$ \cite{Okounkov:2003sp}. We will use the expression from \cite{Okounkov:2003sp},
\begin{equation}
\mathcal{C}(\lambda,\mu,\nu) = q^{\frac{1}{2} \kappa(\lambda) + \frac{1}{2} \kappa(\nu)} S_{\nu^T} (q^{\tilde{\rho}}) \sum_{\eta} S_{\lambda^T/\eta}(q^{\nu+\tilde{\rho}}) S_{\mu/\eta}(q^{\nu^T+\tilde{\rho}})
\end{equation}
with
\begin{equation}
\kappa(\lambda) = \sum_j \left( \lambda_j^2 - (2j-1) \lambda_j \right) = \sum_j \left( \lambda_j^2 - \lambda^{T2}_j \right)
\end{equation}
and "renormalized" Weyl vector of $\mathfrak{sl}(\infty)$
\begin{equation}
\tilde{\rho} = (-1/2, -3/2, -5/2, \ldots)
\end{equation}
For comparison with Schur functions of a finite number of variables let us use the identity
\begin{equation}
S_{\lambda/\mu}(q^{\tilde{\rho}+\nu}) = (-1)^{\sum_j (\lambda_j - \mu_j)} S_{\lambda^T/\mu^T} (q^{-\tilde{\rho}-\nu^T})
\end{equation}
from \cite{Okounkov:2003sp}\footnote{Note that there is an extra transposition in the exponent compared to \cite{Okounkov:2003sp} necessary for this identity to work.} to rewrite the topological vertex in more useful form
\begin{equation}
\mathcal{C}(\lambda,\mu,\nu) = (-1)^{\sum_j (\lambda_j + \mu_j + \nu_j)} q^{\frac{1}{2} \kappa(\lambda) + \frac{1}{2} \kappa(\nu)} S_{\nu} (q^{-\tilde{\rho}}) \sum_{\eta} S_{\lambda/\eta}(q^{-\tilde{\rho}-\nu^T}) S_{\mu^T/\eta}(q^{-\tilde{\rho}-\nu}).
\end{equation}
Specializing now to the case of only two non-empty asymptotic Young diagrams, we have
\begin{equation}
\mathcal{C}(\cdot, \mu^T, \nu) = (-1)^{\sum_j (\mu_j + \nu_j)} q^{\frac{1}{2} \kappa(\nu)} S_{\nu} (q^{-\tilde{\rho}}) S_{\mu}(q^{-\tilde{\rho}-\nu})
\end{equation}
which we can directly compare with (\ref{cnlambdapm}) and we find
\begin{align}
\label{pairingtotv}
\nonumber
\mathcal{P}(\Lambda_+,\Lambda_-) & = (-1)^{\sum_j (\Lambda_{+j} + \Lambda_{-j})} q^{\sum_j( \Lambda_{+j} \Lambda_{-j} + (1/2-j)\Lambda_{+j} - \frac{1}{2} \Lambda_{-j}^2)} \mathcal{C}(\cdot,\Lambda_+^T,\Lambda^-) \\
& = (-1)^{\sum_j (\Lambda_{+j} + \Lambda_{-j})} q^{\sum_j( \Lambda_{+j} \Lambda_{-j} - \frac{1}{2} \Lambda_{+j}^{T2} - \frac{1}{2} \Lambda_{-j}^2)} \mathcal{C}(\cdot,\Lambda_+^T,\Lambda^-)
\end{align}
This means that the character of completely degenerate representations of $\mathcal{W}_{1+\infty}$ that come from representations of $\mathcal{W}_N$ is equal (up to an overall normalization) to the topological vertex with two non-empty asymptotic labels. It is not possible to see representations with three asymptotic labels in $\mathcal{W}_N$, because the truncations to spins up to $N$ corresponds combinatorially to having maximum height of plane partition bounded by $N$. But since there is no such a restriction in $\mathcal{W}_{1+\infty}$, one is lead to conjecture that also the representations which have three asymptotic Young diagrams are completely degenerate representations of $\mathcal{W}_{1+\infty}$ and furthermore that their character is given (up to an overall normalization) by the full topological vertex with three non-empty Young diagrams. We will verify this conjecture at lower levels for the simplest representation which has all three asymptotics nontrivial - the representation $(\Box,\Box,\Box)$ - in section \ref{boxboxbox}.

The topological vertex has an important property which is its cyclicity,
\begin{equation}
\mathcal{C}(\lambda,\mu,\nu) = \mathcal{C}(\mu,\nu,\lambda) = \mathcal{C}(\nu,\lambda,\mu).
\end{equation}
This combined with the expression (\ref{pairingtotv}) for the pairing function lets us express the pairing function $\mathcal{P}(\mu,\nu)$ in terms of the quantum dimensions of the representations
\begin{equation}
\mathcal{P}(\mu,\nu) = q^{-\frac{1}{2}\sum_j (\mu_j-\nu_j)^2} \sum_{\eta,\tau,\sigma} q^{\frac{1}{2}\sum_j (\tau_j^2+\sigma_j^2)} N^{\mu}_{\eta\tau} N^{\nu}_{\eta\sigma} \mathcal{P}(\tau,\cdot) \mathcal{P}(\sigma,\cdot).
\end{equation}
Here the coefficients $N^{\lambda}_{\mu\nu}$ are the Littlewood-Richardson coefficients of $\mathfrak{su}(\infty)$. For Young diagrams with small number of boxes using this formula is quite efficient way of computing the characters.

\subsection{Identification of charges of primary states}
\label{seccharges}

\begin{figure}
\begin{center}
\includegraphics[scale=0.4]{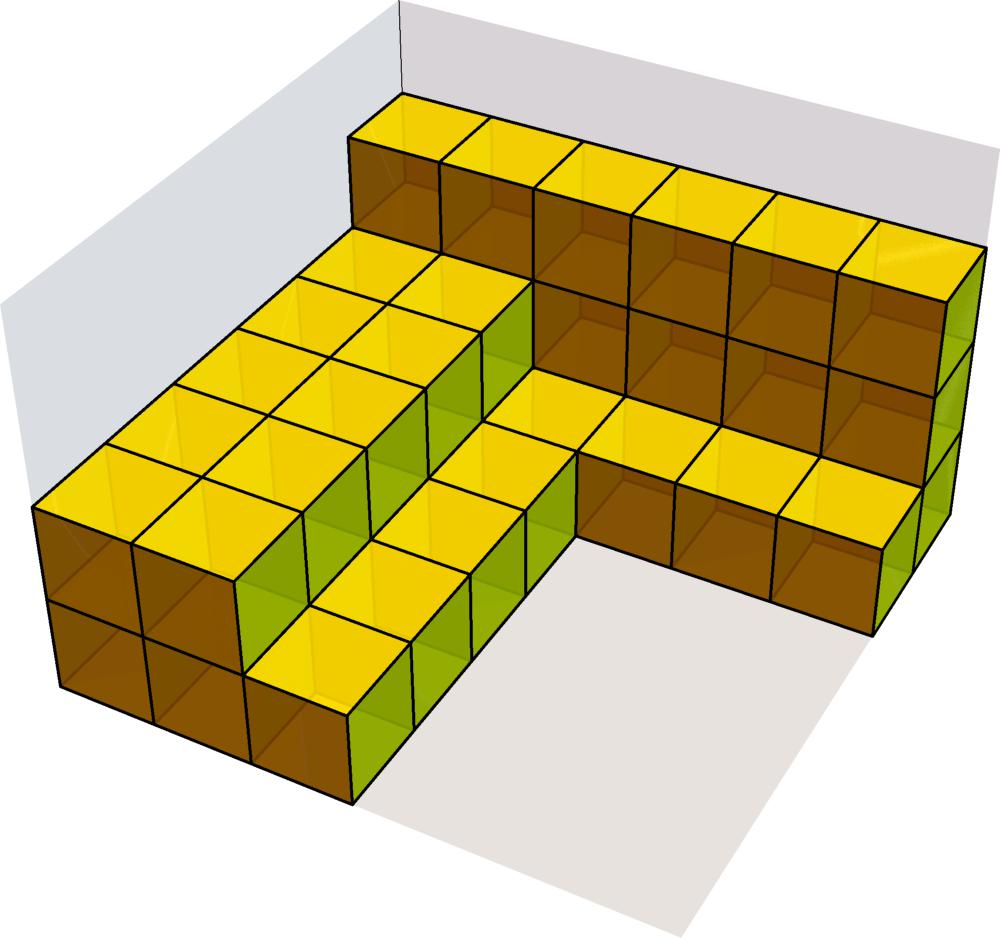}
\end{center}
\caption{The minimal configuration of boxes with asymptotic Young diagrams $(\Lambda_{13},\Lambda_{23}) = \left(\ydiagram{3,2},\ydiagram{2,1,1}\right)$.}
\label{figminconf}
\end{figure}

In this section we will turn to computation the highest weight charges (and in particular of $L_0$ eigenvalue) of the highest weight state of $\mathcal{W}_{1+\infty}$ representation labelled by two Young diagrams. These are the completely degenerate representations that come naturally from the representation theory of $\mathcal{W}_N$. As was described in the previous section, this highest weight state should combinatorially correspond to minimal configuration of boxes which asymptotically approach two Young diagrams, $\Lambda_{23}$ in $x_1 \to \infty$ direction and $\Lambda_{13}$ in $x_2 \to \infty$ direction. An example of such a configuration with asymptotics
\begin{equation}
\Lambda_{23} = \ydiagram{2,1,1}, \quad \quad \quad \Lambda_{13} = \ydiagram{3,2}
\end{equation}
is given in figure \ref{figminconf}.

Before looking for a general formula for $\psi(u)$ charges of the asymptotic configuration $(\Lambda_{13},\Lambda_{23})$, let us look at behaviour of $\psi_1$ and $\psi_2$ charges, which are in the CFT language identified with $J_0$ and $2L_0$. From the commutation relation (Y4') we see that each box in a partition gives a contribution $1$ to $L_0$ eigenvalue and vanishing contribution to $J_0$ eigenvalue. If we try to build an infinite tower of boxes in $x_1$ direction asymptotic to $(\infty,x_2,x_3)$, naively we would expect the $J_0$ eigenvalue of this configuration to vanish while $L_0$ eigenvalue of this configuration should be infinite. But the generating function of $\psi(u)$ charges provides a natural regularization of this infinite sum leaving us with a finite result for the $L_0$ eigenvalue of the highest weight state. Perhaps surprisingly, this regularized sum is precisely equal to the expression for the conformal dimension in $\mathcal{W}_{1+\infty}$ representation theory which is given in terms of the quadratic Casimir operators.

A tower of $n$ boxes in $x_1$ direction at coordinates $(x_2,x_3)$ in $2-3$ plane contributes to $\psi(u)$ by a factor (see \ref{psiproduct})
\begin{align}
\prod_{j=1}^n \varphi(u-j h_1-x_2 h_2-x_3 h_3) & = \frac{(u-x_2 h_2-x_3 h_3)(u-x_2 h_2-x_3 h_3 - h_1)}{(u-x_2 h_2-x_3 h_3 + h_2) (u-x_2 h_2-x_3 h_3 + h_3)} \times \\
\nonumber
& \times \frac{(u-x_2 h_2-x_3 h_3 - n h_1 + h_2) (u-x_2 h_2-x_3 h_3 - n h_1 + h_3)}{(u-x_2 h_2-x_3 h_3 - n h_1) (u-x_2 h_2-x_3 h_3 - n_h1 + h_2 + h_3)}.
\end{align}
Asymptotic expansion of this expression at $u \to \infty$ is
\begin{equation}
1 + \frac{2n\sigma_3}{u^3} + \cdots
\end{equation}
so we see that the contribution of this configuration to charge $\psi_1$ is zero while the contribution to the energy $L_0 = \frac{1}{2}\psi_2$ is $n$ as stated above. Let us now take the limit in the opposite order, first sending $n \to \infty$ at fixed value of $u$ (and other parameters) and only after that expanding at $u \to \infty$. The infinite products becomes simply
\begin{equation}
\prod_{j=1}^{\infty} \varphi(u-j h_1-x_2 h_2-x_3 h_3) = \frac{(u-x_2 h_2-x_3 h_3)(u-x_2 h_2-x_3 h_3 - h_1)}{(u-x_2 h_2-x_3 h_3 + h_2) (u-x_2 h_2-x_3 h_3 + h_3)}
\end{equation}
and the $u \to \infty$ expansion is now
\begin{equation}
\label{singlepile}
1 - \frac{\sigma_3}{h_1 u^2} - \frac{\sigma_3(h_1 + 2x_2 h_2 + 2x_3 h_3)}{h_1 u^3} + \cdots
\end{equation}
We see that two things happened. First of all, the charge $\psi_1$ is now no longer zero, but it gets contribution $-\frac{1}{h_1}$ from the single box at $x_1 \to \infty$. Second, there is a finite non-trivial contribution of this configuration to the $L_0$ eigenvalue.

\paragraph{Computation of $\psi(u)$ for two asymptotic labels}

Now we want to determine $\psi(u)$ for the minimal configuration of boxes with asymptotics $(\Lambda_{13},\Lambda_{23})$ like the one in the figure \ref{figminconf}. The configuration of boxes with two asymptotic Young diagrams $\Lambda_{13}$ and $\Lambda_{23}$ has layers at constant $x_3=j$ as shown in the figure \ref{twoyounglayer}. After combining the individual piles of boxes as in (\ref{singlepile}) in both $x_1$ and $x_2$ directions and removing the boxes in the intersection of two orthogonal piles that we counted twice, we find for the $\psi(u)$ eigenvalues of the full minimal (highest weight) configuration the expression
\begin{multline}
\label{hwpsi}
\psi(u) \ket{\Lambda_{13},\Lambda_{23}} = \frac{u+\psi_0 \sigma_3}{u} \prod_{j>0} \frac{(u-jh_3+h_3)(u-h_1 \Lambda_{13,j} -h_2 \Lambda_{23,j}-jh_3)}{(u-jh_3)(u-h_1\Lambda_{13,j}-h_2\Lambda_{23,j}-jh_3+h_3)} \ket{\Lambda_{13},\Lambda_{23}}.
\end{multline}
Note that in this product almost all the factors cancelled. If only a finite number of $\Lambda_{13,j}$ and $\Lambda_{23,j}$ are nonzero (which we generally assume), the infinite product reduces to a finite product. Furthermore, the dependence on $\Lambda_{13}$ and $\Lambda_{23}$ is only through the combination
\begin{equation}
\Lambda_j \equiv h_1 \Lambda_{13,j} + h_2 \Lambda_{23,j}
\end{equation}
and is in fact a symmetric function of variables $\Lambda_j + h_3 j$.

\begin{figure}[h]
\centering
\begin{tikzpicture}
\draw[thick] (0,2) -- (5,2) -- (5,4) -- (0,4) -- (0,2);
\draw[thick] (0,2) -- (0,0) -- (3,0) -- (3,4);
\draw[thick, decorate, decoration={brace, amplitude=10, raise=4}] (5,4) -- (5,2) node [midway, xshift=28] {$\Lambda_{23,j}$};
\draw[thick, decorate, decoration={brace, amplitude=10, raise=4, mirror}] (0,0) -- (3,0) node [midway, yshift=-20] {$\Lambda_{13,j}$};
\draw[thick, ->] (-0.3,2.5) -- (-0.3,1) node[midway, xshift=-8] {$x_2$};
\draw[thick, ->] (2,4.3) -- (4,4.3) node[midway, yshift=6] {$x_1$};
\end{tikzpicture}
\caption{$j$-th layer of a highest weight configuration of boxes with two asymptotic Young diagrams $\Lambda_{13}$ and $\Lambda_{23}$.}
\label{twoyounglayer}
\end{figure}

In particular, the first few charges are
\begin{equation}
\psi_j \ket{\Lambda_{13},\Lambda_{23}} = \psi_{\Lambda,j} \ket{\Lambda_{13},\Lambda_{23}}
\end{equation}
with
\begin{align}
\nonumber
\psi_{\Lambda,1} & = -\frac{1}{h_1 h_2} \sum_j \Lambda_j \\
\nonumber
\psi_{\Lambda,2} & = -\frac{1}{h_1 h_2} \sum_j \Lambda_j^2 - \frac{h_3}{h_1 h_2} \sum_j (2j-1) \Lambda_j - \psi_0 h_3 \sum_j \Lambda_j \\
\psi_{\Lambda,3} & = -\frac{1}{h_1 h_2} \sum_j \left( \Lambda_j^3 + h_3 (3j-2) \Lambda_j^2 + h_3^2 (3j^2-3j+1) \Lambda_j \right) \\
\nonumber
& + \frac{h_3}{h_1 h_2} \sum_{j<k} \Lambda_j \Lambda_k - \psi_0 h_3 \sum_j \left( \Lambda_j^2 + h_3 (2j-1) \Lambda_j \right).
\end{align}
The $\psi_1$ charge just counts the number of boxes of asymptotic Young diagrams,
\begin{equation}
\psi_{\Lambda,1} = -\frac{1}{h_1} \sum_{j} \Lambda_{23,j} -\frac{1}{h_2} \sum_{j} \Lambda_{13,j}.
\end{equation}
To compare $\psi_{\Lambda,2}$ to $L_0$ eigenvalue in $\mathcal{W}_{\infty}$, we first need to decouple the spin $1$ current. One way of doing it is by shifting the $\psi_1$ eigenvalue to zero using a shift in $u$. Equivalently we can just look at shift-invariant combination
\begin{equation}
\label{l0inf}
L_0^{(\infty)} = \frac{\psi_2}{2} - \frac{\psi_1^2}{2\psi_0}
\end{equation}
which is the Virasoro $L_0$ generator of $\mathcal{W}_{\infty}$ (after the spin $1$ current is decoupled). The eigenvalue of this element on the highest weight eigenstate $(\Lambda_{13},\Lambda_{23})$ is now
\begin{equation}
\label{htolambda}
-\frac{1}{2h_1 h_2} \sum_j \Lambda_j^2 - \frac{h_3}{2h_1 h_2} \sum_j (2j-1) \Lambda_j - \frac{\psi_0 h_3}{2} \sum_j \Lambda_j - \frac{1}{2\psi_0 h_1^2 h_2^2} \sum_{j,k} \Lambda_j \Lambda_k
\end{equation}
which will later be compared to $L_0$ eigenvalue of the highest weight state in $(\Lambda_+,\Lambda_-)$ representations of $\mathcal{W}_N$.

\paragraph{Highest weight charges of $\mathcal{W}_{\infty}$ representations}
Now we want to determine the other highest weight charges of completely degenerate representations of $\mathcal{W}_{\infty}$. We will follow the conventions of \cite{Prochazka:2014gqa}. First of all, we use the free-field representation of $\mathcal{W}_{1+N}$ on $N$ free bosons
\begin{equation}
J_j(z) J_k(w) \sim \frac{\delta_{jk}}{(z-w)^2} + reg.
\end{equation}
in terms of which the $\mathcal{W}_{1+N}$ generators $U_k(z)$ are
\begin{equation}
R(z) = \normord{\prod_{j=1}^N \left(\alpha_0 \partial + J_j(z) \right)} = \sum_{k=0}^N U_k(z) (\alpha_0 \partial)^{N-k}
\end{equation}
(the normal ordering here is not in fact necessary). Defining the highest weight state $\ket{\Lambda}$ by
\begin{align}
\nonumber
J_{j,0} \ket{\Lambda} & = \Lambda_j \ket{\Lambda} \\
J_{j,m} \ket{\Lambda} & = 0, \quad \quad \quad m>0
\end{align}
we can compute the $U_{k,0}$ eigenvalues when acting on $\ket{\Lambda}$. Following \cite{Bouwknegt:1992wg}, we can act with $R(z)$ on $\ket{\Lambda}$ and find that the leading terms are
\begin{equation}
\prod_{j=1}^N \left(\alpha_0 \partial + \Lambda_j z^{-1} \right) = \sum_{k=0}^N u_k(\Lambda) z^{-k} (\alpha_0 \partial)^{N-k}
\end{equation}
where we denoted the $U_{k,0}$ eigenvalues by $u_k(\Lambda)$. Acting now with this differential operator on $z^l, l=0, \ldots, N$ we find a set of equations
\begin{equation}
\label{ulambdaeqn}
\sum_{k=0}^N \alpha_0^{N-k} \left[ l \right]_{N-k} u_k(\Lambda) = \prod_{j=1}^N \left( \Lambda_j + \alpha_0 j - \alpha_0 (N-l) \right)
\end{equation}
Notice that these equations are degree $N$ polynomials in $l$ so if they are satisfied for $l=0,\ldots,N$, they are satisfied for any $l$. One way of solving these equations is by noticing that the left hand side is the binomial transform of the sequence
\begin{equation}
\alpha_0^k k! u_{N-k}(\Lambda)
\end{equation}
which can be easily inverted and we find
\begin{equation}
u_k(\Lambda) = \alpha_0^{k-N} \sum_{l=0}^{N-k} \frac{(-1)^{N-k-l}}{l! (N-k-l)!} \prod_{j=1}^N \left( \Lambda_j + \alpha_0 j - \alpha_0(N-l) \right).
\end{equation}
For example,
\begin{align}
\label{utolambda}
\nonumber
u_1(\Lambda) & = \sum_{j=1}^N \Lambda_j \\
u_2(\Lambda) & = \sum_{j<k} \Lambda_j \Lambda_k - \alpha_0 \sum_j (j-1) \Lambda_j \\
\nonumber
u_3(\Lambda) & = \sum_{j<k<l} \Lambda_j \Lambda_k \Lambda_l - \alpha_0 \sum_{j<k} (j+k-3) \Lambda_j \Lambda_k + \alpha_0^2 \sum_j (j-1)(j-2) \Lambda_j.
\end{align}
A more useful way of solving the equation for $u_k(\Lambda)$ is using the shifted symmetric functions introduced in \cite{ooshifted}. In particular, the authors introduce the shifted elementary symmetric functions
\begin{equation}
e_k^*(x_1, \ldots) \equiv \sum_{j_1 < j_2 < \cdots} (x_{j_1} + k-1) (x_{j_2} + k-2) \cdots x_{j_k}.
\end{equation}
We can define these functions for a finite number of variables, in which case they are polynomials in $x_j$ and satisfy the stability condition
\begin{equation}
e_k^*(x_1, \ldots, x_N) = e_k^*(x_1, \ldots, x_N, x_{N+1} = 0).
\end{equation}
Analogously to the case of symmetric polynomials and symmetric functions \cite{macdonaldsym}, this property allows us to consider also an infinite number of variables. The most important property of the shifted elementary symmetric functions for our purposes is the existence of the generating series
\begin{equation}
\prod_{j \geq 1} \frac{u + x_j - j + 1}{u-j+1} = \sum_{k \geq 0} \frac{e_k^*(x_1, \ldots)}{\left[ u \right]_k}.
\end{equation}
The equation (\ref{ulambdaeqn}) can be written in a similar form,
\begin{equation}
\sum_{k=0}^N \frac{u_k(\Lambda)}{(-\alpha_0)^k \left[u\right]_k} = \prod_{j=1}^N \frac{u - \frac{\Lambda_j}{\alpha_0} - j + 1}{u-j+1}
\end{equation}
with $u = N-l-1$. We see that we can make the identification
\begin{equation}
u_k(\Lambda) = (-\alpha_0)^k e_k^*\left( -\frac{\Lambda_1}{\alpha_0}, -\frac{\Lambda_2}{\alpha_0}, \ldots \right)
\end{equation}
and furthermore we have a nice generating series
\begin{equation}
\mathcal{U}_{\Lambda}(u) \equiv \sum_{k \geq 0} \frac{u_k(\Lambda)}{(-u)(-u + \alpha_0)\cdots(-u + (k-1)\alpha_0)} = \prod_{j \geq 1} \frac{u - \Lambda_j - \alpha_0 j + \alpha_0}{u - \alpha_0 j + \alpha_0}.
\end{equation}

\paragraph{Translation between $u_k(\Lambda)$ and $\psi_k$}
Let us now compare the expressions for highest weight charges in $\mathcal{Y}$ and in $\mathcal{W}_{1+\infty}$. We first identify the parameters
\begin{align}
\label{upsiparam}
\nonumber
h_1 h_2 & = -1 \\
\nonumber
h_3 & = \alpha_0 \\
\psi_0 & = N \\
\nonumber
\Lambda & = h_1 \Lambda_{13} + h_2 \Lambda_{23} = h_1 \Lambda_+ + h_2 \Lambda_-
\end{align}
and under this identification, the generating function of $\psi(u)$ charges (\ref{hwpsi}) becomes
\begin{equation}
\psi_{\Lambda}(u) = \frac{u-N\alpha_0}{u} \left(\prod_{j \geq 1} \frac{u - \Lambda_j - \alpha_0 j}{u - \alpha_0 j}\right) \left(\prod_{j \geq 1} \frac{u - \Lambda_j - \alpha_0 j + \alpha_0}{u - \alpha_0 j + \alpha_0}\right)^{-1}
\end{equation}
or
\begin{empheq}[box=\fbox]{equation}
\label{psitoucharges}
\psi_{\Lambda}(u) = \frac{u-N\alpha_0}{u} \, \frac{\mathcal{U}_{\Lambda}(u-\alpha_0)}{\mathcal{U}_{\Lambda}(u)}
\end{empheq}
By series expanding this function at $u = \infty$ we find explicit relations between Yangian and $\mathcal{W}$-algebra charges (when acting on a highest weight state). For the first few of them, we have
\begin{align}
\label{psitouchargesexp}
\nonumber
\psi_{\Lambda,0} & = N \\
\nonumber
\psi_{\Lambda,1} & = u_1 \\
\psi_{\Lambda,2} & = -2u_2 + u_1^2 - (N-1)\alpha_0 u_1 \\
\nonumber
\psi_{\Lambda,3} & = 3u_3 - 3u_1 u_2 + u_1^3 + 2(N-3)\alpha_0 u_2 - (N-1)\alpha_0 u_1^2 - (N-1)\alpha_0^2 u_1 \\
\nonumber
\psi_{\Lambda,4} & = -4u_4 + 4u_1 u_3 + 2u_2^2 - 4u_1^2 u_2 + u_1^4 - 3(N-6)\alpha_0 u_3 + (3N-8)\alpha_0 u_1 u_2 \\
\nonumber
& \quad - (N-1)\alpha_0 u_1^3 \quad + 2(3N-7)\alpha_0^2 u_2 - (N-1)\alpha_0^2 u_1^2 - (N-1)\alpha_0^3 u_1
\end{align}
This set of relations is triangular, so we can also determine $U$-charges in terms of $\psi$-charges uniquely, remembering that $u_0 \equiv 1$.

\paragraph{Energy}
Let us now compare the energy eigenvalue of the highest weight state in $(\Lambda_+,\Lambda_-)$ representation. First of all, we have the spectral shift invariant $\psi_2$ eigenvalue (\ref{l0inf})
\begin{equation}
h_{\infty} = \frac{\psi_{\Lambda,2}}{2} - \frac{\psi_{\Lambda,1}^2}{2\psi_{\Lambda,0}} = - u_2 - \frac{(N-1)\alpha_0}{2} u_1 + \frac{N-1}{2N} u_1^2
\end{equation}
which exactly matches the action of the zero mode of $T_{\infty}$ on the highest weight state. Plugging in the explicit expressions (\ref{utolambda}), we find
\begin{equation}
h_{\infty}(\Lambda_+,\Lambda_-) = \frac{1}{2} \sum_j \Lambda_j^2 + \frac{\alpha_0}{2} \sum_j (2j-1) \Lambda_j - \frac{N \alpha_0}{2} \sum_j \Lambda_j - \frac{1}{2N} \sum_{j,k} \Lambda_j \Lambda_k
\end{equation}
which is precisely (\ref{htolambda}) if we use the identifications (\ref{upsiparam}). For comparison with $\mathcal{W}_{\infty}$, we can use (\ref{lambdatoh}) together with (\ref{upsiparam}) and write
\begin{align}
\label{l0lplm}
\nonumber
h_{\infty}(\Lambda_+,\Lambda_-) & = \frac{\lambda_2}{2} \left[ \sum_j \Lambda_{+,j} - \frac{1}{\lambda_1} \sum_j \Lambda_{+,j}^2 - \frac{1}{\lambda_3} \sum_j (2j-1) \Lambda_{+,j} + \frac{1}{\lambda_1 \lambda_3} \sum_{j,k} \Lambda_{+,j} \Lambda_{+,k} \right] \\
\nonumber
& \quad + \frac{\lambda_1}{2} \left[ \sum_j \Lambda_{-,j} - \frac{1}{\lambda_2} \sum_j \Lambda_{-,j}^2 - \frac{1}{\lambda_3} \sum_j (2j-1) \Lambda_{-,j} + \frac{1}{\lambda_2 \lambda_3} \sum_{j,k} \Lambda_{-,j} \Lambda_{-,j} \right] \\
& \quad - \sum_j \Lambda_{+,j} \Lambda_{-,j} + \frac{1}{N} \sum_j \Lambda_{+,j} \Lambda_{-,j}
\end{align}
which is indeed the same formula as (\ref{hwn}) if we put
\begin{align}
\nonumber
\lambda_1 & = N \frac{p-p^\prime}{p^\prime} \\
\lambda_2 & = N \frac{p^\prime-p}{p} \\
\nonumber
\lambda_3 & = N.
\end{align}

\subsection{Schiffmann-Vasserot linearization}
Since both $\mathcal{Y}$ and the Schiffmann-Vasserot \cite{schifvas} algebra obtained as a limit of the spherical double affine Hecke algebras of $GL(n)$ both act on the equivariant cohomology of the instanton moduli spaces and interpolate between $\mathcal{W}_N$ algebras, one might suspect that they are in fact the same algebra. In this section we will introduce a useful non-linear transformation of the Baxter algebra generators that makes the connection between $\mathcal{Y}$ and $\mathbf{SH^c}$ manifest \cite{schifvas}.

The transformation that we are interested in linearizes the commutation relations between the elements of the Baxter subalgebra $\psi_j$ and $e_j$ and $f_j$ generators. Let us first define formal $x_k$ variables via
\begin{equation}
\psi(u) = 1 + \sigma_3 \sum_{j=0}^{\infty} \frac{\psi_j}{u^{j+1}} = \left( 1 + \sigma_3 \sum_{j=0}^{\infty} \frac{c_j}{u^{j+1}} \right) \prod_k \varphi(u-x_k).
\end{equation}
The $x_k$ variables play an auxiliary role, analogous to the role of the Chern roots in the splitting principle in the theory of characteristic classes. Up to a $c_j$-dependent prefactor, we wrote the generating function $\psi(u)$ in the product from analogous to the form of its action on (\ref{psiproduct}). As a second step, we define a new operators $\chi_j$ as the Newton power sum functions of these $x_k$. 

Parameters $c_k$ for $k \geq 2$ are arbitrary and will parametrize the transformation. Let us find a more explicit expression for this transformation. We have
\begin{align}
\prod_k \varphi(u-x_k) & = \exp \left[ \sum_k \log\frac{(u-x_k+h_1)(u-x_k+h_2)(u-x_k+h_3)}{(u-x_k-h_1)(u-x_k-h_2)(u-x_k-h_3)} \right] \\
\nonumber
& = \exp \left[ \sum_{l=1}^{\infty} \sum_k \sum_{m=0}^l \frac{1}{l u^l} {l \choose m} x_k^m \left(h_1^{l-m} + h_2^{l-m} + h_3^{l-m}\right) \left(1-(-1)^{l-m}\right) \right]
\end{align}
and as we explained above, we identify $\chi_k$ with the Newton power sum functions
\begin{equation}
\sum_k x_k^m \leftrightarrow \chi_k
\end{equation}
We can thus write the full form of the transformation
\begin{multline}
\label{svtrans}
1 + \sigma_3 \sum_{j=0}^{\infty} \frac{\psi_j}{u^{j+1}} = \left( 1 + \sigma_3 \sum_{j=0}^{\infty} \frac{c_j}{u^{j+1}} \right) \times \\
\times \exp \left[ \sum_{l=1}^{\infty} \sum_{m=0}^l \frac{1}{l u^l} {l \choose m} \chi_m P_{l-m}(h_1,h_2,h_3) \left(1-(-1)^{l-m}\right) \right].
\end{multline}
Here $P_l(x_1,x_2,x_3)$ is the $l$-th Newton power sum polynomial in three variables $x_1$, $x_2$, $x_3$. For this mapping to make sense as a formal power series expansion at $u \to \infty$, we must correctly choose $c_0$ and $c_1$ in order to get a correct value of the central elements $\psi_0$ and $\psi_1$. This means that we need to require
\begin{align}
c_0 & = \psi_0 \\
c_1 & = \psi_1.
\end{align}
All other $\chi_j$ charges are determined by expanding the relation (\ref{svtrans}). For the first few of them we find
\begin{align}
\nonumber
\chi_0 & = \frac{1}{2} \left( \psi_2 - c_2 \right) \\
\chi_1 & = \frac{1}{6} \left( \psi_3 - \sigma_3 \psi_0 \psi_2 + \sigma_3 \psi_0 c_2 - c_3 \right) \\
\nonumber
\chi_2 & = \frac{1}{12} \Big( \psi_4 - \sigma_3 \psi_0 \psi_3 - \sigma_3 \psi_1 \psi_2 + \sigma_3^2 \psi_0^2 \psi_2 + \sigma_2 \psi_2 \\
\nonumber
& \quad - \sigma_2 c_2 - \sigma_3^2 \psi_0^2 c_2 + \sigma_3 \psi_1 c_2 + \sigma_3 \psi_0 c_3 - c_4 \Big).
\end{align}
By construction, these new generators of the algebra have simple commutation relations with $e_j$ and $f_j$ generators (as well as with themselves),
\begin{align}
\nonumber
\left[ \chi_j, e_k \right] & = e_{j+k} \\
\left[ \chi_j, f_k \right] & = -f_{j+k} \tag{SV} \\
\nonumber
\left[ \chi_j, \chi_k \right] & = 0
\end{align}
as can also be verified for small $j$ directly. In the context of the representations of $\mathcal{Y}$, $\chi_j$ generators are additive in the boxes of the plane partitions (just as the generating function $\psi(u)$ was $\varphi$-multiplicative): from the commutation relation (SV) and action of $e_j$ generators (\ref{eansatz}) we find
\begin{equation}
\chi_j(\Lambda+\Box) = \chi_j(\Lambda) + h_{\Box}^j.
\end{equation}

Commutation relations (SV) (which followed from (Y4), (Y4'), (Y5), (Y5') and (Y0)) let us reduce the commutation relations (Y1) and (Y2) to only $\mathrm{(Y1)}_{00}$ or $\mathrm{(Y2)}_{00}$. In fact, using the symmetry
\begin{equation}
\mathrm{(Y1)}_{jk} = \mathrm{(Y1)}_{kj}
\end{equation}
and
\begin{equation}
0 = \left[ \chi_l, \mathrm{(Y1)}_{jk} \right] = \mathrm{(Y1)}_{j+l,k} + \mathrm{(Y1)}_{j,k+l}
\end{equation}
following from (SV) using the Jacobi identity, we see that only $\mathrm{(Y1)}_{00}$ is independent (and similarly for $\mathrm{(Y2)}_{00}$). A similar argument shows that also only one of identities (Y6) and (Y7), namely $\mathrm{(Y6)}_{000}$ and $\mathrm{(Y7)}_{000}$ is enough to generate all of them if we use (SV).

\paragraph{Comparison of $\mathcal{Y}$ and $\mathbf{SH^c}$}

Tsymbaliuk's presentation of $\mathcal{Y}$ \cite{tsymbaliukrev} makes the connection to $\mathbf{SH^c}$ evident. Definition 1.31 of \cite{schifvas} is very reminiscent of Yangian commutation relations, namely if we identify
\begin{align}
D_{1,k} & \sim e_k \\
D_{-1,k} & \sim f_k \\
E_{k} & \sim \psi_k \\
D_{0,k} & \sim \chi_{k-1}
\end{align}
we see that relations (1.70-1.72) of \cite{schifvas} are exactly of the form of (SV) and (Y3). (Y0) is satisfied in $\mathbf{SH^c}$ by construction. The relations in the subalgebra $\mathcal{Y}^+$, (Y1) and (Y6) are not explicitly written down in \cite{schifvas}, but they are given in \cite{arbesschif}. Equations (3.3) and (3.4) of \cite{arbesschif} are precisely the identities $\mathrm{(Y2)}_{00}$ and $\mathrm{(Y6)}_{000}$ which as explained above generate all (Y2) and (Y6) by application of (SV).

\subsection{Map between affine Yangian and $U$-basis of $\mathcal{W}_{1+\infty}$}
In this section we will represent the elements $e_0, f_0$ and $\psi_3$ generating $\mathcal{Y}$ in terms of $U$-basis mode operators of $\mathcal{W}_{1+\infty}$. In order to do so, we will compare the representation of both algebras on $N$ free bosons using the coproduct. Representing $\mathcal{Y}$ on one free boson and using the coproduct to combine $N$ of these representations to a $N$-boson representation of $\mathcal{Y}$ is the most basic way of producing interesting representations \cite{schifvas:2011, feigin2012quantum}.

\paragraph{$N$-boson representation of $\mathcal{Y}$ using $\Delta$}
We will use the coproduct $\Delta$ (\ref{newcoproduct}) to induce a representation of $\mathcal{Y}$ on $N$ free bosons from the one-boson representations of section \ref{freebosonrep}. We will map the parameters in the same way as in (\ref{upsiparam}). The generating function of the charges $\psi(u)$ of the vacuum state of one free boson is
\begin{equation}
\frac{u+\psi_0\sigma_3}{u} = \frac{u-\alpha_0}{u}
\end{equation}
We want the coproduct to preserve the charges of the highest weight state in the vacuum representation. In order to achieve this, we must make a spectral shift (change the $\widehat{\mathfrak{u}(1)}$ charge of the highest weight state). One possible way of doing this follows from the identity
\begin{equation}
\frac{u+N\psi_0\sigma_3}{u} = \frac{u+\psi_0\sigma_3}{u} \frac{u+2\psi_0\sigma_3}{u+\psi_0\sigma_3} \cdots \frac{u+N\psi_0\sigma_3}{u+(N-1)\psi_0\sigma_3}.
\end{equation}
This means that we can take the $j$-th free boson with $\widehat{\mathfrak{u}(1)}$-charge shifted from zero to
\begin{equation}
\psi_1^{(j)} = -(j-1) \sigma_3 \psi_0^{(j)2}
\end{equation}
To summarize, in order to find a suitable $N$ boson representation of $\mathcal{Y}$, we will proceed with the following steps:
\begin{enumerate}
\item find an expression for iterated coproduct $\Delta^{(N)}$ using the expressions (\ref{newcoproduct}) and (\ref{newcoproducts})
\item in $j$-th factor perform the spectral shift transformation (\ref{spectralshifteq}) with $q$ in the $j$-th factor equal to $-(j-1) \sigma_3 \psi_0^{(j)}$
\item represent the individual factors on free boson Fock space following section \ref{freebosonrep}
\end{enumerate}
Let us follow this procedure in the case of $\psi_0$. First of all, by iterating the formula for coproduct (\ref{newcoproducts}), we find  that the coproduct on $\psi_0$ is simply additive,
\begin{equation}
\Delta^{(N)}(\psi_0) = \sum_{j=1}^N \psi_0^{(j)}
\end{equation}
The spectral transformation has no effect since $\psi_0$ is invariant under it. The free boson representation of section (see section \ref{seccharges}) has $\psi_0 = 1$ so the final result is
\begin{equation}
\psi_0^{tot} = N
\end{equation}
consistently with conventions used in section \ref{seccharges}, see (\ref{upsiparam}).

Let us now turn to $\psi_1$. The iterated coproduct is now
\begin{equation}
\Delta^{(N)}(\psi_1) = \sum_{j=1}^N \psi_1^{(j)} + \sigma_3 \sum_{j<k} \psi_0^{(j)} \psi_0^{(k)}
\end{equation}
After the spectral shift we find
\begin{equation}
\sum_{j=1}^N \left( \psi_1^{(j)} - (j-1) \sigma_3 \psi_0^{(j)} \right) + \sigma_3 \sum_{j<k} \psi_0^{(j)} \psi_0^{(k)}.
\end{equation}
Representing the individual factors in terms of the free boson of section \ref{freebosonrep} we find finally
\begin{equation}
\psi_1^{tot} = \sum_{j=1}^N a_0^{(j)}.
\end{equation}
Proceeding similarly with other lower spin generators, we find
\begin{eqnarray}
\nonumber
e_0^{tot} & = & \sum_{j=1}^N a_{-1}{(j)} \\
\nonumber
e_1^{tot} & = & \frac{1}{2} \sum_{j=1}^N \sum_{r \in \mathbbm{Z}} a_{r}^{(j)} a_{-r-1}^{(j)} \\
f_0^{tot} & = & -\sum_{j=1}^N a_{1}^{(j)} \\
\nonumber
f_1^{tot} & = & -\frac{1}{2} \sum_{j=1}^N \sum_{r \in \mathbbm{Z}} a_{r+1}^{(j)} a_{-r}^{(j)} + \alpha_0 \sum_{j=1}^N (N+1-2j) a_1^{(j)} \\
\nonumber
\psi_2^{tot} & = & \sum_{j=1}^N \left( a_0^{(j)2} - \alpha_0 (N+1-2j) a_0^{(j)} + 2\sum_{r>0} a_{-r}^{(j)} a_r^{(j)} \right) \\
\end{eqnarray}
and most importantly
\begin{eqnarray}
\label{psi3tot}
\psi_3^{tot} & = & \sum_{j=1}^N \Big[ a_0^{(j)3} + 6 a_0^{(j)} \sum_{r>0} a_{-r}^{(j)} a_r^{(j)} + 3 \sum_{r,s>0} (a^{(j)}_{-r-s} a^{(j)}_r a^{(j)}_s + a^{(j)}_{-r} a^{(j)}_{-s} a^{(j)}_{r+s}) \\
\nonumber
& & -3\alpha_0 \sum_{r>0} r a^{(j)}_{-r} a^{(j)}_r - \frac{\alpha_0}{2} (2N+3-6j) a^{(j)2}_0 - \alpha_0 (2N+3-6j) \sum_{r>0} a^{(j)}_{-r} a^{(j)}_{r} \\
\nonumber
& & + \alpha_0^2 (N+1-2Nj+3j^2-3j) a_0^{(j)} \Big] - \frac{\alpha_0}{2} \sum_{j,k} a^{(j)}_0 a^{(k)}_0 - 6\alpha_0 \sum_{j>k} \sum_{r>0} r a^{(j)}_{-r} a^{(k)}_r.
\end{eqnarray}
For reference, we used the following formula for iterated coproduct of $\psi_3$ which can be derived from (\ref{newcoproduct}) and (\ref{newcoproducts}):
\begin{eqnarray}
\nonumber
\Delta^{(n)}(\psi_3) & = & \sum_{j=1}^n \psi_3^{(j)} + \sigma_3 \sum_{j \neq k} \psi_0^{(j)} \psi_2^{(k)} + \sigma_3 \sum_{j < k} \psi_1^{(j)} \psi_1^{(k)} + \sigma_3^2 \sum_{j<k, l \neq j,k} \psi_0^{(j)} \psi_0^{(k)} \psi_1^{(l)} \\
\nonumber
& & + \sigma_3^3 \sum_{j<k<l<m} \psi_0^{(j)} \psi_0^{(k)} \psi_0^{(l)} \psi_0^{(m)} + 6 \sigma_3 \sum_{j<k} \sum_{m>0} m J_m^{(j)} J_{-m}^{(k)}.
\end{eqnarray}

\paragraph{$N$-boson representation of $\mathcal{W}_{1+\infty}$}
The results of the previous section should now be compared to the representation of $\mathcal{W}_{1+\infty}$ on $N$ free bosons. Following the conventions of \cite{Prochazka:2014gqa}, we find
\begin{align}
\nonumber
U_{1,n} & = \sum_{j=1}^N a^{(j)}_n \\
U_{2,n} & = \sum_{1 \leq j < k \leq N} \sum_{m \in \mathbbm{Z}} a^{(j)}_m a^{(k)}_{n-m} - \alpha_0 \sum_{j=1}^N (j-1) (n+1) a^{(j)}_n \\
\nonumber
U_{3,0} & = \sum_{1 \leq j < k < l \leq N} \sum_{r+s+t=0} a^{(j)}_r a^{(k)}_s a^{(l)}_t - \alpha_0 \sum_{1 \leq j < k \leq N} \sum_{r \in \mathbbm{Z}} (j-1) (-r+1) a^{(j)}_{-r} a^{(k)}_r \\
\nonumber
& \quad - \alpha_0 \sum_{1 \leq j < k \leq N} \sum_{r \in \mathbbm{Z}} (k-2) (r+1) a^{(j)}_{-r} a^{(k)}_r + \alpha_0^2 \sum_{j=1}^N (j-1)(j-2) a^{(j)}_0
\end{align}
and in particular the zero mode of the total stress-energy tensor is
\begin{equation}
L_0 = \frac{1}{2} \sum_{j=1}^N a_0^{(j)2} + \sum_{j=1}^N \sum_{r>0} a_{-r}^{(j)} a_r^{(j)} - \frac{\alpha_0}{2} \sum_{j=1}^N (N+1-2j) a^{(j)}_0.
\end{equation}
We thus see immediately that
\begin{align}
\nonumber
\psi_2^{tot} & = 2 L_0 = -2 U_{2,0} -(N-1)\alpha_0 U_{1,0} + U_{1,0}^2 + 2\sum_{k>0} U_{1,-k} U_{1,k} \\
e_0^{tot} & = U_{1,-1} \\
\nonumber
f_0^{tot} & = -U_{1,1}.
\end{align}
It remains to identify $\psi_3$. In terms of $U$-fields, we define
\begin{empheq}[box=\fbox]{align}
\label{psi3tou}
\nonumber
\psi_3^{\mathcal{U}} & = 3 U_{3,0} - 3U_{1,0} U_{2,0} - 3 \sum_{m>0} \left( U_{1,-m} U_{2,m} + U_{2,-m} U_{1,m} \right) \\
\nonumber
& \quad + U_{1,0}^3 + 6 U_{1,0} \sum_{m>0} U_{1,-m} U_{1,m} \\
& \quad + 3 \sum_{m,n>0} \left[ U_{1,-m-n} U_{1,m} U_{1,n} + U_{1,-m} U_{1,-n} U_{1,m+n} \right] \\
\nonumber
& \quad - (N-1) \alpha_0 U_{1,0}^2 - (2N-3) \alpha_0 \sum_{m>0} U_{1,-m} U_{1,m} \\
\nonumber
& \quad - (N-1)\alpha_0^2 U_{1,0} + 2(N-3) \alpha_0  U_{2,0} -3\alpha_0 \sum_{m>0} m U_{1,-m} U_{1,m}
\end{empheq}
and expressing this in terms of bosonic oscillators, we find
\begin{align}
\label{coprodu}
\nonumber
\psi_3^{\mathcal{U}} & = \sum_{j=1}^N \Big[ a_0^{(j)3} + 6 a_0^{(j)} \sum_{r>0} a_{-r}^{(j)} a_r^{(j)} + 3 \sum_{r,s>0} (a^{(j)}_{-r-s} a^{(j)}_r a^{(j)}_s + a^{(j)}_{-r} a^{(j)}_{-s} a^{(j)}_{r+s}) \\
\nonumber
& \quad -3\alpha_0 \sum_{r>0} r a^{(j)}_{-r} a^{(j)}_r - \frac{\alpha_0}{2} (2N+3-6j) a^{(j)2}_0 - \alpha_0 (2N+3-6j) \sum_{r>0} a^{(j)}_{-r} a^{(j)}_{r} \\
& \quad + \alpha_0^2 (N+1-2Nj+3j^2-3j) a_0^{(j)} \Big] - \frac{\alpha_0}{2} \sum_{j,k} a^{(j)}_0 a^{(k)}_0 - 6\alpha_0 \sum_{j>k} \sum_{r>0} r a^{(j)}_{-r} a^{(k)}_r \\
\nonumber
& = \psi_3^{tot}
\end{align}
which is exactly the same expression as (\ref{psi3tot}). This identification passes many further non-trivial tests:
\begin{itemize}
\item It is easy to see that acting on the highest weight state, we get the correct charges as in (\ref{psitouchargesexp}).
\item For $N=1$ this reduces to one free boson representation studied before in section \ref{freebosonrep}.
\item Furthermore, one can compute the eigenvalues of this operator in the vacuum representation using the explicit commutation relations of $\mathcal{W}_{1+\infty}$ directly as in \cite{Prochazka:2014gqa} and find an exact match with expectations at least up to level $6$.
\item As one final check, one can determine $e_1$ and $f_1$ using $e_0, f_0$ and $\psi_3^{\mathcal{U}}$ and find
\begin{align}
e_1 & = -U_{2,-1} + \sum_{m>0} U_{1,-m} U_{1,m-1} = L_{-1} \\
f_1 & = U_{2,1} - \sum_{m>0} U_{1,1-m} U_{1,m} + (N-1)\alpha_0 U_{1,1} = -L_1
\end{align}
and these are precisely the generators of Virasoro subalgebra of Yangian found in section \ref{secvirasoro}.
\end{itemize}
Note that is was necessary to use the correct coproduct (\ref{newcoproduct}). Had we used the naive version (\ref{delta0}) (which is not a homomorphism of algebras), we would have missed the last term in (\ref{psi3tou}) which is the main non-trivial ingredient of the map.

\subsection{Representation $(\Box,\Box,\Box)$}
\label{boxboxbox}
Now we will illustrate the results obtained thus far on the completely degenerate representation labelled by $(\Box,\Box,\Box)$. This representation is the simplest one that has all three non-trivial asymptotic Young diagram labels and so in particular does not come from a completely degenerate representation of $\mathcal{W}_N$. We will compute the first few terms of the character in two different ways. First using the topological vertex and the combinatorial counting the boxes, and second by determining the rank of the Shapovalov form using the explicit formula for commutation relations of $\mathcal{W}_{1+\infty}$ in the quadratic $U$-basis.

\begin{figure}
\begin{center}
\includegraphics[scale=0.4]{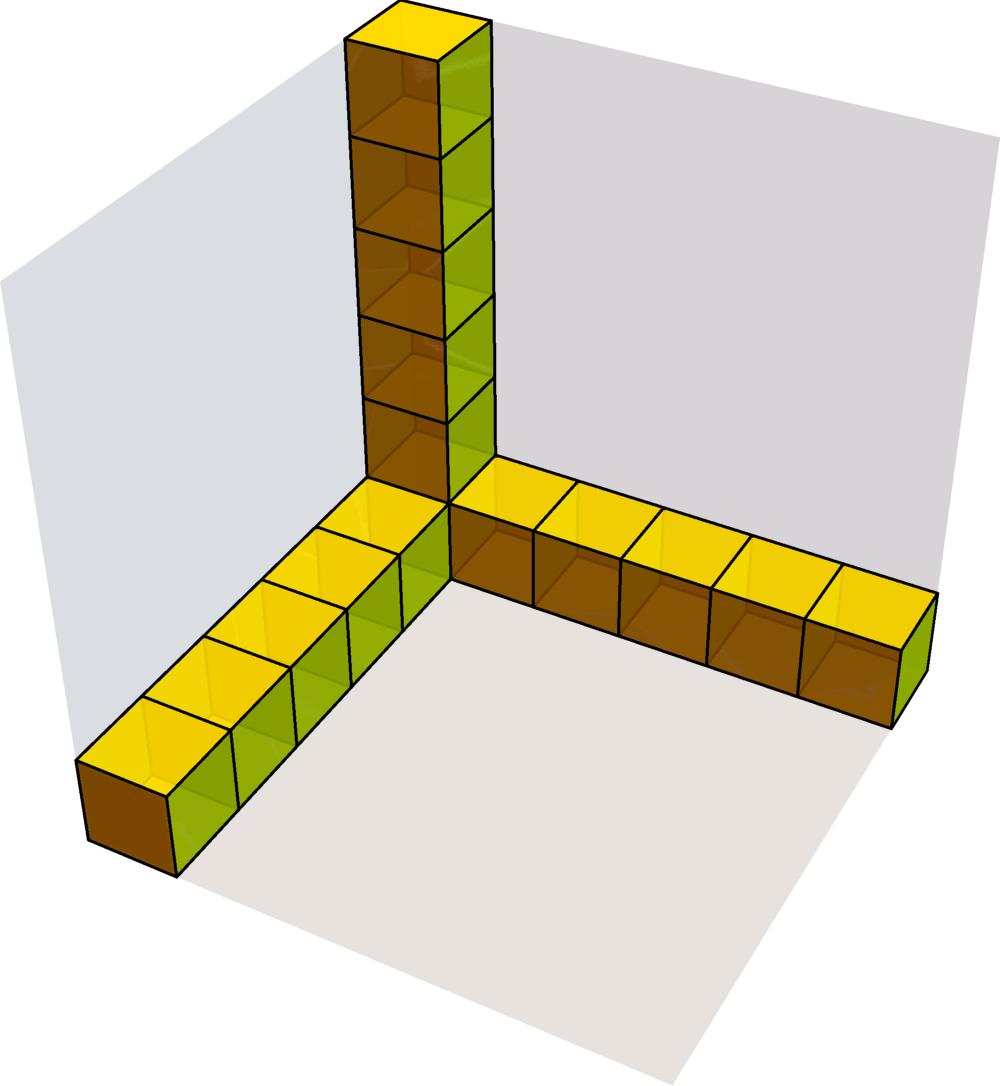}
\end{center}
\caption{A configuration of boxes representing the highest weight state in $(\Box,\Box,\Box)$ representation of $\mathcal{W}_{1+\infty}$. There are $3$ possible ways of adding a box, corresponding to $3$ states at level $1$. There are $9$ ways of adding $2$ boxes.}
\label{bbbhw}
\end{figure}

The highest weight state of this representation has three infinite columns of boxes along $x_1$, $x_2$ and $x_3$ axes as illustrated in figure \ref{bbbhw}. From this configuration of boxes, we can compute the generating function of $\psi$-charges of the highest weight state
\begin{align}
\nonumber
\psi_{(\Box,\Box,\Box)}(u) & = \frac{u+\psi_0 \sigma_3}{u} \varphi(u) \prod_{j=1}^{\infty} \Big[ \varphi(u-j h_1) \varphi(u-j h_2) \varphi(u-j h_3) \Big] \\
& = \frac{u+\psi_0 \sigma_3}{u} \frac{u^3}{(u+h_1)(u+h_2)(u+h_3)}.
\end{align}
We need to translate this to $U$-charges as in (\ref{psitoucharges}). It is easy to check that the equation
\begin{equation}
\frac{u^3}{(u+h_1)(u+h_2)(u+h_3)} = \frac{u^3}{(u^2+\sigma_3 u-1)(u-\sigma_3)} = \frac{\mathcal{U}_{(\Box,\Box,\Box)}(u+\sigma_3)}{\mathcal{U}_{(\Box,\Box,\Box)}(u)}
\end{equation}
determines the $U$-charges uniquely to be
\begin{equation}
u_j(\Box,\Box,\Box) = \frac{1+j \sigma_3^2}{j! \sigma_3^j} \prod_{k=1}^{j-2} \Big(1-k(k+1)\sigma_3^2\Big).
\end{equation}
Using now this formula for the charges of the highest weight state, we can determine the number of states in the irreducible representation at lower levels by computing the rank of the Shapovalov form. Using the explicit commutation relations of \cite{Prochazka:2014gqa}, we find
\begin{equation}
\Tr q^{L_0 - h_{(\Box,\Box,\Box)}} = 1 + 3q + 9q^2 + \cdots
\end{equation}
On the other hand, the topological vertex predicts
\begin{equation}
\Tr q^{L_0 - h_{(\Box,\Box,\Box)}} = \frac{1-q+q^2-q^3+q^4}{(1-q)^3} \prod_{j=1}^{\infty} \frac{1}{(1-q^j)^j} = 1 + 3q + 9q^2 + 22q^3 + 52q^4 + \ldots
\end{equation}
so we find a nice match at these low levels. This is a non-trivial verification of the various ingredients that went into the computation, namely
\begin{itemize}
\item the identification of $(\Box,\Box,\Box)$ as a degenerate representation of $\mathcal{W}_{1+\infty}$; at level $1$ we find $3$ states, so there is indeed an infinite number of null states at level $1$
\item the conjecture that the states in this representation are combinatorially given by plane partitions with asymptotics $(\Box,\Box,\Box)$, or equivalently that the character of this representation is given by the topological vertex
\item the identification of the highest weight $\psi(u)$ charge being given by the product (\ref{psiproduct})
\item the map between $\psi(u)$ charges and $u_j$ charges (\ref{psitoucharges})
\item the commutation relations of $\mathcal{W}_{1+\infty}$ generators as found in \cite{Prochazka:2014gqa}
\end{itemize}

\subsection{Non-generic values of parameters}
\label{specializations}

Up to now we have only considered irreducible representations of $\mathcal{W}_{1+\infty}$ or $\mathcal{Y}$ for generic values of parameters $\lambda_j$ or $h_j$. If we specialize these parameters, the representations on partitions can become reducible and additional singular vectors may appear. This was studied in the case of the vacuum representation of $\mathcal{W}_{1+\infty}$ in \cite{Prochazka:2014gqa}. If we specialize to the vacuum representation, it is a well-known fact that for $\mathcal{W}_N$ the vacuum representation becomes degenerate for values of the central charge which correspond to the $\mathcal{W}_N$ minimal models. In the case of $\mathcal{W}_{1+\infty}$ there is two-dimensional space of parameters, so the analogue of the discrete set of $\mathcal{W}_N$ minimal models are discrete codimension $2$ specializations of the parameters.

The null states in the vacuum representation of $\mathcal{W}_{1+\infty}$ are associated to rectangles in the coordinate planes of the plane partition space. The null state associated to a rectangle of dimensions $a_k \times a_j$ in $x_j-x_k$ plane appears at level
\begin{equation}
a_j a_k
\end{equation}
if
\begin{equation}
0 = (\lambda_j-a_k+1)(\lambda_k-a_j+1)-(a_j-1)(a_k-1) = \lambda_j \lambda_k - (a_j-1) \lambda_j - (a_k-1) \lambda_k,
\end{equation}
see \cite{Prochazka:2014gqa}. Translating this to $h_j$ parameters, this condition is
\begin{empheq}[box=\fbox]{equation}
0 = \psi_0 h_1 h_2 h_3 + a_1 h_1 + a_2 h_2 + a_3 h_3 = h_{\Box} + \psi_0 \sigma_3.
\end{empheq}
where $h_{\Box}$ are the weighted coordinates of the box $(a_1,a_2,a_3)$ of the rectangle that is furthest from the origin. At least one of $a_j$ must be equal to $1$. For example, at the level $2$ the Kac determinant contains a factor
\begin{equation}
(\psi_0 h_1 h_2+1)(\psi_0 h_1 h_3+1)(\psi_0 h_2 h_3+1).
\end{equation}
The three zeros as functions of $\psi_0$ correspond to three possible null states at level $2$, which are just the states
\begin{equation}
\ket{\ydiagram{2}}, \ket{\ydiagram{1,1}}, \ket{\begin{ytableau} \scriptstyle 2 \end{ytableau}}
\end{equation}
constructed previously in section \ref{vacrep}. In fact, from the inner products of the level $2$ basis (\ref{vaclevel2inner}) and the orthogonality of this basis we see that the state $\ket{\ydiagram{2}}$ becomes null exactly for
\begin{equation}
\psi_0 h_2 h_3 + 1 = 0
\end{equation}
and similarly for the other two states. At the next level, the situation is analogous and we have new null states if
\begin{equation}
0 = (\psi_0 h_1 h_2 + 2) (\psi_0 h_1 h_3 + 2) (\psi_0 h_2 h_3 + 2).
\end{equation}
These are the vanishing conditions associated to states
\begin{equation}
\ket{\ydiagram{3}}, \ket{\ydiagram{1,1,1}}, \ket{\begin{ytableau} \scriptstyle 3 \end{ytableau}}
\end{equation}
Interesting things happens at level $4$ where we can have for the first time a null state associated to a rectangle that is not a column: apart from the descendants of the lower level states, the new null states appear at level $4$ if
\begin{equation}
0 = (\psi_0 h_1 h_2 + 3) (\psi_0 h_1 h_3 + 3) (\psi_0 h_2 h_3 + 3) (\psi_0 h_1 h_2 - 1)(\psi_0 h_1 h_3 - 1)(\psi_0 h_2 h_3 - 1).
\end{equation}
The first three factors are associated to the states
\begin{equation}
\ket{\ydiagram{4}}, \ket{\ydiagram{1,1,1,1}}, \ket{\begin{ytableau} \scriptstyle 4 \end{ytableau}}
\end{equation}
while the other three factors are associated to the states
\begin{equation}
\ket{\ydiagram{2,2}}, \ket{\begin{ytableau} \scriptstyle 2 & \scriptstyle 2 \end{ytableau}}, \ket{\begin{ytableau} \scriptstyle 2 \\ \scriptstyle 2 \end{ytableau}}.
\end{equation}

Let us emphasize again that although there are null states associated to blocks of size $a_1 \times a_2 \times a_3$ with at least one of $a_j$ being one, nothing special happens for blocks which have all dimension different from $1$. Although this was checked only up to level $8$ in \cite{Prochazka:2014gqa} and later extended to level $9$, there are two reasons why this should hold at all levels:
\begin{itemize}
\item computing the Kac determinant up to level $8$ would already show the null state appearing associated to block of dimension $2 \times 2 \times 2$ - no such null state is present
\item the new null states would presumably produce new minimal models when intersecting with $\mathcal{W}_N$ curves \cite{Prochazka:2014gqa}; it is a well-known fact from the representation theory of $\mathcal{W}_N$ that the $\mathcal{W}_N$ minimal models are parametrized by two integers (which we can associate to dimensions of rectangles discussed above)
\end{itemize}

\begin{figure}
\begin{center}
\includegraphics[scale=0.5]{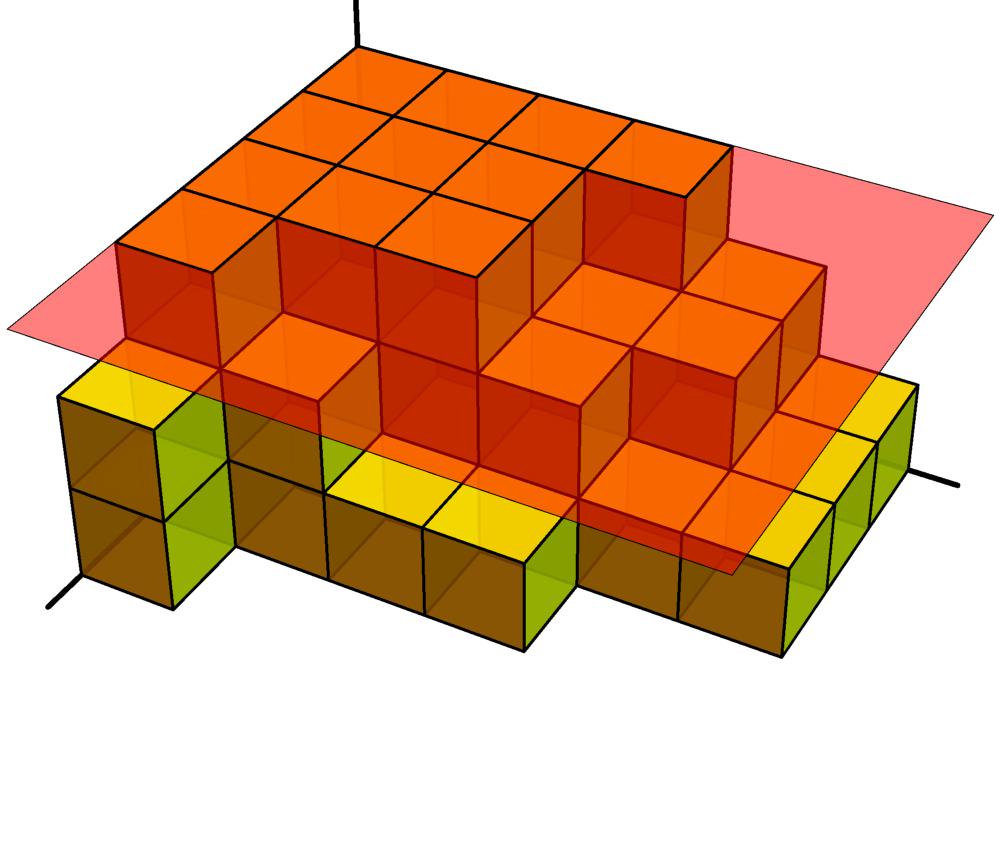}
\end{center}
\caption{An example of a level $51$ state in the vacuum representation of $\mathcal{Y}$ with parameters satisfying $\lambda_3=3$ (this is when $\mathcal{W}_{1+\infty}$ truncates to $\widehat{\mathfrak{u}(1)} \times \mathcal{W}_3$. Because of this constraint, the boxes are not allowed to grow in $x_3$ direction higher than to height $3$, which is shown by the red plane.}
\label{h3null}
\end{figure}

\paragraph{Null states associated to $\mathcal{W}_N$ truncation}
We can say more about the null states if an equation of the form
\begin{equation}
0 = \psi_0 h_1 h_2 + N
\end{equation}
is satisfied and otherwise the parameters are generic (we thus have a codimension $1$ specialization in the $h_j$-parameter space). In terms of $\lambda_j$ parameters, this is just the statement that
\begin{equation}
\lambda_3 = N
\end{equation}
and this in turn is the well-known condition for truncation from $\mathcal{W}_{1+\infty}$ to $\widehat{\mathfrak{u}(1)} \times \mathcal{W}_N$. The vacuum character in this case is given by the function counting plane partitions which have height at most $N$. This is consistent with the fact that the first state that violates this condition (a column of $N+1$ boxes in $x_3$ direction) becomes null. In the case of $N = 3$ this truncation is illustrated in the figure \ref{h3null}.

\paragraph{Codimension $2$ example - Lee-Yang minimal model}

\begin{figure}
\begin{center}
\includegraphics[scale=0.3]{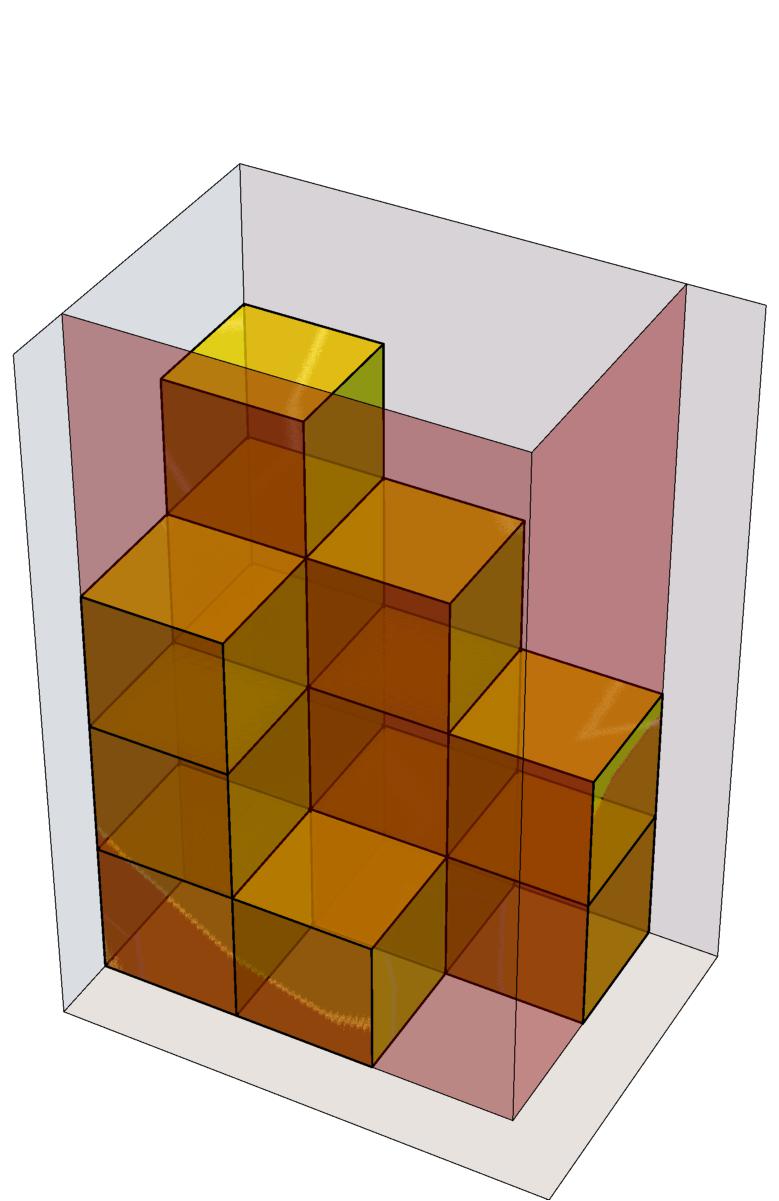}
\end{center}
\caption{An example of an allowed state at level $13$ in the vacuum representation if we have a geometric restriction by two planes.}
\label{h23null}
\end{figure}

In the case that we impose two independent conditions on $\lambda_j$ parameters, we arrive at one of the points of the discrete set of $\mathcal{W}_{1+\infty}$ minimal models. A simple example of this for a choice of the parameters
\begin{equation}
\lambda_1 = 3, \quad \lambda_2 = 2, \quad \lambda_3 = -\frac{6}{5}.
\end{equation}
Recall \cite{Prochazka:2014gqa} that $\lambda_j$ parameters are restricted to satisfy
\begin{equation}
\frac{1}{\lambda_1} + \frac{1}{\lambda_2} + \frac{1}{\lambda_3} = 0
\end{equation}
so we are only free to choose two of them \cite{Gaberdiel:2012ku,Prochazka:2014gqa}. In this case, the central charge of $\mathcal{W}_{\infty}$ subalgebra is
\begin{equation}
c_{\infty} = c_{1+\infty} - 1 = (\lambda_1-1)(\lambda_2-1)(\lambda_3-1) = -\frac{22}{5}
\end{equation}
so this corresponds to the Virasoro minimal model describing the Lee-Yang singularity \cite{DiFrancesco:1997nk}. The parameters of $\mathcal{Y}$ are determined to be
\begin{align}
\nonumber
h_1 & = -2 \\
\nonumber
h_2 & = -3 \\
h_3 & = 5 \\
\nonumber
\psi_0 & = \frac{1}{5}
\end{align}
(up to a rescaling $\alpha$ which we fixed by this choice). From the discussion of the null states above, one could naively expect that the states of the vacuum representation in this model are given by plane partitions with restrictions in two directions: there can be at most $3$ boxes on top of each other in $x_1$ direction and at most $2$ boxes in $x_2$ direction. This expectation is however not correct. It is easy to count the plane partitions with these geometric restrictions at lower levels. The projection in $x_3$ direction of the plane partition bounded in $x_1$ and $x_2$ directions looks like
\begin{center}
\begin{tikzpicture}[scale=0.50]
\draw[step=1.4cm,help lines,black] (0,0) grid (4.2,2.8);
\node at (0.7,0.6) {$n_{12}$};
\node at (2.1,0.6) {$n_{22}$};
\node at (3.5,0.6) {$n_{32}$};
\node at (0.7,2) {$n_{11}$};
\node at (2.1,2) {$n_{21}$};
\node at (3.5,2) {$n_{31}$};
\end{tikzpicture}
\end{center}
where $n_{jk}$ are the corresponding heights of the columns in $x_3$ direction. The counting function of these generalized partitions is
\begin{align}
\label{counting23}
\mathcal{P}\left( \ydiagram{3,3}, \cdot \right) & = \frac{1}{(1-q)(1-q^2)^2(1-q^3)^2(1-q^4)} \\
\nonumber
& \simeq 1 + q + 3q^2 + 5q^3 + 9q^4 + 13q^5 + 22q^6 + 30q^7 + 45q^8 + 61q^9 + 85q^{10} + \ldots
\end{align}
Note that the counting function is given by $q$-dimension (\ref{hookform}) for the diagram
\begin{equation}
\ydiagram{3,3}
\end{equation}
This holds more generally for counting of the plane partitions that fit into a rectangle of dimensions $a_1 \times a_2$. An example of a state of this form is given in figure \ref{h23null}.

The vacuum character of the Lee-Yang model is on the other hand given by (\ref{wncharacter}). We can consider it as $N=2$ minimal model or $N=3$ minimal model and the result is the same. There are also more compact formulas for this character like the one given in \cite{Fukuda:2015ura},
\begin{align}
\chi_{\mathrm{boson}}(q) \cdot \chi_{\mathrm{LY},0}(q) & \sim \frac{1}{\prod_{j=1}^{\infty} (1-q^j)} \frac{1}{\prod_{j=0}^{\infty} (1-q^{5j+2})(1-q^{5j+3})} \\
\nonumber
& \simeq 1 + q + 3q^2 + 5q^3 + 9q^4 + 14q^5 + 24q^6 + 36q^7 + 57q^8 + 84q^9 + 126q^{10} + \ldots
\end{align}
Comparing this to (\ref{counting23}) we see that there are \emph{more} states in the irreducible vacuum representation corresponding to the Lee-Yang model than the number of plane partitions limited by the two planes independently. This means that imposing the two algebraic conditions at the same time leads to a weaker constraint than imposing both geometric constraints independently. At the level of the states, there is a state at level $5$ of Lee-Yang that does not satisfy naive geometric restrictions. Using the explicit action of the box creation operators (\ref{eansatz}) together with (\ref{esol}) that we use to see if a given state is null or not, we find in fact two extra non-zero states at level $5$,
\begin{equation}
\ket{\ydiagram{4,1}} \quad \quad \mathrm{and} \quad \quad \ket{\ydiagram{3,1,1}}
\end{equation}
that violate the naive geometric constraint of fitting into the $3 \times 2$ rectangle. Using the Shapovalov form, one finds that in irreducible representation these two states are in fact proportional,
\begin{equation}
\ket{\ydiagram{4,1}} \sim \ket{\ydiagram{3,1,1}}.
\end{equation}
This only makes sense if in particular all the $\psi(u)$ charges of these two states are the same. That this is true is guaranteed by the important fact that the plane partition space now became periodic,
\begin{equation}
(h_1,h_2,h_3) \cdot (x_1 + \lambda_1, x_2, x_3) = (h_1,h_2,h_3) \cdot (x_1, x_2 + \lambda_2, x_3).
\end{equation}
From the point of view of $\mathcal{Y}$ the boxes at coordinates $(4,1,1)$ and $(1,3,1)$ are indistinguishable and should be considered as the same. We see that we are allowed to leave the rectangle
\begin{equation}
\ydiagram{3,3}
\end{equation}
but we must satisfy both of the conditions for a proper plane partition at the same time: to be allowed to add a box to $(4,1,x_3) \simeq (1,3,x_3)$ there must already be a box at both $(3,1,x_3)$ \emph{and} $(1,2,x_3)$. It turns out that the periodic extension of this rule is all that is needed to properly count all the state in this representation of $\mathcal{W}_{1+\infty}$.

To summarize, we can restrict (with no loss of generality because of the periodicity) to the plane partitions that are bounded in $x_2$ direction by $2$ rows. The states of the vacuum representation are then given by generalized partitions
\begin{center}
\begin{tikzpicture}[scale=0.50]
\draw[step=1.4cm,help lines,black] (0,0) grid (12.6,2.8);
\node at (0.7,0.6) {$n_{12}$};
\node at (2.1,0.6) {$n_{22}$};
\node at (3.5,0.6) {$n_{32}$};
\node at (4.9,0.6) {$n_{42}$};
\node at (6.3,0.6) {$n_{52}$};
\node at (7.7,0.6) {$n_{62}$};
\node at (9.1,0.6) {$n_{72}$};
\node at (10.5,0.6) {$n_{82}$};
\node at (12,0.6) {$\cdots$};
\node at (0.7,2) {$n_{11}$};
\node at (2.1,2) {$n_{21}$};
\node at (3.5,2) {$n_{31}$};
\node at (4.9,2) {$n_{41}$};
\node at (6.3,2) {$n_{51}$};
\node at (7.7,2) {$n_{61}$};
\node at (9.1,2) {$n_{71}$};
\node at (10.5,2) {$n_{81}$};
\node at (12,2) {$\cdots$};
\end{tikzpicture}
\end{center}
with the usual plane partition restrictions
\begin{align}
\nonumber
n_{j,1} & \geq n_{j,2} \\
n_{j,1} & \geq n_{j+1,1} \\
\nonumber
n_{j,2} & \geq n_{j+1,2}
\end{align}
and further conditions coming from the periodicity restrictions explained above:
\begin{equation}
n_{j,1} \geq n_{j-3,2}.
\end{equation}
The level of the state represented by this diagram is just
\begin{equation}
\sum_{j,k} n_{jk}
\end{equation}
Equivalently, we could use the periodicity conditions in the other way and instead consider the partitions
\begin{center}
\begin{tikzpicture}[scale=0.50]
\draw[step=1.4cm,help lines,black] (0,0) grid (4.2,8.4);
\node at (0.8,0.6) {$\cdots$};
\node at (2.2,0.6) {$\cdots$};
\node at (3.6,0.6) {$\cdots$};
\node at (0.7,2.0) {$n_{71}$};
\node at (2.1,2.0) {$n_{81}$};
\node at (3.5,2.0) {$n_{91}$};
\node at (0.7,3.4) {$n_{42}$};
\node at (2.1,3.4) {$n_{52}$};
\node at (3.5,3.4) {$n_{62}$};
\node at (0.7,4.8) {$n_{41}$};
\node at (2.1,4.8) {$n_{51}$};
\node at (3.5,4.8) {$n_{61}$};
\node at (0.7,6.2) {$n_{12}$};
\node at (2.1,6.2) {$n_{22}$};
\node at (3.5,6.2) {$n_{32}$};
\node at (0.7,7.6) {$n_{11}$};
\node at (2.1,7.6) {$n_{21}$};
\node at (3.5,7.6) {$n_{31}$};
\end{tikzpicture}
\end{center}
which is equivalent to the previous one. Relabelling $n_{41} \to n_{13}$ etc. which follows from the periodicity shows immediatelly that we are again getting the same conditions. The last thing that we need to do now is to count these generalized partitions. The result at lower levels is
\begin{equation}
1 + q + 3q^2 + 5q^3 + 9q^4 + 14q^5 + 24q^6 + 36q^7 + 57q^8 + 84q^9 + 126q^{10} + \ldots
\end{equation}
which exactly matches the vacuum character of the minimal model describing the Lee-Yang singularity.

We can rephrase the previous result in terms of partitions associated to a partially ordered set \cite{Fukuda:2015ura}. The system of inequalities above can be rewritten in terms of a poset (the arrows are pointing in the direction of non-increasing value)
\begin{center}
\begin{tikzpicture}
\matrix(m)[matrix of math nodes, row sep=2em, column sep=2em]
{& \mathbf{n_{21}} & \mathbf{n_{31}} & \mathbf{n_{41}} & \mathbf{n_{51}} & \mathbf{n_{61}} & \cdots \\
\mathbf{n_{11}} \\
& \mathbf{n_{12}} & \mathbf{n_{22}} & \mathbf{n_{32}} & \mathbf{n_{42}} & \mathbf{n_{52}} & \cdots \\};
\path[line width=0.35mm,->](m-2-1) edge (m-3-2);
\path[line width=0.35mm,->](m-2-1) edge (m-1-2);
\path[line width=0.35mm,->](m-3-2) edge (m-3-3);
\path[line width=0.35mm,->](m-3-2) edge (m-1-4);
\path[line width=0.35mm,->](m-3-3) edge (m-3-4);
\path[line width=0.35mm,->](m-3-3) edge (m-1-5);
\path[line width=0.35mm,->](m-3-4) edge (m-3-5);
\path[line width=0.35mm,->](m-3-4) edge (m-1-6);
\path[line width=0.35mm,->](m-3-5) edge (m-3-6);
\path[line width=0.35mm,->](m-3-5) edge (m-1-7);
\path[line width=0.35mm,->](m-3-6) edge (m-3-7);
\path[line width=0.35mm,->](m-1-2) edge (m-1-3);
\path[line width=0.35mm,->](m-1-2) edge (m-3-3);
\path[line width=0.35mm,->](m-1-3) edge (m-1-4);
\path[line width=0.35mm,->](m-1-3) edge (m-3-4);
\path[line width=0.35mm,->](m-1-4) edge (m-1-5);
\path[line width=0.35mm,->](m-1-4) edge (m-3-5);
\path[line width=0.35mm,->](m-1-5) edge (m-1-6);
\path[line width=0.35mm,->](m-1-5) edge (m-3-6);
\path[line width=0.35mm,->](m-1-6) edge (m-1-7);
\path[line width=0.35mm,->](m-1-6) edge (m-3-7);
\end{tikzpicture}
\end{center}
and this is precisely one of the results of \cite{Fukuda:2015ura}. Everything in this section generalizes straightforwardly to the case of minimal models where two of $\lambda_j$ parameters are coprime integers, which were called the minimal models with level-rank duality in \cite{Fukuda:2015ura}.

\paragraph{Non-trivial primary}

It is easy to generalize the considerations of this section to the irreducible completely degenerate representations different from the vacuum representation. Minimal models studied in \cite{Fukuda:2015ura} have the property that there is only one allowed asymptotic Young diagram. For the choice of parameters considered in this section, we have bounding planes at $x_1 = 3$ and $x_2 = 2$ and the non-trivial asymptotic growth is allowed only in the $x_3$ direction. The primaries of this model are parametrized by Young diagrams that fit into the $3 \times 2$ rectangle. The number of partitions that fit in $\lambda_1 \times \lambda_2$ rectangle is equal to
\begin{equation}
{\lambda_1 + \lambda_2 \choose \lambda_1} = {\lambda_1 + \lambda_2 \choose \lambda_2}
\end{equation}
(one is just counting say south or west going paths from one corner to the opposite one). In the Lee-Young case this would lead to $10$ possible primaries. But there are further identifications\footnote{By identification we mean here that there are primaries represented by different paths, which only differ by the value of the central element $\psi_1$ (the $\widehat{\mathfrak{u}(1)}$ charge). All the highest weight charges of $\mathcal{W}_{\infty}$ are the same. $\psi(u)$ function these highest weight states differ only by the spectral shift.}, following from the periodicity conditions as explained above. In the south-west path picture, we are allowed to choose the origin of the path arbitrarily. In the Lee-Yang case we have a $5$-fold identification and the result is that there are just $2$ primaries in the Lee-Yang model. The situation is depicted in the following figure:
\begin{center}
\begin{tikzpicture}[scale=0.50]
\filldraw[fill=yellow!60] (0,1) rectangle (1,2);
\filldraw[fill=yellow!60] (4,1) rectangle (5,2);
\filldraw[fill=yellow!60] (5,1) rectangle (6,2);
\filldraw[fill=yellow!60] (4,0) rectangle (5,1);
\filldraw[fill=yellow!60] (8,1) rectangle (9,2);
\filldraw[fill=yellow!60] (9,1) rectangle (10,2);
\filldraw[fill=yellow!60] (10,1) rectangle (11,2);
\filldraw[fill=yellow!60] (8,0) rectangle (9,1);
\filldraw[fill=yellow!60] (9,0) rectangle (10,1);
\filldraw[fill=yellow!60] (12,1) rectangle (13,2);
\filldraw[fill=yellow!60] (13,1) rectangle (14,2);
\filldraw[fill=yellow!60] (16,1) rectangle (17,2);
\filldraw[fill=yellow!60] (17,1) rectangle (18,2);
\filldraw[fill=yellow!60] (18,1) rectangle (19,2);
\filldraw[fill=yellow!60] (16,0) rectangle (17,1);
\filldraw[fill=yellow!60] (4,5) rectangle (5,6);
\filldraw[fill=yellow!60] (4,4) rectangle (5,5);
\filldraw[fill=yellow!60] (8,5) rectangle (9,6);
\filldraw[fill=yellow!60] (9,5) rectangle (10,6);
\filldraw[fill=yellow!60] (8,4) rectangle (9,5);
\filldraw[fill=yellow!60] (9,4) rectangle (10,5);
\filldraw[fill=yellow!60] (12,5) rectangle (13,6);
\filldraw[fill=yellow!60] (13,5) rectangle (14,6);
\filldraw[fill=yellow!60] (14,5) rectangle (15,6);
\filldraw[fill=yellow!60] (12,4) rectangle (13,5);
\filldraw[fill=yellow!60] (13,4) rectangle (14,5);
\filldraw[fill=yellow!60] (14,4) rectangle (15,5);
\filldraw[fill=yellow!60] (16,5) rectangle (17,6);
\filldraw[fill=yellow!60] (17,5) rectangle (18,6);
\filldraw[fill=yellow!60] (18,5) rectangle (19,6);
\draw[help lines] (0,4) grid (3,6);
\draw[thick,red] (3,6) -- (0,6) -- (0,4);
\draw[help lines] (4,4) grid (7,6);
\draw[thick,red] (7,6) -- (5,6) -- (5,4) -- (4,4);
\draw[help lines] (8,4) grid (11,6);
\draw[thick,red] (11,6) -- (10,6) -- (10,4) -- (8,4);
\draw[help lines] (12,4) grid (15,6);
\draw[thick,red] (15,6) -- (15,4) -- (12,4);
\draw[help lines] (16,4) grid (19,6);
\draw[thick,red] (19,6) -- (19,5) -- (16,5) -- (16,4);
\draw[help lines] (0,0) grid (3,2);
\draw[thick,red] (3,2) -- (1,2) -- (1,1) -- (0,1) -- (0,0);
\draw[help lines] (4,0) grid (7,2);
\draw[thick,red] (7,2) -- (6,2) -- (6,1) -- (5,1) -- (5,0) -- (4,0);
\draw[help lines] (8,0) grid (11,2);
\draw[thick,red] (11,2) -- (11,1) -- (10,1) -- (10,0) -- (8,0);
\draw[help lines] (12,0) grid (15,2);
\draw[thick,red] (15,2) -- (14,2) -- (14,1) -- (12,1) -- (12,0);
\draw[help lines] (16,0) grid (19,2);
\draw[thick,red] (19,2) -- (19,1) -- (17,1) -- (17,0) -- (16,0);
\node at (3.5,1) {$\simeq$};
\node at (7.5,1) {$\simeq$};
\node at (11.5,1) {$\simeq$};
\node at (15.5,1) {$\simeq$};
\node at (3.5,5) {$\simeq$};
\node at (7.5,5) {$\simeq$};
\node at (11.5,5) {$\simeq$};
\node at (15.5,5) {$\simeq$};
\node at (-2,1) {$h=-\frac{1}{5}:$};
\node at (-2,5) {$h=0:$};
\end{tikzpicture}
\end{center}

This means that up to identifications, we have only one primary different from the vacuum. With no loss of generality we can choose the corresponding asymptotic Young diagram to be $\Box$ in $x_3$ direction. All the state counting goes exactly as above, except for the fact that now we have effectively $n_{11} = \infty$ and the level associated to the diagram must be shifted not to include this infinite constant: the projection of the plane partition
\begin{center}
\begin{tikzpicture}[scale=0.50]
\draw[step=1.4cm,help lines,black] (0,0) grid (12.6,2.8);
\node at (0.7,0.6) {$n_{12}$};
\node at (2.1,0.6) {$n_{22}$};
\node at (3.5,0.6) {$n_{32}$};
\node at (4.9,0.6) {$n_{42}$};
\node at (6.3,0.6) {$n_{52}$};
\node at (7.7,0.6) {$n_{62}$};
\node at (9.1,0.6) {$n_{72}$};
\node at (10.5,0.6) {$n_{82}$};
\node at (12,0.6) {$\cdots$};
\node at (0.7,2) {$\infty$};
\node at (2.1,2) {$n_{21}$};
\node at (3.5,2) {$n_{31}$};
\node at (4.9,2) {$n_{41}$};
\node at (6.3,2) {$n_{51}$};
\node at (7.7,2) {$n_{61}$};
\node at (9.1,2) {$n_{71}$};
\node at (10.5,2) {$n_{81}$};
\node at (12,2) {$\cdots$};
\end{tikzpicture}
\end{center}
must satisfy the same inequalities as above and the corresponds to a state at level
\begin{equation}
\sum_{\substack{j,k \\ (j,k) \neq (1,1)}} n_{jk}.
\end{equation}
Counting these states, we find the character
\begin{equation}
1 + 2q + 4q^2 + 7q^3 + 13q^4 + 21q^5 + 35q^6 + 54q^7 + 84q^8 + 126q^9 + 188q^{10} + \ldots
\end{equation}
which needless to say is equal to the correct character \cite{Fukuda:2015ura} up to an overall normalization,
\begin{equation}
\chi_{\mathrm{boson}}(q) \cdot \chi_{\mathrm{LY},-\frac{1}{5}}(q) \sim \frac{1}{\prod_{j=1}^{\infty} (1-q^j)} \frac{1}{\prod_{j=0}^{\infty} (1-q^{5j+1})(1-q^{5j+4})}.
\end{equation}
We can again associate a poset to the system of inequalities implied by the box counting problem and we find the diagram
\begin{center}
\begin{tikzpicture}
\matrix(m)[matrix of math nodes, row sep=2em, column sep=2em]
{& \mathbf{n_{21}} & \mathbf{n_{31}} & \mathbf{n_{41}} & \mathbf{n_{51}} & \mathbf{n_{61}} & \cdots \\
\mathbf{\infty} \\
& \mathbf{n_{12}} & \mathbf{n_{22}} & \mathbf{n_{32}} & \mathbf{n_{42}} & \mathbf{n_{52}} & \cdots \\};
\path[line width=0.35mm,->](m-2-1) edge (m-3-2);
\path[line width=0.35mm,->](m-2-1) edge (m-1-2);
\path[line width=0.35mm,->](m-3-2) edge (m-3-3);
\path[line width=0.35mm,->](m-3-2) edge (m-1-4);
\path[line width=0.35mm,->](m-3-3) edge (m-3-4);
\path[line width=0.35mm,->](m-3-3) edge (m-1-5);
\path[line width=0.35mm,->](m-3-4) edge (m-3-5);
\path[line width=0.35mm,->](m-3-4) edge (m-1-6);
\path[line width=0.35mm,->](m-3-5) edge (m-3-6);
\path[line width=0.35mm,->](m-3-5) edge (m-1-7);
\path[line width=0.35mm,->](m-3-6) edge (m-3-7);
\path[line width=0.35mm,->](m-1-2) edge (m-1-3);
\path[line width=0.35mm,->](m-1-2) edge (m-3-3);
\path[line width=0.35mm,->](m-1-3) edge (m-1-4);
\path[line width=0.35mm,->](m-1-3) edge (m-3-4);
\path[line width=0.35mm,->](m-1-4) edge (m-1-5);
\path[line width=0.35mm,->](m-1-4) edge (m-3-5);
\path[line width=0.35mm,->](m-1-5) edge (m-1-6);
\path[line width=0.35mm,->](m-1-5) edge (m-3-6);
\path[line width=0.35mm,->](m-1-6) edge (m-1-7);
\path[line width=0.35mm,->](m-1-6) edge (m-3-7);
\end{tikzpicture}
\end{center}
which is just like the one in \cite{Fukuda:2015ura} if we leave out the $\infty$ at $n_{11}$ position.

\section{Summary and discussion}

In this article we have studied the connection between the non-linear $\mathcal{W}_{1+\infty}$ introduced in \cite{Gaberdiel:2012ku} and the Yangian $\mathcal{Y}$ of affine $\mathfrak{gl}(1)$ introduced in \cite{schifvas,Maulik:2012wi,tsymbaliukrev} and translated many properties between the two pictures. Each of the two descriptions of the algebra has its advantages. The Yangian picture is very useful for studying the representation theory. It gives us a natural infinite-dimensional commutative subalgebra that can be explicitly diagonalized in the representation spaces. The representation theory in the Yangian picture reduces to a large extent to the combinatorics of box-counting. The null states in completely degenerate representations that appear if we tune the parameters of the algebra have nice combinatorial interpretation in terms of the configurations of boxes. $\mathcal{W}_{1+\infty}$ on the other hand is a chiral algebra of the conformal field theory. This means that it automatically brings the notions of the local fields and their correlation functions to the picture. The Heisenberg and Virasoro algebra are obvious subalgebras of $\mathcal{W}_{1+\infty}$. In this picture, we can also directly study the fusion, the action of the modular group coming from the geometry of the underlying Riemann surface.

\paragraph{Interpolating algebra, integrability}
Both $\mathcal{W}_{1+\infty}$ and the Schiffmann-Vasserot algebra $\mathbf{SH^c}$ were constructed as interpolating algebras for a discrete series of algebras associated to the Dynkian diagram $A_{N-1}$. In the case of $\mathcal{W}_{1+\infty}$ these are the $\mathcal{W}_N$ algebras, while for $\mathbf{SH^c}$ they are the spherical degenerate double affine Hecke algebras of $GL(N)$. It would be interesting to see if also the usual Yangians associated to $GL(N)$ can be related in natural way in $\mathcal{Y}$. Because of the boson-fermion equivalence and $GL(\infty)$ picture of fermion CFT, it seems plausible that such construction could go through. Yangians or $\mathcal{W}$-algebras are often underlying symmetry algebras of integrable models. It would be nice to see a bigger algebra such as $\mathcal{Y}$ unifying various classes of integrable models of $A_N$ type. For the connection of $\mathcal{Y}$ to spin chains see \cite{Maulik:2012wi,Zhu:2015nha,Feigin:2015raa}.

\paragraph{CFT properties}
Given a new class of representations $(\Lambda_{12},\Lambda_{23},\Lambda_{31})$ of $\mathcal{W}_{1+\infty}$, it would be interesting to study their fusion, the modular transformation properties of their characters and other basic conformal field theory properties. In particular, the construction of modular-invariant partition functions of $\mathcal{W}_{1+\infty}$ would be useful in the context of Gaberdiel-Gopakumar duality.

\paragraph{New minimal models}
The algebra $\mathcal{W}_{1+\infty}$ is expected to have a larger class of minimal models than those coming from $\mathcal{W}_N$ \cite{Prochazka:2014gqa}. It would be interesting to understand the characters of their representations, fusion, modular transformation properties. The plain partition picture can be generalized to all $\mathcal{W}_N$ minimal models and all the $\mathcal{W}_N$ minimal model characters have a simple box-counting interpretation. But the generalization of this to the new $\mathcal{W}_{1+\infty}$ does not seem to be so straightforward.

\paragraph{Topological strings}
Apart from the topological vertex, there is also its refinement which does not treat all three directions in the plane partition space equally - the refined topological vertex \cite{Awata:2005fa, Iqbal:2007ii}. A special direction is chosen along one of the three coordinate axes. There is a continuous generalization of the choice of the special direction in the form of the index vertex \cite{Nekrasov:2014nea}. It would be nice to see if any of these refinements have any simple $\mathcal{W}_{1+\infty}$ interpretation. Since $\mathcal{Y}$ has the infinite-dimensional abelian subalgebra $\mathcal{B}$, one could easily think of large class of refinements of the characters themselves.

\section*{Acknowledgments}
I would like to thank Rajesh Gopakumar, Renann Lipinski-Jusinskas, Mat\v{e}j Kudrna, Yutaka Matsuo, Elli Pomoni, Sylvain Ribault, Olivier Schiffmann, Alessandro Torrielli, Alexander Tsymbaliuk and especially to Vasily Pestun and Martin Schnabl for useful discussions.

I am grateful to Institut des Hautes Études Scientifiques for hospitality during the early stages of this work. I thank to the Centro de Ciencias de Benasque Pedro Pascual for providing stimulating environment while this work was in progress.

This research was supported by the Grant Agency of the Czech Republic under the grant P201/12/G028 and in parts by the DFG Transregional Collaborative Research Centre TRR~33 and the DFG cluster of excellence Origin and Structure of the Universe.

\bibliography{winfyang}

\end{document}